\documentclass[12pt]{article}

\usepackage{graphicx}
\usepackage{amssymb,amsmath,bm,color}
\newcommand{\fsi}{FSI }
\newcommand{\degr}{$^o\,$}
\newcommand{\rts}{ \sqrt s}
\usepackage{epstopdf}

\begin{document}

\thispagestyle{empty}

\title{Search for Quasi Bound $\eta$ Mesons}


\author{\textbf{H. Machner\footnote{hartmut.machner@uni-due.de}}\\Fachbereich Physik, Univ. Duisburg-Essen,\\ Lotharstr. 1, 47048 Duisburg, Germany}

\maketitle

\begin{abstract}
The search for a quasi bound $\eta$ meson in atomic nuclei is reviewed. This tentative state is studied theoretically as well as experimentally. The theory starts from elastic $\eta$ nucleon scattering which is derived from production data within some models. From this interaction the $\eta$ nucleus interaction is derived. Model calculations predict binding energies and widths of the quasi bound state. Another method is to derive the $\eta$ nucleus interaction from excitation functions of $\eta$ production experiments. The $s$ wave interaction is extracted from such data via final state interaction theorem. We give the derivation of $s$ wave amplitudes in partial wave expansion and in helicity amplitudes and their relation to observables. Different experiments extracting the final state interaction are discussed as are production experiments. So far only three experiments give evidence for the existence of the quasi bound state: a pion double charge exchange experiment, an effective mass measurement, and a transfer reaction at recoil free kinematics with observation of the decay of the state.
\end{abstract}



\section{Introduction}\label{sec:Introduction}

Hadrons are composite particles which interact via the strong
interaction. They can be grouped into two classes according to their
valence quark substructure: into mesons ($q\bar q$) and baryons
($qqq$). Hence there are three different types of interactions:
\begin{itemize}
  \item baryon-baryon
  \item meson-baryon
  \item meson-meson.
\end{itemize}
In the nucleon-nucleon case, the interaction leads to bound or
almost bound states and we have many different nuclei. The same is
true if a nucleon is replaced by a hyperon. However, the
baryon-baryon interaction in matter differs from the free one. Also
three-body forces seem to play a role there \cite{Hammer13}. For
such forces the next type of interaction is involved: $\pi$-nucleon
interactions with excitations of the $\Delta$ nucleon resonance
\cite{Weinberg92}. The third type is again dominated by intermediate
resonances. Low energy $\pi-\pi$ scattering is a good testing ground
for chiral perturbation theory \cite{Colangelo01}. In this review we
are dealing with the second type of interactions namely
meson-nucleon where the meson is bound in nuclei. Such interactions
can be studied by scattering of the mesons. Especially rich
information can be gained from bound states. Here we will
concentrate on the $\eta$ meson. Why is the $\eta$ meson special? It
has the same quantum numbers as the $\pi^0$: $J^\pi = 0^-$. But it
differs in isospin ($T=0$ instead of $T=1$) as well as in mass:
547.862$\pm$0.018 MeV compared to 134.9766$\pm$0.0006 MeV
\cite{PDG10}. This may point to a different internal structure.
Since the $\eta$ can decay both via the strong as well as the
electromagnetic interaction its lifetime $\tau_\eta=(5.02\pm
0.19)\times 10^{-19}$ s is almost two orders of magnitude shorter
than the neutral pion: $\tau_{\pi^0}=(8.52\pm 0.18)\times 10^{-17}$
s. Both mesons have of course a $q\bar q$ structure which has to be
different to assure the different masses. One believes that both
mesons are mixtures of the pure quantum states \cite{Coon86},
\cite{Magiera00}. A mixing angle of $6\pm 5$ mrad has been reported
\cite{Abdel-Bary03} which is much smaller than the $\eta-\eta'$
mixing angle \cite{Cheung84}. Since the interaction between the
hadrons is a residual one one may expect the same strength,
independent of the different meson type. The scattering lengths of
the pion-nucleon interaction is rather small and as a result the
strong interaction shift in the $1s$ state of pionic atoms is
repulsive. Contrary to this, the $\eta$-nucleon interaction at small
momenta is attractive and rather strong. This was first pointed out
by Bhalerao and Liu \cite{Bhalerao85} and later applied by Haider
and Liu \cite{Haider_Liu86} to predict quasi bound $\eta$ mesons in
atomic nuclei for mass numbers $A\geq 12$. In the following text we
apply the standard sign convention in meson physics
\cite{Goldberger-Watson} for the $s$ wave scattering parameters
\begin{equation} p \cot \delta_0 = \frac{1}{a} + \frac{1}{2} r_0
p^2\,, \label{definition} \end{equation} with $p$ the $\eta$
momentum, $\delta_o$ the $s$ wave phase shift, $a$ the scattering
length and $r_0$ the effective range. For a real attractive
potential $a_{r} < 0$ means binding. Contrary to the $\pi-N$ systems
where the scattering length is real at very small energies here the
$\eta N\to \pi N$ channel is always open and hence the scattering
length is complex. From such large  values for the scattering length
$a(\eta N)$, Haider and Liu\cite{Haider_Liu86} have shown that
$\eta$ can be bound in nuclei with  A $\ge$ 12. Other groups have
also found similar results \cite{Garcia-Recio02, Hayano99,
Sofianos97, Friedman13}. In the following text we frequently use the
term bound state instead of the more strict quasi bound state. This
is common in the literature.

The text is organised as follows. First we will discuss some theoretical approaches and predictions. Then we will review experiments searching for the existence of such $\eta$ nucleus quasi bound states.

\section{Theoretical Considerations}\label{sec:Theoret-Considerations}

\subsection{Model Calculations for Quasi Bound States}\label{sub:Model-Calcula-Quasi-Bound-States}

\subsubsection{Prelude}\label{sub:Prelude}

\begin{figure}[h!]
\begin{center}
\includegraphics[width=0.8\textwidth]{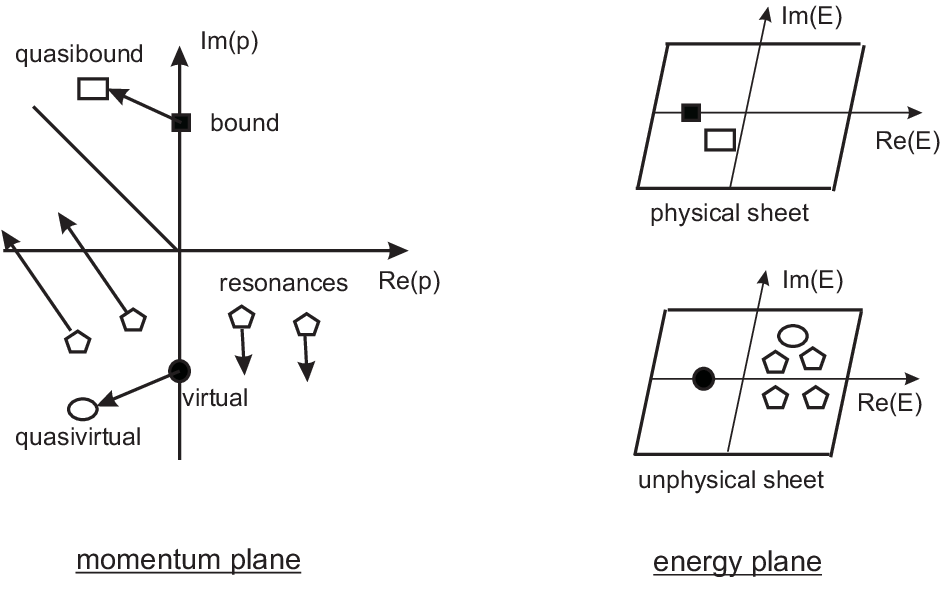}
\caption{The complex momentum plane (left) and
the $E$ planes (right).
Energy and momentum are non relativistically connected via $E=p^2/2\mu$ with $p=p_r+ip_i$ being complex. For $p_i>0$ we have the physical sheet and for $p_i<0$ the unphysical sheet. From the definition we get $E_r=(p_r^2-p_i^2)/2\mu$ and $E_i=p_r p_i/\mu$. Symbols at the end of the arrows (full symbols) are states for a real potential. The arrows indicate how these states move with increasing imaginary potential.}
\label{Fig:complex_plane}
\end{center}
\end{figure}

A state is called a bound state in the usual sense when the sum of its constituent masses is larger than the mass of the composite. In non relativistic quantum mechanics binding is represented by an attractive potential and the state is a solution of the radial Schr\"{o}dinger equation. These solutions lie on the imaginary axis in the momentum plane with $\text{Im}(p)=p_i>0$ (see Fig. \ref{Fig:complex_plane}).
However, a possible $\eta$ bound state is not stable since always the interaction
\begin{equation}
\eta+N\to \pi+N'
\end{equation}
with a nucleon $N$ is possible. If the $\eta$ bound state was in a $s$ state the energy of the final state is
\begin{equation}
m_\eta+m_N-B_\eta=m_\pi+m_{N'}+T_\pi+T_{N'}
\end{equation}
with $T$ the kinetic energies in the final state and $B_\eta$ the binding energy. Here we have neglected Fermi motion of the nucleon and the recoil of the residual nucleus. This leads to $T_\pi\approx 317$~MeV and $T_{N'}\approx 47.3$~MeV where we have assumed a binding energy of $10$ MeV. These energies are clearly too large for the two final state particles to stay in the nucleus. The resonance $S_{11}(1535)$ dominates the $\eta+N$ interaction even close to threshold ($\sqrt s_0=1487$~MeV), since the width of the resonance is wide: its half width is $\approx 75$~MeV \cite{PDG10} and thus reaches down to the threshold. We will come back to this point in Section \ref{sssec:eta-N-scatter}. Because of the possible decay of the state it is a quasi bound state and this fact is accounted for by a complex potential. A quasi bound state is located in the second quadrant of the complex momentum plane in Fig. \ref{Fig:complex_plane}. For a real potential resonances are in the third and fourth quadrant. When the imaginary potential is switched on they move down in the fourth quadrant and move up in the third quadrant and may reach for sufficiently strong imaginary potential the second quadrant \cite{Cassing82}. The boundary between a resonance and a quasi bound state is the 45\degr line with the quasi bound state above this line. The task is now to produce a complex potential for elastic scattering $\eta N\to \eta N$, construct from this a complex $\eta A\to \eta A$ potential and then search for poles in the upper part of the second quadrant.

\subsubsection{$\eta N$ scattering}\label{sssec:eta-N-scatter}

The $\eta$-nucleon scattering length $a(\eta N) $ or more generally
the matrix $T(\eta N\to \eta N)$  is quite poorly known. The reason
is that the lifetime of the $\eta$ is too short to produce $\eta$
beams. So $a(\eta N)$ or $T(\eta N\to \eta N)$ has to be extracted
in rather indirect ways. The inputs are production cross sections of
$\pi^-p\to \eta n$ and $\gamma p\to \eta p$ reactions. Also decays
into the channels $\gamma N$, $\pi N$, $\pi\pi N$ and $\eta N$ were
considered. The major mechanism that generates the imaginary part of
$a(\eta A)$ is the reaction $\eta A_i\to N^*(A - 1) \to \pi A_f$,
where $N^*$ is the nucleon resonance $N^*(1535)$ with a strong
coupling to both the $\eta$ and the pion \cite{PDG10},
\cite{Wycech95}.

\begin{figure}[h!]
\begin{center}
\includegraphics[width=0.6\textwidth]{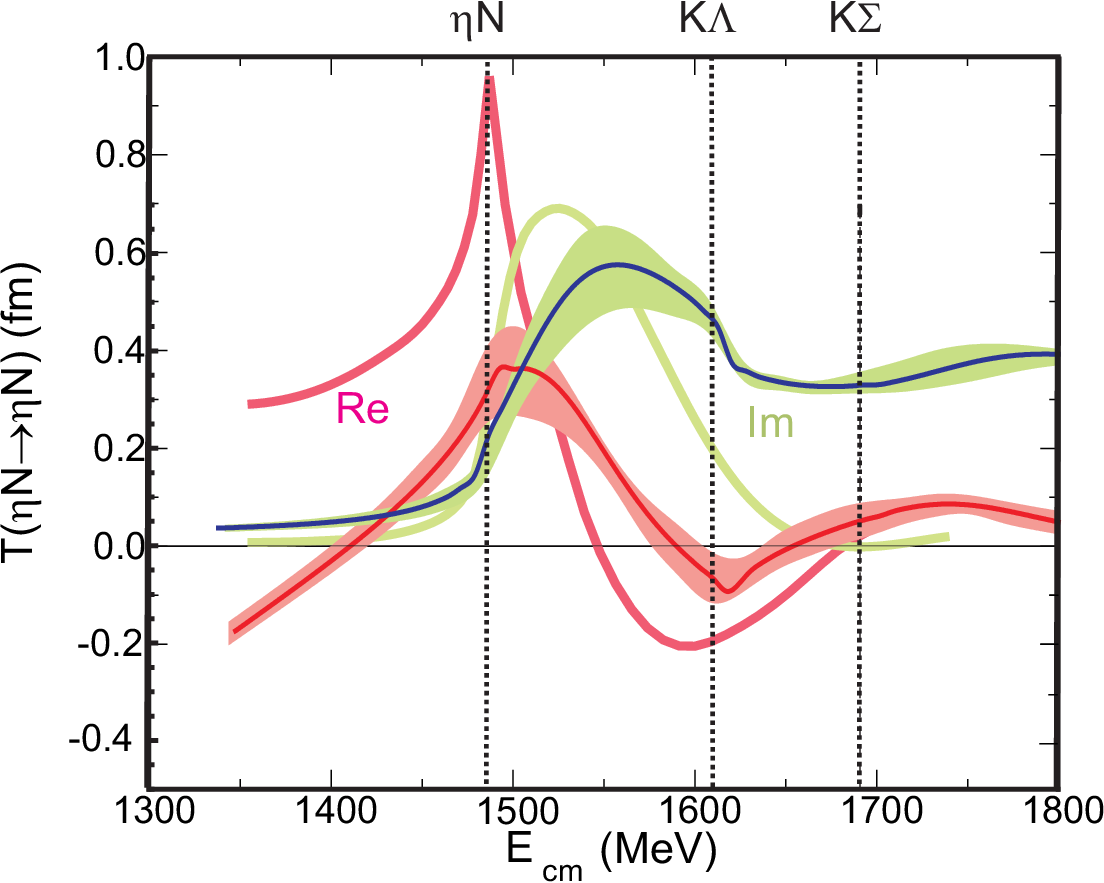}
\caption{The energy dependence of the real and imaginary part of the
elastic $\eta$-nucleon $T$-matrix. The curves with the 1$\sigma$
error band are from Mai et al. \cite{Mai12}, the curves without
error bands are from Green and Wycech \cite{Green05}. The vertical
lines indicated thresholds for free $\eta N\to \eta N$  scattering, $\eta N\to K\Lambda$ and $\eta N\to K\Sigma$ reactions.}
\label{Fig:T-matrix}
\end{center}
\end{figure}

The approach usually applied is the $K$ matrix approach with
different inputs, say $\pi N\to\pi N$, $\pi N\to \eta N$, and
$\gamma N\to \eta N$. The $T$ matrix for the different channels
$i,j$ can be expanded into
\begin{equation}\label{equ:K-matrix}
T_{i,j} = K_{i,j} +i\sum_m K_{i,m} Q_m T_{m,j}
\end{equation}
with $Q$ being the diagonal matrix of the c.m. momenta in each channel.  Nucleon resonances which couple to the $\eta N$ channel are in the energy range of interest. These resonances were accounted for by
\begin{equation}\label{equ:two-resonances}
K_{i,j}=B_{i,j} + \sum_{res}\frac{\sqrt{\gamma_i\gamma_j}}{(E_{0,res}-E)}\,,
\end{equation}
with $B_{i,j}$ a background, $\gamma_{i,j}$ the couplings of the
resonances to the corresponding channel and $E_{0,res}$ the
resonance position. The background matrix changes the bare pole
positions to those obtained in scattering experiments. The
parameters $B_{i,j}$, $\gamma_i$, and $E_0$ are fit parameters.
Further information is found in Ref. \cite{Green97}. An energy
dependence of a so deduced elastic $T$ matrix is shown further down
with other results in Fig. \ref{Fig:T-matrix}.

\begin{figure}[ht]
\begin{center}
\includegraphics[width=1.0\textwidth]{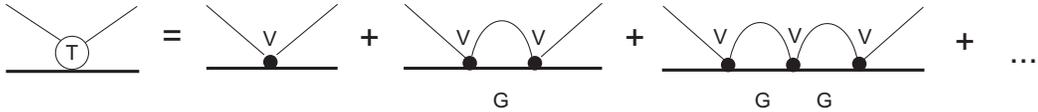}
\caption{Symbolical representation of the Bethe--Salpeter equation.}
\label{Fig:BSE}
\end{center}
\end{figure}

Another method is based on chiral unitarity. The $T$-matrix within this model is given by
\begin{equation}
T(\eta N\to \eta N) = \left(1-VG\right)^{-1}V\,,
\label{equ:T-chiral}
\end{equation}
where $G$ is the energy dependent diagonal matrix of loop functions,
and $V$ the kernel matrix, respectively. The loop functions describe
the propagation of intermediate states in the medium. The result is
obtained by solving a Bethe--Salpeter equation which is shown
schematically in Fig. \ref{Fig:BSE}.
The potential $V$ is derived from the lowest order chiral Lagrangian
containing SU(3) flavour symmetry. The $\eta N$ scattering amplitude
can be calculated by considering the coupled channels: $\pi^- p$,
$\pi^0 n $, $\eta n$,$K^0\Lambda$,
$K^+\Sigma^-$,$K^0\Sigma^0$,$\pi^0\pi^- p$, and $\pi^+\pi^- n$. It
should be noted that the same  matrix inversion can be used to solve
the multi-channel T-matrix from a Lippmann--Schwinger equation
iterating the potential $V$ to all orders \cite{Kaiser97}. While the
Bethe---Salpeter equation is a four dimensional integral equation
and thus covariant, the Lippmann--Schwinger equation is usually
three dimensional. For the connection between the Bethe--Salpeter
equation and the Lippmann--Schwinger equation see \cite{Pilkuhn79}.
In \cite{Kaiser97} only the radial equation is used with the
integration performed over the relative momentum of the off-shell
meson-baryon pair in intermediate channels. The obtained
multi-channel $S$-matrix $S_{ij}=\delta_{ij}-2i\sqrt{p_ip_j}T_{ij}$
is in this case strictly unitary in the subspace of open channels.
The total $s$ wave cross section for a transition $(i\to j)$ is
$\sigma_{i,j}=4\pi p_i/p_j|T_{i,j}|^2$.

In Table \ref{tab:Scattering_length} we compile results for the complex scattering length $a(\eta N)$. In some cases not only the scattering length but also the effective range $r_0$ is given and also shown in Table \ref{tab:Scattering_length}.

\begin{table}[ht!]
  \centering
  \caption{Selection of elastic $\eta N$ scattering length $a(\eta N)$ and effective range parameters $r_0$ where given by the authors.}\label{tab:Scattering_length}
  \begin{tabular}{|l|l|l|}
    \hline
$a(\eta N)$      (fm) &$r_0$ (fm)           & Reference \\ \hline
 & & \\
0.219$_{-0.068}^{+0.047}$  +  i0.235$_{-0.055}^{+0.148}$ & & chiral model I \cite{Mai12}\\
0.25 + i0.16          &                     & isobar model \cite{Bennhold91}\\
0.27 + i0.22          &                     & isobar model \cite{Bhalerao85}\\
$\leq$0.3             &                     & $pn\to d\eta$ \cite{Grishina00}\\
0.378$_{-0.101}^{+0.092}$  +  i0.201$_{-0.036}^{+0.043}$& & chiral model II \cite{Mai12}\\
0.41  +  i0.56 & &      global \\ & & energy dependent \cite{Arndt05}\\
0.46(9) + i0.18(3)    &                     &  K matrix \cite{Briscoe02}\\
0.476 + i0.279        &                     & electro production \cite{Tiator94}\\
0.487 + i0.171        &  -6.06 - i0.177     & K matrix \cite{Feuster98}\\
0.51 + i0.21          &                     & K matrix \cite{Sauermann95,Deutsch97}\\
0.55(20) + i0.30      &                     & $\pi^-p\to\eta n$ \cite{Wilkin93}\\
0.577 + i0.216        &  -2.807 - i0.057    & K matrix \cite{Feuster98}\\
0.621(40) + i0.306(34)&                     & S-wave resonance \cite{Abaev96}\\
0.68 + i0.24          &                     & coupled channel \\ & &  $S$ matrix \cite{Kaiser95}\\
0.75(4) + i0.27(3)    & -1.50(13)-i0.24(4)  & K matrix \cite{Green97}\\
$\leq$0.75            &                     & $\eta d$ scattering,
\\ & &
AGS equations\cite{Rakityansky01}\\
0.87 + i0.27          &                     & $\eta d\to \eta d$\cite{Green99}\\
0.91(6) + i0.27(2)    & -1.33(15) - i0.30(2)& K matrix, solution A \cite{Green05}\\
0.91(3) + i0.29(4)    &                     & coupled channel \cite{Batinic98}\\
0.980 + i0.37         &                     & isobar model \cite{Arima92}\\
0.991 + i0.347        & -2.081 - i0.81      & K matrix \cite{Penner02}\\
1.05 + i0.27          &                     & coupled K matrices \cite{Green99}\\
1.03 +i0.49 &  & rel. coupled channel \cite{Lutz02} \\
1.14 + i0.31          &                     & coupled K matrices \cite{Arndt05}\\
    \hline
  \end{tabular}
\end{table}

Arndt et al. \cite{Arndt05} derived from the optical theorem a bound
of
\begin{equation}\label{equ:optical} a_i \geq 0.172\pm 0.009\text{
fm}.
\end{equation}
All results compiled in Table \ref{Tab:eta-N}
fulfill this criterion except the one of Ref. \cite{Bennhold91}.

\begin{figure}[h!]
\begin{center}
\includegraphics[width=0.60\textwidth]{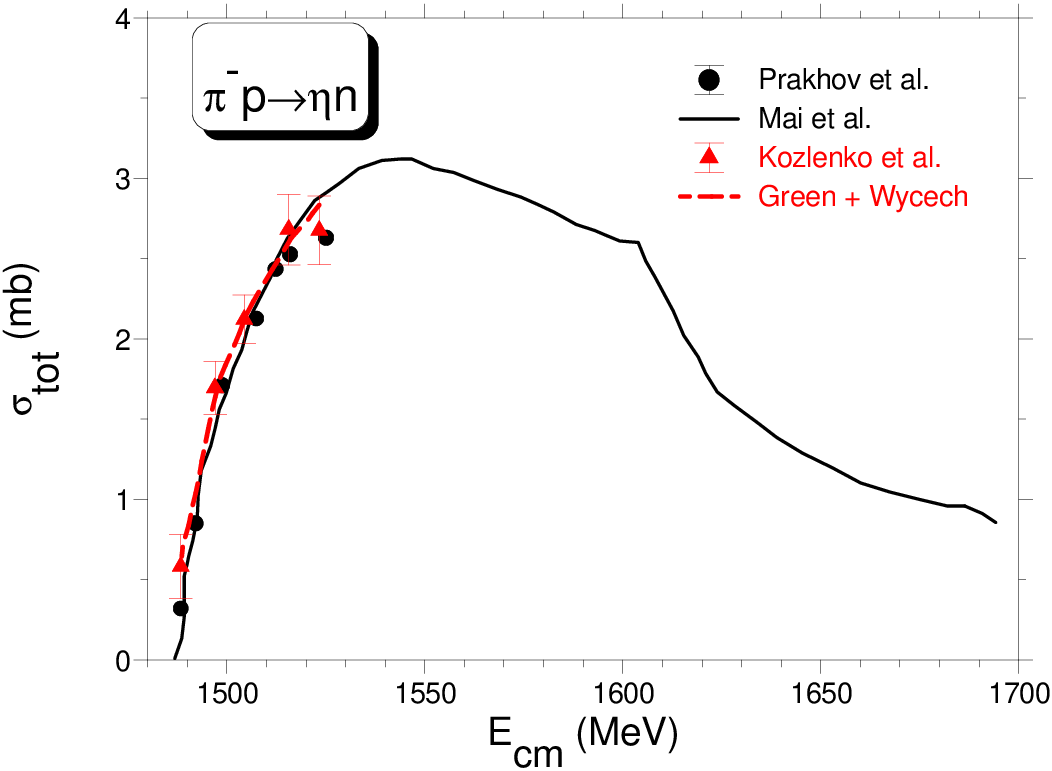}
\caption{Excitation functions for the reaction $\pi^-p\to \eta n$. Data are indicated by dots with error bars, fits by Mai et al. \cite{Mai12} as solid curve, by Green and Wycech \cite{Green05} as dashed line. }
\label{Fig:pi_p_eta_n}
\end{center}
\end{figure}

In the following we will discuss the methods leading to very small
and very large values of the scattering length. Green and Wycech
\cite{Green05} applied the $K$-matrix formalism treating all
available data at the time of their publication (i. e. their
parameter set A). They considered two nucleon resonances which
couple to the $\eta N$ channel: the previously mentioned $N^*(1535)$
and the $N^*(1650)$. An effective range expansion is fitted to the
so derived $T$ matrix by
\begin{equation}\label{equ:T-matrix}
\frac{1}{T(\eta N\to\eta N)} +ip_\eta= \frac{1}{a}+\frac{1}{2}r_0p_\eta^2+sp_\eta^4\,.
\end{equation}
Here $p_\eta$ denotes the $\eta$ momentum in the centre of mass system. The obtained values for the scattering length and effective range are given in Table \ref{tab:Scattering_length}. For the last parameter in Eq. \eqref{equ:T-matrix} they found $s=[-0.15(1)-i0.04(1)]\text{ fm}^3$. The resulting energy dependence is shown in Fig. \ref{Fig:T-matrix}. The maximum of the real part occurs at the $\eta N$ threshold whereas the maximum of the imaginary part is close to the centroid of the $S_{11}(1535)$ resonance. In the same figure also the results from an analysis within the chiral unitary model and solving the Bethe-Salpeter equation \cite{Mai12} are shown. The $N^*(1535)$  and $N^*(1650)$ resonances were dynamically generated. The input were the hadron masses. The $S$-wave cross sections of the $\pi N\to \pi N$ and $\pi N\to \eta N$ were fitted.
In a next step we compare the most important inputs in the two
analyses. In Fig. \ref{Fig:pi_p_eta_n} the input data and the fits
are compared with each other.

\begin{figure}[h!]
\begin{center}
\includegraphics[width=0.60\textwidth]{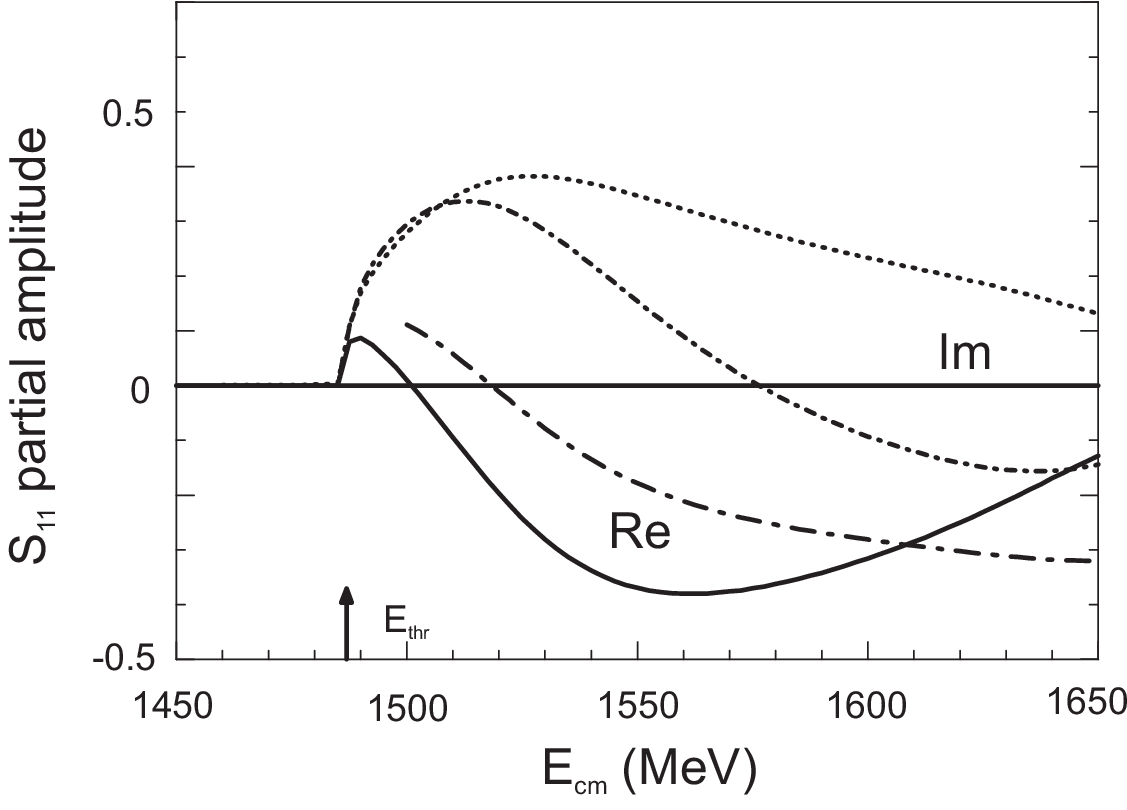}
\caption{$S_{11}$ partial amplitude for $\pi^-+p\to\eta+n$ (adopted from Ref. \cite{Arndt05}). Dash-dotted (dotted) curves show the real (imaginary) parts of amplitudes
corresponding to fit G380. Solid (short-dash-dotted) lines represent the real (imaginary) parts of amplitudes corresponding to the Fit A. All amplitudes are dimensionless, i.e. the phase-space factors were normalised to unity at the resonance position. }
\label{Fig:Arndt}
\end{center}
\end{figure}
In the energy region where data exist there is practically no
difference. Nevertheless the different models lead to different
$T$-matrices. Similar findings were reported by Arndt et al.
\cite{Arndt05}. Here we discuss the $S_{11}$ partial amplitude,
which is the most important one for the formation of $\eta$ bound
states. Arndt et al. extracted these amplitudes in two different
ways. They considered $K$ matrices for a background and two
resonances: $S_{11}(1535)$ with two poles and $D_{13}(1520)$ (see
also the discussion above close to Eq. (\ref{equ:two-resonances})).
Their fit A included the Crystal Ball data as well as their own pion
data. The fit GW380 is a global fit.  While for elastic $\pi$
scattering the results are practically identical they differ for
$\pi^-+p\to\eta+n$ (see Fig. \ref{Fig:Arndt}). Consequently they
differ also for elastic $\eta$ scattering.  For the $K$-matrix
analysis, which is close to the one from Ref. \cite{Green05} they
determined $a(\eta N)= 1.14+i0.31$ fm. However, the global fit
yields $a(\eta N)=0.41+i0.56$ fm. This is the only result where
$a_i>a_r$. and thus raises doubts on its validity. We may conclude
that the variations due to models are larger than due to different
inputs.

\subsubsection{The $\eta$ Nucleus Potential and Search for Quasi Bound States}\label{sub:Optical_A-Potential}

The standard approach is to construct from the $\eta$-nucleon scattering length an optical potential for the $\eta$-nucleus interaction with $A$ the mass number of the nucleus, and then to solve a wave equation with this potential \cite{Haider_Liu86}, \cite{Friedman13}, \cite{Wycech95}, \cite{Wilkin93}. The complex optical potential is given by
\begin{equation}\label{equ:Optical}
U_\text{opt} = V+iW=-\frac{2\pi}{\mu} T(\eta N\to \eta N) A \rho(r)
\end{equation}
with $\mu$ the reduced $\eta N$ mass, $T(\eta N\to \eta N)$ the $\eta$-nucleon transition matrix and $\rho$ the nuclear density. We will call this relation as the $T\rho$ approximation. In the impulse approximation the relation
\begin{equation}\label{equ:Tscatterin}
a(\eta) = T(\eta N\to \eta N,\rts_0)
\end{equation}
with $\rts_0 =m_\eta+m_N$ holds. The quantity $a$ is the effective range. It should be mentioned here that the interest is in a bound state and hence one needs to  know the $T$ matrix at
\begin{equation}\label{eq:negative-energy}
\sqrt{s} = \sqrt{s_0}-B_\eta \,,
\end{equation}
i.e. below threshold (see Fig. \ref{Fig:T-matrix}).

Haider and Liu \cite{Haider_Liu86} solved the integral equation
\begin{equation}\label{equ:Haider-Liu}
\frac{\vec{p}_f^2}{2\mu}\Psi(\vec{p}_f) + \int d^3p_i <\vec{p}_f|U_\text{opt}|\vec{p}_i>\Psi(\vec{p}_i) = E \psi(\vec{p}_f)
\end{equation}
by the inverse iteration method. This equation was derived from the four dimensional Bethe-Salpeter equation which can be rewritten as two coupled equation \cite{Liu80}. One of these leads then to Eq. \eqref{equ:Haider-Liu}. This equation has the appearance of a Schr\"{o}dinger equation with relativistic kinematics. Since no non-relativistic approximations had been introduced in its derivation it has some covariance features. The main advantage of working with such an approach is that the $\eta$ nucleus interaction $U_{\eta A}$ can be related to the elementary $\eta N$ process by unambiguous kinematical transformations. The first-order microscopic $\eta$-nucleus optical potential
has the form
\begin{equation}
<\vec{p}_f\mid U_\text{opt}\mid \vec{p}_i> = \nonumber
\end{equation}
\begin{equation}
\sum_{j}\int d{\vec{Q}}
<\vec{p}_f,-(\vec{p}_f+\vec{Q})\mid T(\sqrt{s_{j}},\eta N\rightarrow\eta N)\mid \vec{p}_i, -(\vec{p}_i+\vec{Q})> \nonumber \end{equation}\begin{equation}\label{equ:pot}
 \times  \phi^{*}_{j}(-\vec{p}_f-{\vec{Q}})
\phi_{j}(-\vec{p}_i-\vec{Q})\ ,
\end{equation}
where the off-shell $\eta N$ interaction
$T(\eta N\rightarrow\eta N)$ is weighted by the product of the nuclear wave functions $\phi^{*}_{j}\phi_{j}$ corresponding to having the nucleon $j$ at the momenta $-(\vec{p}_i+\vec{Q})$ and $-(\vec{p}_f+\vec{Q})$ before and after the collision, respectively. The $\sqrt{s_{j}}$ is the $\eta N$ invariant mass and is equal to the total energy in the c.m. frame of the $\eta$ and the nucleon $j$.
The solution of Eq. \eqref{equ:Haider-Liu} is the complex eigenenergy
\begin{equation}\label{equ:eigenenergy}
E_\eta=-B_\eta-i\frac{\Gamma_\eta}{2}
\end{equation}
with $B_\eta >0$ and $\Gamma_\eta >0$ the binding energy and width, respectively.
The integration in Eq. \eqref{equ:pot} is performed over all Fermi momenta $\vec{Q}$. Results for the full off-shell calculation are given in Ref. \cite{Haider_Liu02}.

In a further approximation the scattering amplitude $T(\eta N\to\eta N)$ is taken out of the integral in Eq. (\ref{equ:pot}) \cite{Haider_Liu02}. This is called the factorisation approximation. Then $\vec{Q}$ is no longer unique. Assuming that the interacting nucleon is at rest before and after the interaction leads to
\begin{equation}\label{equ:FA}
<\vec{Q}> = \frac{A-1}{2A}(\vec{p}_f-\vec{p}_i).
\end{equation}
In this approximation one can choose a mean energy $\Delta$ of the struck nucleon in the Fermi sea. In Table \ref{tab:Eigenenergies} we compare the eigenenergies (i.e. binding energies and widths)  within these two models. Both models give results which agree to each other when $\Delta$=30 MeV is assumed. This can be understood by noting that the average
nuclear binding and Fermi motion amount to about a 30~MeV downward shift  of the $\eta$-nucleon interaction
energy. Smaller assumptions for $\Delta$ result in larger widths. The increase of binding energy and width with mass number $A$ is shown in Figs. \ref{Fig:Predictions_BE} and \ref{Fig:Predictions_Gamma}.

\begin{table}[ht]
\caption{Binding energies $B_\eta$ and widths $\Gamma_\eta$ for $\eta$ mesic nuclei in $1s$ states for the full off-shell model Eq. (\ref{equ:pot}) and the factorisation approximation Eq. (\ref{equ:FA}) with $\Delta$=30 MeV. The $\eta N$ interaction parameters were taken from \cite{Haider_Liu02}.}\label{tab:Eigenenergies}
\centering
\begin{tabular}{|c|c|c|c|c|}
\hline
nucleus& \multicolumn{2}{|c|}{ full off-shell calculation } & \multicolumn{2}{|c|}{ factorisation approximation } \\
  & $B_\eta$ (MeV) & $\Gamma_\eta$ & $B_\eta$ (MeV) & $\Gamma_\eta$ \\ \hline
$^{12}$C & 1.19 & 3.67 & 1.10 & 4.10 \\
$^{26}$Mg & 6.39 & 6.60 & 7.11 & 7.46 \\
$^{90}$Zr & 14.80 & 8.87 & 16.29 & 9.84 \\
$^{208}$Pb & 18.46 & 10.11 & 18.96 & 10.22 \\ \hline
\end{tabular}
\end{table}
In conclusion not only the $\eta N$ interaction at threshold but also below threshold is important, since the average interaction is below threshold.

Instead of a three dimensional self-consistent equation like Eq. \eqref{equ:Haider-Liu} other authors (\cite{Garcia-Recio02}, \cite{Hayano99} and \cite{Friedman13}) made use of the Klein-Gordon equation (KGE)
\begin{equation}\label{equ:KG}
\left\{ \nabla^2 + \left(E_\eta^2-m_\eta^2 \right) -\Pi_\eta[\text{Re}(E_\eta),\rho] \right\}\Psi =0\,,
\end{equation}
which they solved self consistently. Here $E_\eta=m_\eta-B_\eta-i\Gamma/2$. The last term in the bracket $\Pi_\eta[\text{Re}(E_\eta),\rho]$ is the self-energy which is related to the optical potential
\begin{equation}
\Pi_\eta[\text{Re}(E_\eta),\rho]=2(m_\eta-B_\eta) U_\text{opt}.
\end{equation}
It is energy and density dependent with $\rho=\rho_p+\rho_n$, $\rho_p$ the proton density and $\rho_n$ the neutron density. In some works the binding energy is ignored since $B_\eta<<m_\eta$.

Refs. \cite{Sibirtsev04} and \cite{Niskanen13}  searched  for poles in the homogeneous Lippman--Schwinger equation (in coordinate space) associated with bound states and their eigenvalues by varying the potential parameters. The radial wave function is given by
\begin{equation}\label{equ:Lippman-Schwinger}
R_l(r) = -ik \frac{2\mu_{\eta A}}{\hbar ^2}\int_0^\infty j_l(kr_<)h_l^{(1)}(kr_>) U_\text{opt}(r')R_l(r')r'^2dr'\,,
\end{equation}
which is equivalent to the Schr\"{o}dinger equation. Here $\mu_{\eta A}$ denotes the reduced mass of the $\eta A$ system. The Green function arguments are $r_<$,  $r_>$ the smaller and larger of $r$ and $r'$ and $k=\sqrt{2\mu_{\eta A} E /\hbar^2}$.

In Refs. \cite{Belyaev95}, \cite{Rakityansky96}, \cite{Sofianos97} bound states in light nuclei were calculated by the finite rank approximation (FRA). Within this approximation the motion of the $\eta$ and of the nucleons inside the nucleus are treated separately. The internal dynamics of the nucleus enters only via the nuclear wave function. Finally Faddeev-type few body equations were solved to calculate the $\eta$-nucleus $T$ matrix.
Formally, FRA starts from the Hamiltonian
\begin{equation}
H=H_0+U+H_A
\end{equation}
where $H_0$ is the kinetic energy operator of the $\eta$ nucleus motion, i. e. the free Hamiltonian, $U$ is the sum of the $\eta N$ potentials, and $H_A$ is the total Hamiltonian of the nucleus. $H_A$ has a spectral decomposition consisting of bound states as well as of continuum states which reads as
\begin{equation}
H_A=\sum_n {\cal E}^A_n |\Psi^A_n\rangle\langle \Psi^A_n|
+\int E|\Psi^A_E\rangle\langle \Psi^A_E|\,dE\,,
\label{equ:spectral}
\end{equation}
where $|\Psi^A_n\rangle $ are the bound--state eigenfunctions of  $H_A$ with ${\cal E}_n$ being the corresponding energies. In FRA the continuum states were neglected. Light nuclei with $2\leq A\leq 4$ have only one bound state, the ground state. In this case Eq. (\ref{equ:spectral}) reduces to
\begin{equation}\label{equ:FRA}
H_A\approx {\cal E}_0^A|\Psi_0^A><\Psi_0^A|\,.
\end{equation}
Inserting this equation into the transition matrix one gets a Lippman -- Schwinger equation. Another method is to expand the $T$ matrix into a Faddeev-type decomposition leading to integro-differential equations. Finally the parameters of $a_{\eta N}$ were varied until poles were found. Within this approach $\eta$ mesic nuclei exist for  $^4$He. But for sufficiently large $a_{\eta N}$ even the deuteron can bind the $\eta$ \cite{Shevchenko00}.

The same method together with time delay analysis was applied in \cite{Kelkar06a}  to derive the position of $\eta$ bound states in $^3$He and in Ref. \cite{Kelkar06} to derive positions of $\eta$ bound states in {$^2$H} and {$^4$He}. However, the results for the scattering lengths $a_{\eta A}$ violate the conditions Eq. (\ref{equ:cond_2}) and related relations.

Different authors added or modified the above approaches for instance by improving the optical model or introducing final state interactions. However, the main input is still the $T$ matrix or the $\eta$-nucleon scattering length.

\subsubsection{The $\eta$ self-energy}\label{sssec:eta-self-energy}

A lot of theoretical work dealt with improvements of the lowest order optical potential (\ref{equ:Optical}). They calculate the behaviour of the $\eta$ scattering in the nuclear medium. In general there are two different models. Both are based on the observation that the $\eta N$ system couples strongly to the $N^*(1535)$ resonance: $\eta N\to N^*(1535)\to \eta N$ \cite{Chiang91}, \cite{Jido02}. One can evaluate the self-energy by using the $N^*$ dominance hypothesis. In the limit of small $\eta$ momentum and considering the lowest $N^*-N$ hole excitation one obtains \cite{Jido02}
\begin{equation}
\Pi_\eta(\omega_\eta,\rho) = \frac{g_\eta^2 \rho}{m_\eta+B_\eta + m_N^*(\rho)-m_{N^*}^*(\rho)+i\Gamma_{N^*}(\omega,\rho)/2}+\text{ (cross terms)}\,.
\label{equ:self-energy}
\end{equation}
\begin{figure}[ht]
\begin{center}
\includegraphics[height=0.3\textheight]{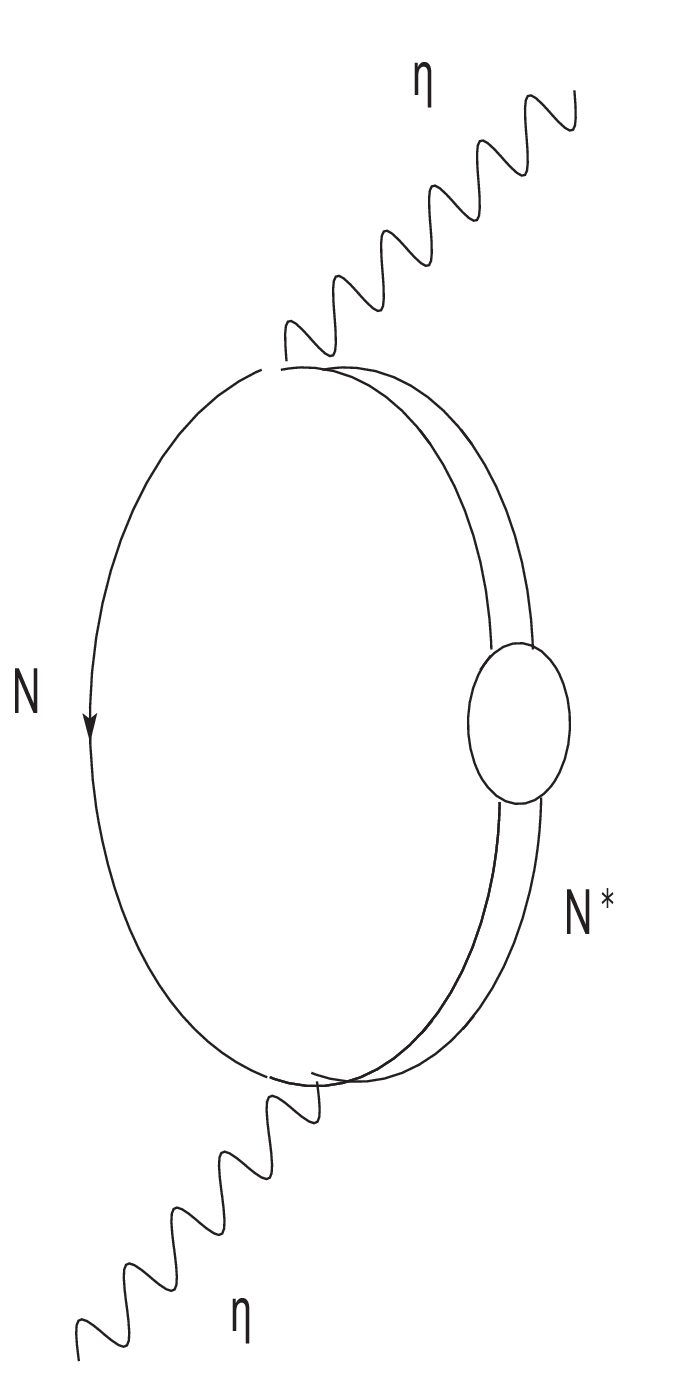}
\hspace {1cm}
\includegraphics[height=0.3\textheight]{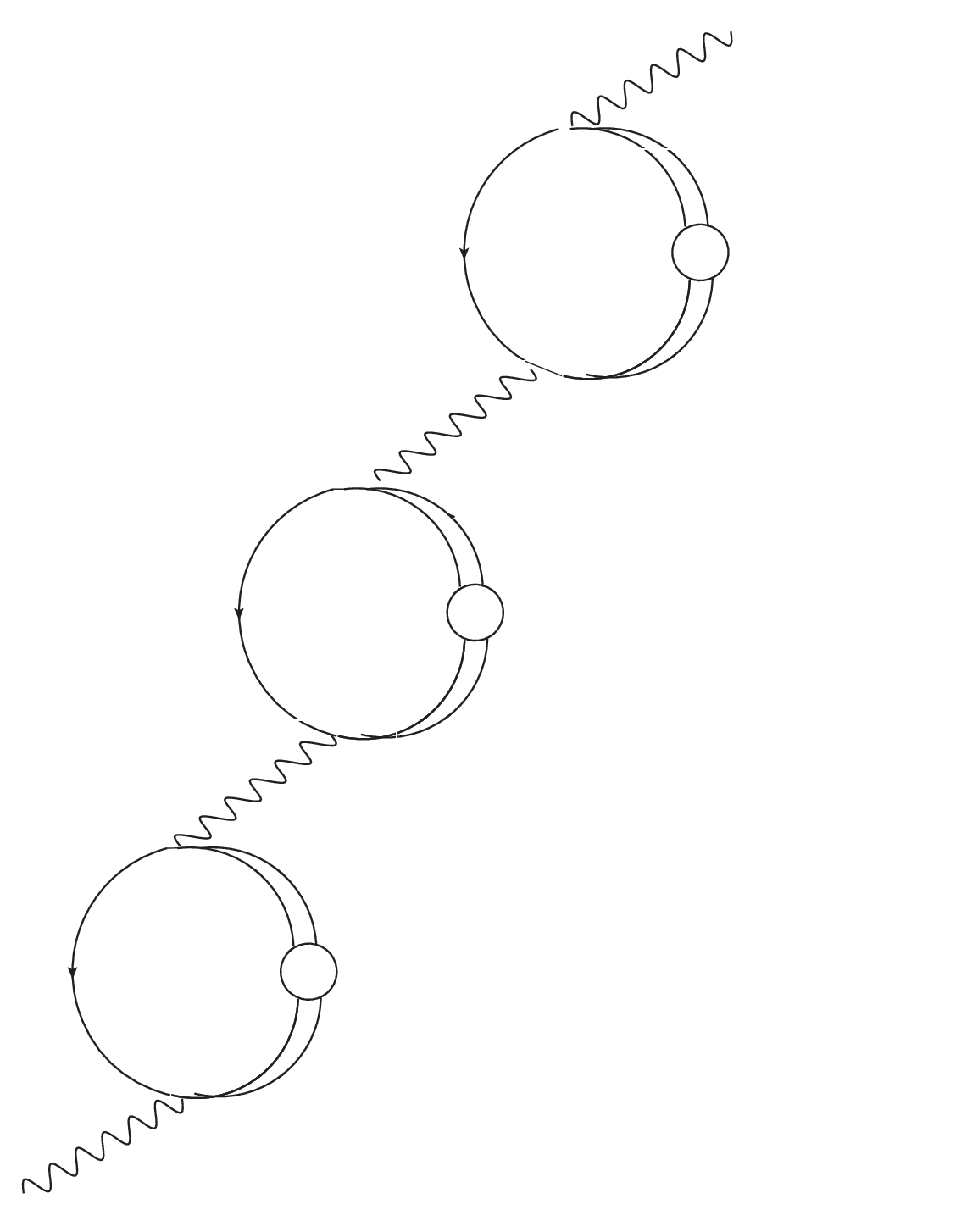}
\caption{Left Panel: Diagrammatic representation of the $\eta$ self-energy in the nucleus through excitation of a $N$-hole. The circle in the $N^*$ is the $N^*$ self-energy. Right panel: Consecutive absorption and emission of the $\eta$ indicating the binding.  }
\label{Fig:N_star}
\end{center}
\end{figure}

Here $g_\eta$ is the S-wave $\eta N N^*$ coupling. From the partial width $\Gamma_{N^*\to \eta N}\cong 75$~MeV \cite{PDG10} one gets at tree level $g_\eta\cong 2.0$. $m_N^*(\rho)$ and $m_{N^*}^*(\rho)$ are the effective masses in the medium.

The difference between the two baryon masses can be evaluated by the assumption that due to partial restoration of the chiral symmetry it decreases with increasing density \cite{Nagahiro09}
\begin{equation}
m_N^*(\rho)-m_{N^*}^*(\rho) =\left(1-C\frac{\rho}{\rho_0}\right)
\left(m_N-m_{N^*}\right)\,.
\label{equ:chiral-doublet}
\end{equation}
The parameter $C$ is in the order of 0.1 to 0.3. This model is called the chiral doublet model \cite{Nagahiro09}, i.e. the $N^*$ is the chiral partner of the nucleon. This model is sketched in Fig. \ref{Fig:N_star}.

A second model is the chiral unitary model in which the $N^*$ is a dynamically generated resonance in the coupled channel meson-baryon scattering \cite{Inoue02}, \cite{Garcia-Recio05}, \cite{Bruns11}. For a discussion of the nature of the $N^*$ see Ref. \cite{Kaiser97}.

In this model the in medium $T$ matrix is given by
\begin{equation}
  T(P^0, \vec P ~; \rho) =
  \left[ 1 - V(\rts) G(P^0, \vec P ~; \rho) \right]^{-1}  V(\rts)
  \label{equ:T_Medium}
\end{equation}
where $(P^0, \vec P)$ is the 4-momentum of the system and $\rho$ is the density of the medium. The kernel $V$ has nothing to do with the medium and is therefore the same as for the free case. The matter effects enter through the loop functions $G$ \cite{Garcia-Recio02}. The self-energy of the $\eta$ meson is
 evaluated in nuclear matter at various densities $\rho$, as a
 function of the $\eta$ energy, $p^0$, and its momentum, $\vec p$, in
 the nuclear matter frame.  It is calculated by means of
\begin{equation}
     \Pi_{\eta}(P^0, \vec P~; \rho)
     =
     4 \! \int^{p_F} \! \!  \frac{d^3 \vec p_n }{(2\pi)^3}~
     T(P^0, \vec P ~; \rho)
 \label{eqn:selfint}
\end{equation}
where $\vec p_n$ is the momentum of the nucleon and $p_F$ is the Fermi momentum at nuclear density $\rho$. With these inputs one can then solve the Klein--Gordon equation as stated above. It was shown that in this approach the $N^*$ resonance in medium barely moved with respect to the one in vacuum \cite{Inoue02}.

We now compare the optical potentials $U(r)=V(r)+iW(r)$ developed within the above discussed models.
\begin{figure}[h!]
\begin{center}
\includegraphics[width=0.6\textwidth]{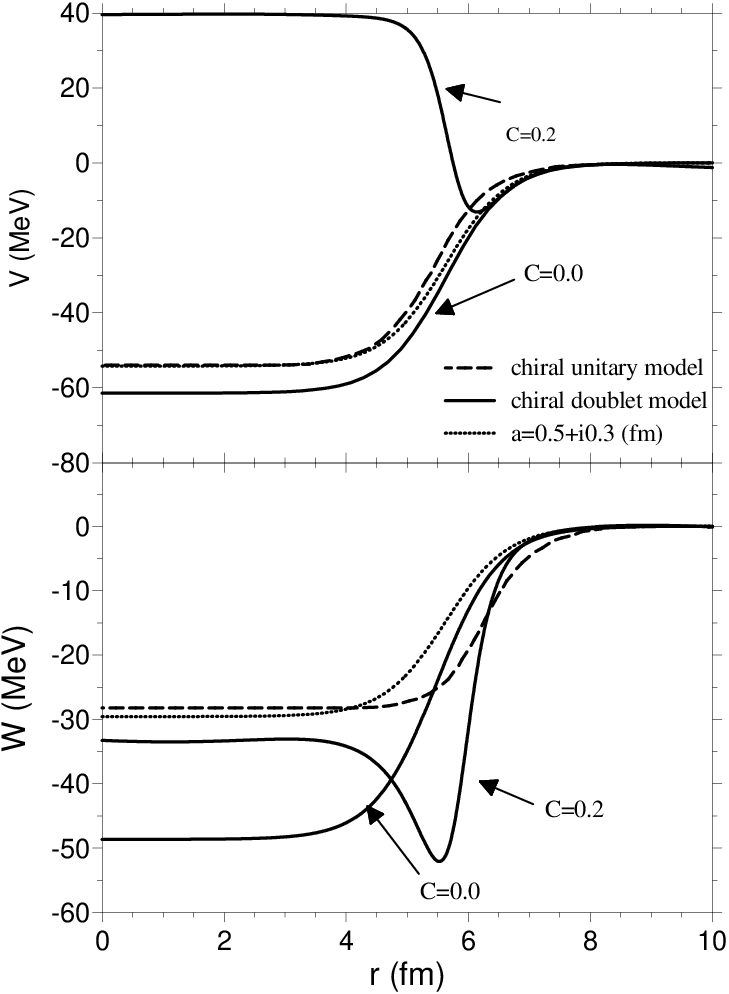}
\caption{Real part $V$ and imaginary part $W$ of optical potential $U$ for the case of $\eta$ meson interaction with $^{132}$Xe. The dotted curves are the $T\rho$ approximation Eq. (\ref{equ:Optical}) with scattering length from \cite{Wilkin93}. The result for the chiral unitary model are scaled from Ref. \cite{Inoue02}. The radial dependencies for the chiral doublet model are shown as solid curves with two choices of $C$ indicated next to the appropriate curve.}
\label{Fig:Potentials}
\end{center}
\end{figure}
The real and imaginary parts are shown in Fig. \ref{Fig:Potentials}. Here we restrict the discussion to three models: the $T\rho$ approximation Eq. (\ref{equ:Optical}), the chiral unitary model Eq. (\ref{equ:T_Medium}) \cite{Inoue02} and the chiral doublet model Eq. (\ref{equ:chiral-doublet})\cite{Jido02}. All calculations are for the heavy nucleus $^{132}$Xe. For the $T\rho$ approximation we have applied a scattering length $a=0.5+i0.3$ fm \cite{Wilkin93}. The result coincides almost with the chiral unitary model result except for the nuclear surface region. Here the difference between both models is larger in the case of the imaginary part than for the real part. For the chiral doublet model we show two calculations \cite{Jido02}, one for $C=0$, i. e. without effects of chiral restoration and one with $C=0.2$. While the first choice is close to the $T\rho$ approximation and the chiral unitary model, although with a much deeper imaginary potential, it looks dramatically different for the second choice. The real part is repulsive in the interior and only a small pocket in the surface region is attractive and allows binding. However, the shallowness of the real part together with the deep imaginary part does not favour bindings for light nuclei. Jido et al. \cite{Jido02} report binding energies for $L=0$ and $L=1$. These values are compiled in Table \ref{Tab_Xe132} for $^{132}\text{Xe}\otimes\eta$.
\begin{table}[h!]
\centering
\caption{Binding energies $B_\eta$  and widths $\Gamma_\eta$ for the $^{132}\text{Xe}\otimes\eta$ nucleus for the chiral doublet model (from Ref. \cite{Jido02}). }\label{Tab_Xe132}
\begin{tabular}{lllll}
\hline
$C$& \multicolumn{2}{c}{ $L=0$ } & \multicolumn{2}{c}{ $L=1$ } \\
  & $B_\eta$ (MeV) & $\Gamma_\eta$ (MeV) & $B_\eta$ (MeV) & $\Gamma_\eta$ (MeV) \\
\hline
0.0 & 38.4 & 39.6 & 30.5 & 43.8 \\
0.2 & 41.2 & 49.0 & 33.0 & 55.0 \\ \hline
\end{tabular}
\end{table}
If this model describes nature it will be extremely hard to find experimentally such bound states since the full width at half maximum $\Gamma$ is always larger than the position. While a sharp peak can be simply distinguished from an underlying background it is much more difficult to distinguish between a broad structure and the background. As a result the centroid and width of such a structure is deduced with a larger uncertainty than it will be the case for a narrow structure.

One feature of chiral restoration in nuclear matter is the shift of the meson masses $m(\rho)$ compared to the free meson mass $m(0)$. Waas and Weise \cite{Waas97} as well as Inoue and Oset \cite{Inoue02} derived a relation
\begin{equation}
\frac{m(\rho)}{m(0)} = 1-0.05 \frac{\rho}{\rho_0}.
\label{equ:meson_mass}
\end{equation}
\begin{figure}[h!]
\begin{center}
\includegraphics[width=0.6\textwidth]{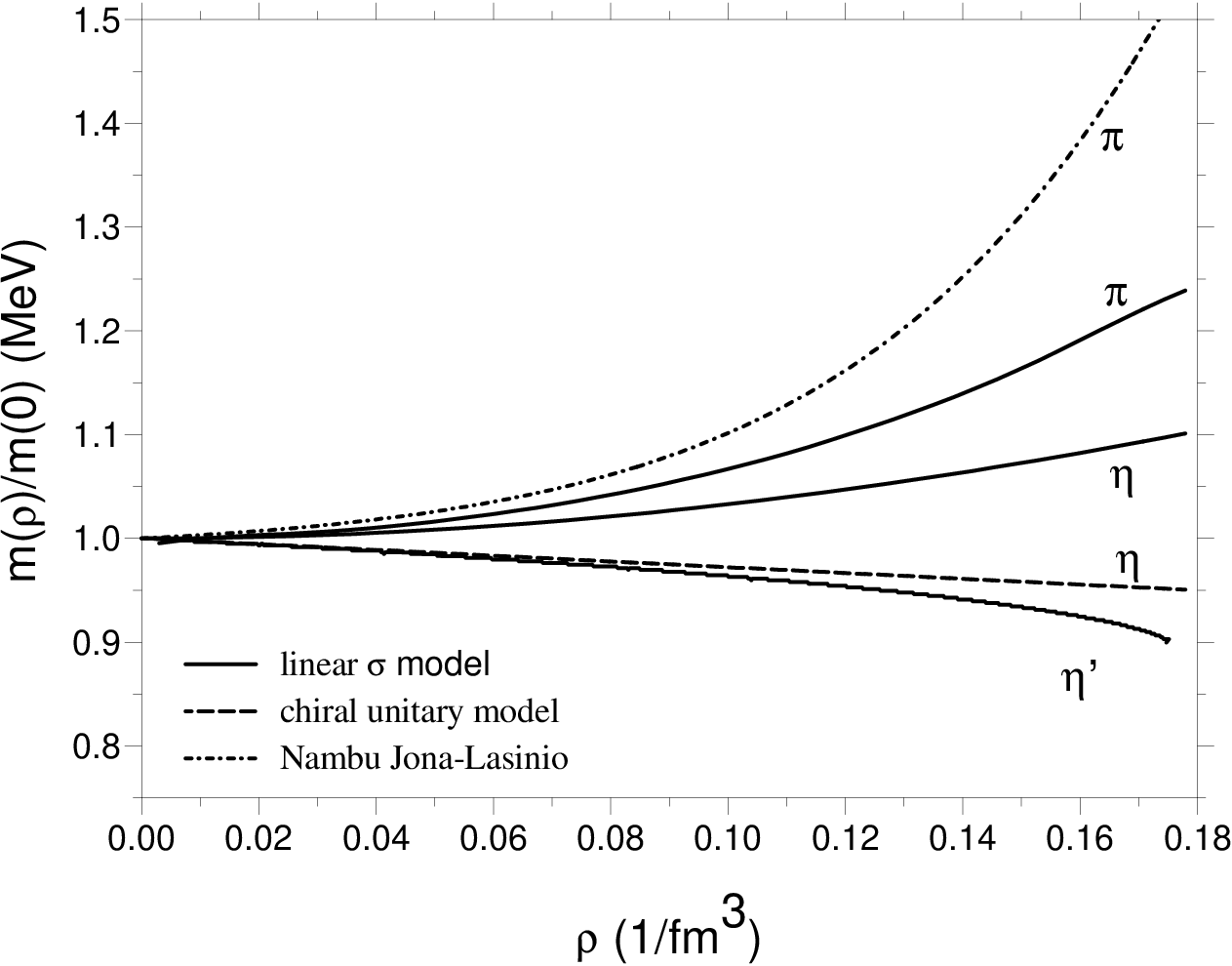}
\caption{The dependence of the masses of $\pi$, $\eta$ and $\eta'$  on the nuclear density. The solid curves are the linear $\sigma$ model \cite{Sakai13} while the dashed curve is the relation (\ref{equ:meson_mass}) for the $\eta$ mass. In addition the density dependence of the $\pi$ in the Nambu--Jona--Lasinio model \cite{Bernard87} is shown as dashed-dotted curve.}
\label{Fig:Meson_Mass}
\end{center}
\end{figure}
This corresponds to a 5\% reduction of the $\eta$ mass at normal nuclear density~$\rho_0$. This relation is shown in Fig. \ref{Fig:Meson_Mass} together with the result from a calculation within a linear $\sigma$ model \cite{Sakai13}. The latter has the opposite sign and predicts a 10\% increase. It will be interesting to see whether experiments can solve this discrepancy. In the figure the density dependence of the pion is also shown within the linear $\sigma$ model and a Nambu--Jona--Lasinio model \cite{Bernard87}. In this case the masses within the two models increase with density. It is interesting to note that for vector mesons all calculations, to the best of our knowledge, show a decrease of the mass with increasing density \cite{Bernard87}, \cite{Hatsuda},  \cite{Klingl99}. $K^+$ increase while $K^-$ decrease strongly with increasing density \cite{Waas97}.

Another model is the quark meson coupling model \cite{Guichon88}. A recent description can be found in \cite{Wittenbury13}. Within this model the $\eta$ wave function taken in the MIT bag model is used to couple the quark and antiquark fields in the meson to the $\sigma$ field in the nucleus \cite{Tsushima00}. The $\eta$ mass decrease  to 50-100 MeV in matter, depending on the $\eta-\eta'$ mixing angle. The elastic $\eta$ nucleon scattering strongly depends on the singlet component of the $\eta$.

\subsubsection{Positions and Widths of $\eta$ Bound States}\label{sub:Positio-Widths-eta-Bound-States}

Since there is a large scatter in the inputs to the calculations, the final results on the bound states vary largely. Also different forms of the nuclear density were applied. Niskanen and Machner \cite{Niskanen13} used a modified harmonic oscillator and Fermi distributions as density profiles. Haider and Liu \cite{Haider_Liu02} used a hollow exponential, a Fermi distribution with three parameters, a modified harmonic oscillator, and a harmonic oscillator for different nuclei. Friedman et al. \cite{Friedman13} made use of a Fermi gas model. Wilkin \cite{Wilkin93} approximated the $^3$He density by a Gaussian.

\begin{figure}[ht]
\begin{center}
\includegraphics[width=0.6\textwidth]{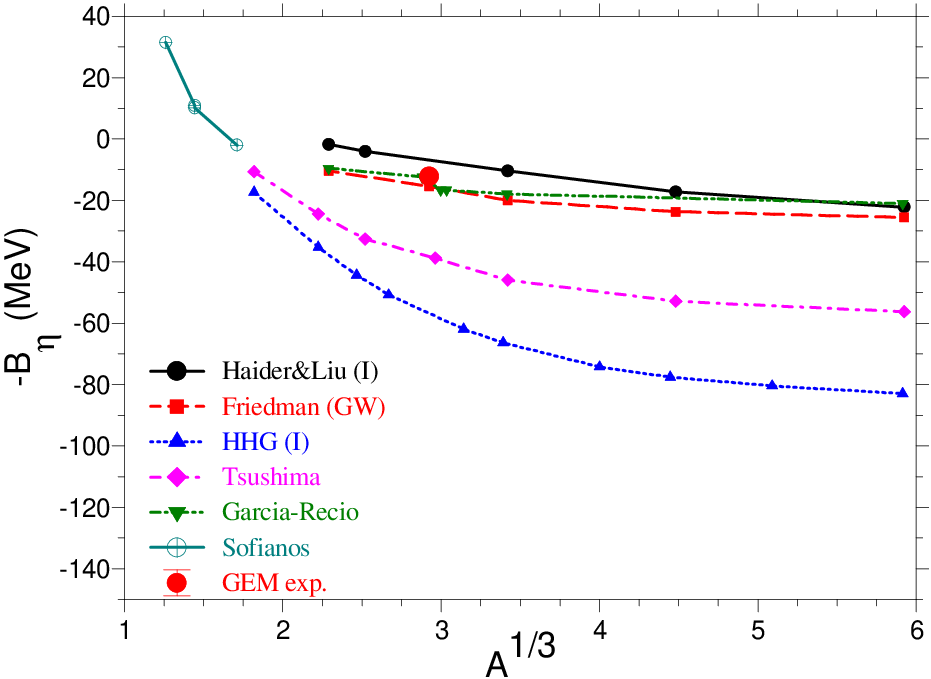}
\caption{Predictions for binding energies. The symbols indicate the published values. The curves are given to guide the eye. The thick full point with error bar is the experimental results of \cite{Budzanowski09}.}
\label{Fig:Predictions_BE}
\end{center}
\end{figure}

\begin{figure}[ht]
\begin{center}
\includegraphics[width=0.6\textwidth]{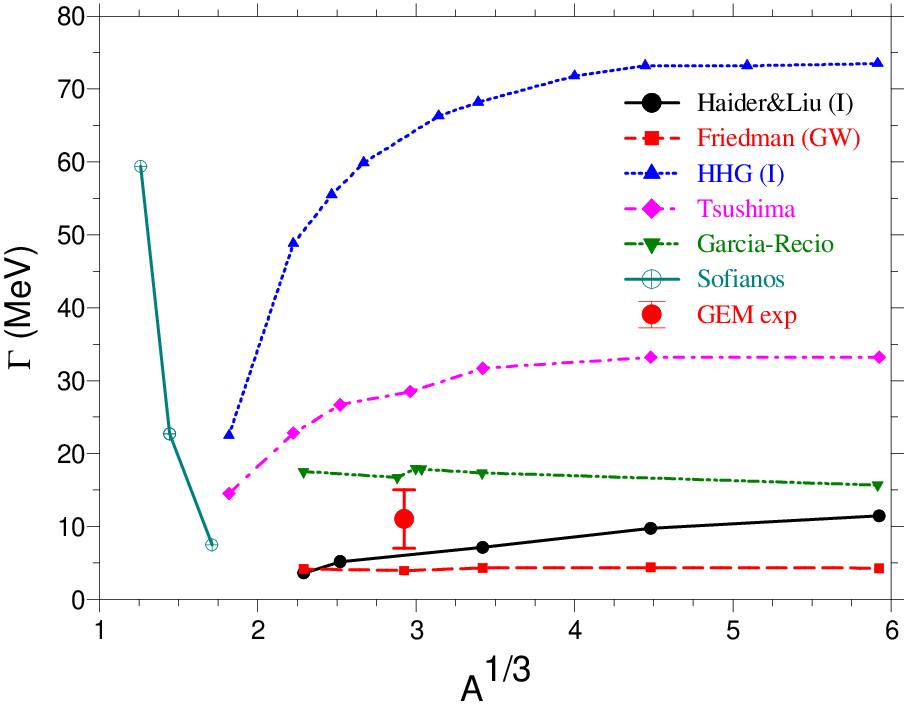}
\caption{Predictions for widths. The symbols indicate the published values. The curves are given to guide the eye. The thick full point with error bar is the results of the experiment from Ref. \cite{Budzanowski09}.}
\label{Fig:Predictions_Gamma}
\end{center}
\end{figure}

We will now compare the mass dependence of the binding energy and the width for some selected approaches. The input is given in Table \ref{Tab:eta-N}.
\begin{table}[ht]
\centering
\caption{Table of $\eta$ nucleon scattering length values applied in various model calculations yielding bound states.}\label{Tab:eta-N}
\begin{tabular}{|c|c|c|}
\hline
Ref. & $a$ (fm) & method \\
\hline
Haider + Liu (set I) Ref. \cite{Haider_Liu86} & $0.28+ i 0.19$ & Schr\"{o}dinger like \\
Friedman et al. (GW model) Ref. \cite{Friedman13} & 0$.96+ i 0.24$& KGE\\
Hayano et al. Ref.\cite{Hayano99} & $0.718 +i 0.263$&KGE\\
Garcia-Recio et al. \cite{Garcia-Recio02} & $0.264+i 0.245$&KGE \\
Sofianos et al. \cite{Sofianos97}(to bind $^4$He) & $>$ $0.47 +i 0.3$ &FRA\\
\hline
\end{tabular}
\end{table}
In addition an alternative method namely the quark meson coupling was applied by Tsushima \cite{Tsushima00}. The results of all calculations are shown in Figs. \ref{Fig:Predictions_BE} and \ref{Fig:Predictions_Gamma}.
All calculations show the same qualitative behaviour: the binding becomes stronger and the width wider for increasing nucleus mass. However, there are large qualitative differences.

\begin{figure}[ht]
\begin{center}
\includegraphics[width=0.6\textwidth]{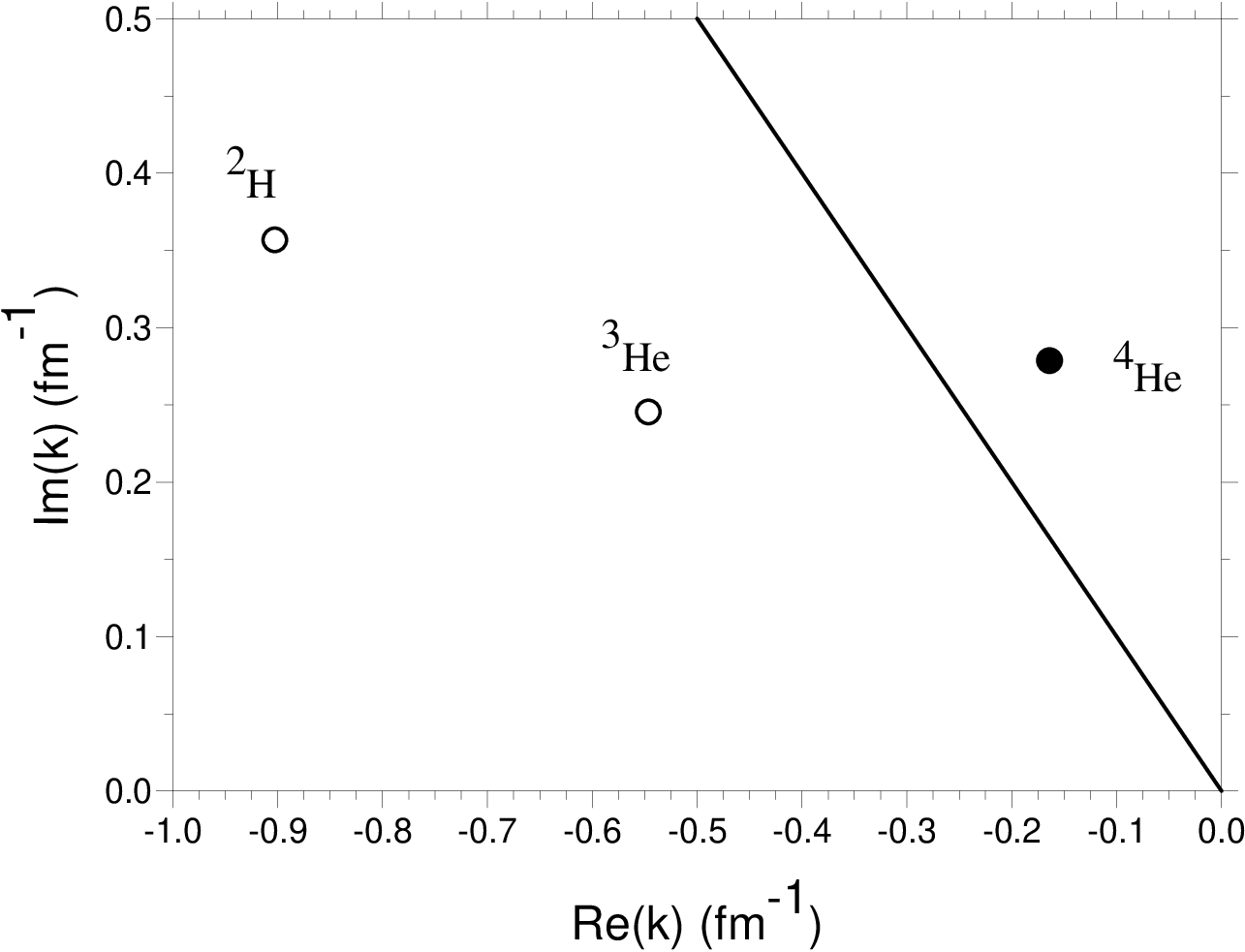}
\caption{The position of poles in the complex plane for the indicated nuclei. The calculations in the framework of FRA were performed with a scattering length $a_{\eta N} = (0.55 + i0.30)$~fm. Open symbols are resonances, the full dot indicates a quasi bound state (from \cite{Sofianos97}).}
\label{Fig:complex_Sofianos}
\end{center}
\end{figure}

It is interesting to have a closer look to the predictions for the very light nuclei. Sofianos and Rakityansky \cite{Sofianos97} calculated for a scattering length $a_{\eta N}=(0.55 + i0.30)$ fm, which is the scattering length extracted by Wilkin \cite{Wilkin93} from production data, the pole positions in the complex momentum plane, which are shown in Fig. \ref{Fig:complex_Sofianos}.
They applied the framework of FRA (see above). The two lighter nuclei do not bind the $\eta$ and form only resonances, whereas $^4$He does. When $\text{Re}(a_{\eta N})$ increases, all the poles move up and
to the right, and when a resonance pole crosses the diagonal it becomes a quasi bound pole. An overview of pole values $B_\eta$ and $\Gamma_\eta/2$ for $\eta$ light nuclei systems is given in \cite{Kelkar13}.

\subsection{Final State Interactions}
A possible way to extract the properties of a bound state is to extract the $\eta$ nucleus scattering length from the final state interaction \cite{Goldberger-Watson}. One has to measure an excitation function of a
reaction
\begin{equation} ^{Z_1}A_1+^{Z_2}A_2\to
\{^{Z_1}A_1+^{Z_2}A_2\}_{gs}+\eta
\end{equation}
with $\{^{Z_1}A_1+^{Z_2}A_2\}_{gs}$ the fused nuclear system in its ground state and the $\eta$ relative to that in a $s$ state. One can either measure the $\eta$ or the fused nuclear system. The measurement of the decay of the $\eta$ into photons does not allow the conclusion that the nuclear system is in its ground state due to the limited resolution in the two photon detection. Instead one measures the four vector of the fused nucleus together with particle identification and reconstructs the properties of the $\eta$. This, however, limits the method to light nuclei.

The method is to extract the effective range parameters from the matrix element
\begin{gather}\label{equ:cross_section}
|f_s|^2=\frac{d\sigma_s}{d\Omega}\frac{p_i}{p_f}
\end{gather}
with $p$ the momenta in the incident and final state in the cm system and $d\sigma_s/d\Omega$ the $s$ wave part of the cross section, as will be discussed in the next section.  The parameters scattering length $a$ and effective range $r_0$ have to be complex since always the channel $\eta+N\to\pi+N$ is open. Because the square of $a$ is fitted to the data the sign of $a_r$ can not be found from such measurements. The case with more than one $s$ wave will be discussed below. Criteria for a bound state with and without effective range will be discussed already in section \ref{sub:Niskanen}.

One can naively assume that the $s$ wave part of the cross section close to threshold is just
\begin{equation}
\label{equ:s-wave-part}
\frac{d\sigma_s}{d\Omega} = \frac{\sigma_{tot}}{4\pi}\,.
\end{equation}
However, often other waves than just the $s$ wave contribute to the total cross section even close to threshold. In this case the decomposition of the total cross section into partial waves is possible from the knowledge of  spin observables in addition to cross sections.
In the following paragraphs we will give some theoretical prerequisites allowing to extract the $s$ wave contribution from measurements.

\subsubsection{Relation between pole parameters and effective range parameters}\label{sub:Niskanen}

In the simplest approach the $s$ wave amplitude is related to the
scattering length via
\begin{equation}
f_s(p)=\frac{f_B}{1-iap}
\label{equ:scattering_length}
\end{equation}
with $f_B$ the production amplitude.
$f_s(p)$ has a pole in the complex plane that occurs for
\begin{equation}
p_0=-\frac{i}{a}=\frac{(-i)a^*}{|a|^2}\,.
\label{equ:pole_position}
\end{equation}
With \begin{equation}
E=\frac{p_{0}^2}{2\mu_{\eta A}}
\end{equation} we find
\begin{equation}\label{equ:B-pole}
B_\eta =  \frac{a_r^2-a_i^2}{2\mu_{\eta A}|a|^4}
\end{equation}
and
\begin{equation}\label{equ:Gamma-pole}
-\frac{\Gamma_\eta}{2} =  \frac{2a_ra_i}{2\mu_{\eta A}|a|^4}\,.
\end{equation}
For $B_\eta >0$ follows from Eq. (\ref{equ:B-pole})
\begin{equation}\label{equ:cond_1}
|a_r|>|a_i|
\end{equation}
and from Eq. (\ref{equ:Gamma-pole})
\begin{equation}
a_r a_i<0\,.
\end{equation}
Unitarity requires $a_i>0$ and therefore
\begin{equation}\label{equ:cond_2}
a_r<0\,.
\end{equation}

When the \fsi expansion is extended to the second term [see Eq. (\ref{definition})] the effective range $r_0$ enters
\begin{equation}\label{equ:effective_range}
f_s(p)=\frac{f_B}{1-iak+\frac{1}{2}ar_0}\,.
\end{equation}
Recently Sibirsev et al. \cite{Sibirtsev04a} gave a more complete condition applicable for this case:
\begin{equation}
 \text{Re} [a^3(a^\ast - r_0^\ast)] > 0.
 \label{equ:cond_full}
\end{equation}

\begin{figure}[h!]
\centering
\includegraphics[width=0.6\textwidth]{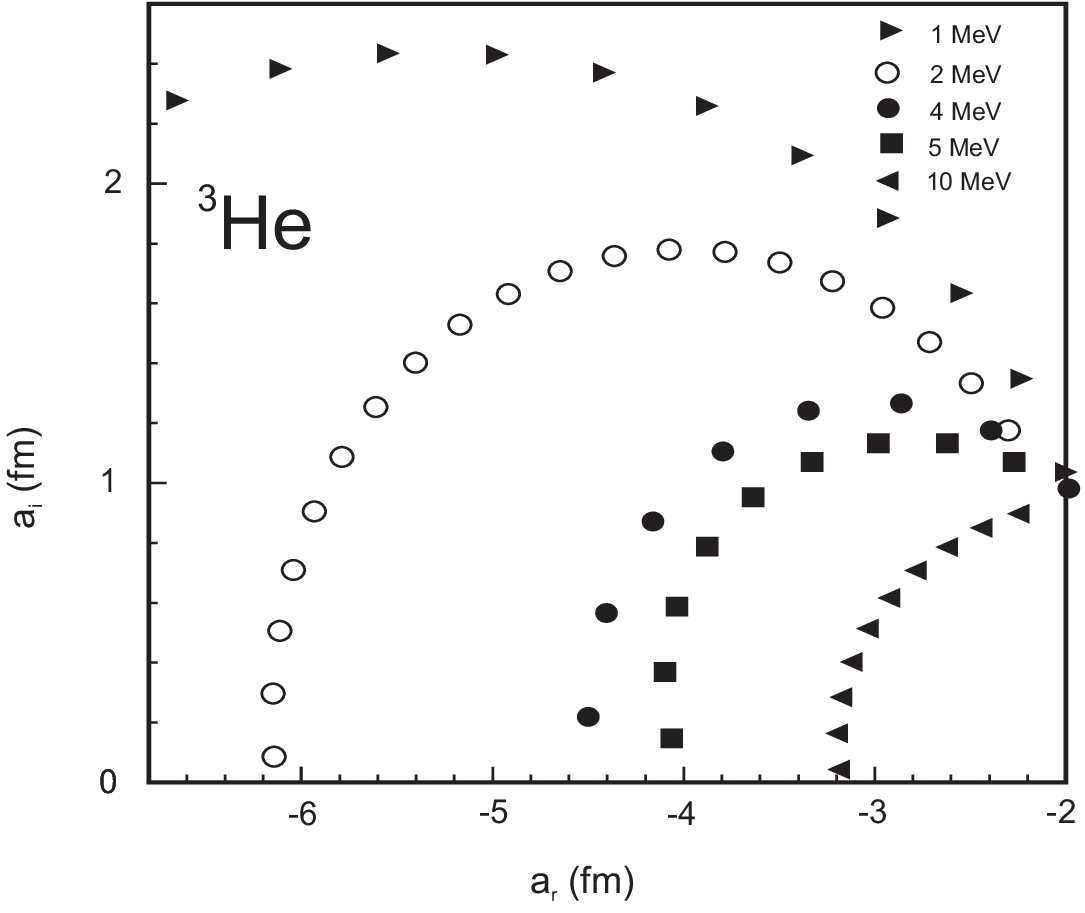}
\caption{The $s$-wave binding energy
$B_\eta  $ contours for 1, 2, 4, 5, and 10 MeV  in the complex $a_{\rm R},a_{\rm I}$ plane \cite{Sibirtsev04}.}
\label{Fig:He3-real}
\end{figure}

\begin{figure}[h!]
\centering
\includegraphics[width=0.6\textwidth]{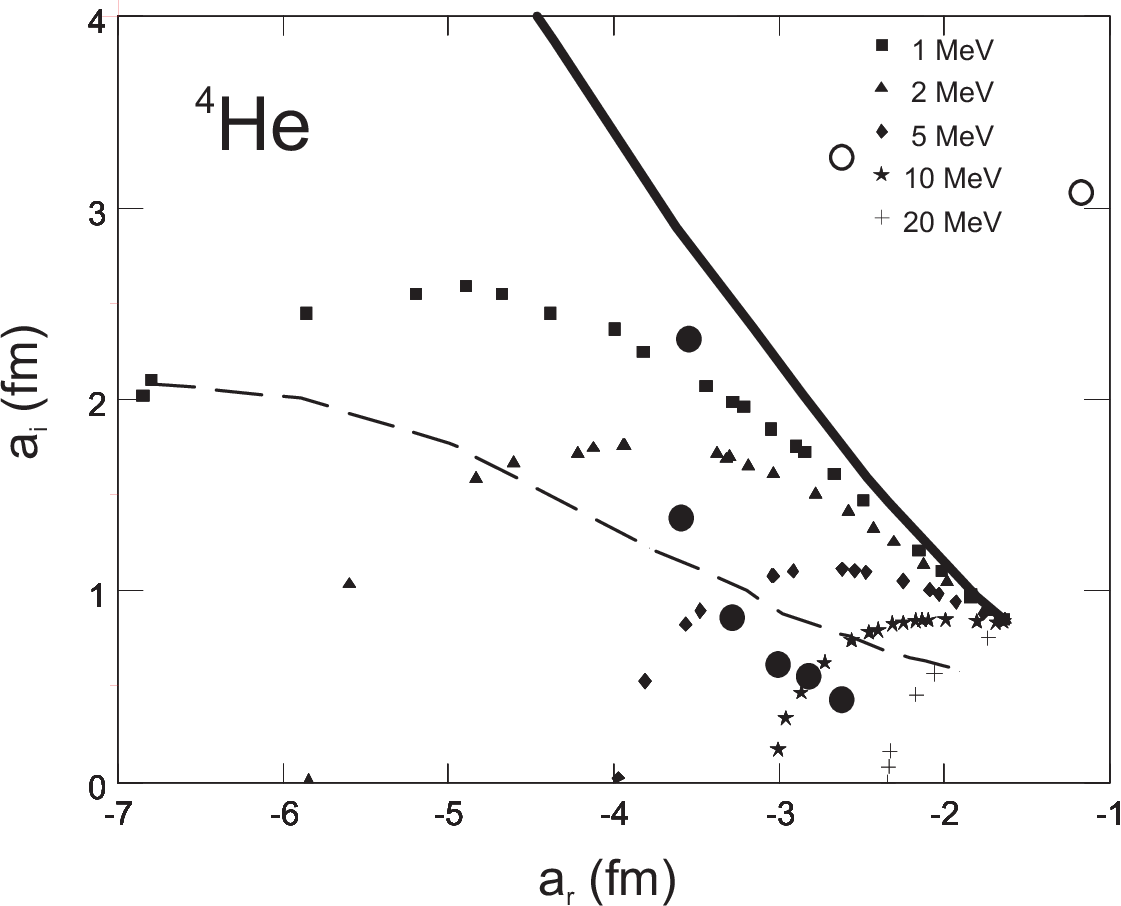}
\caption{The $s$-wave binding energy
$B_\eta  $ contours for 1, 2, 5, 10 and
20 MeV (see legend) in the complex $(a_{\rm R},a_{\rm I})$ plane (adapted from Ref. \cite{Niskanen13}).
The line shows the zero energy, {\it i.e.} above it there
is no binding as explained in the text. Below the dashed curve the relation $|B_\eta|>|\Gamma_\eta/2|$ is valid. The thick filled circles are possible bound states, the open circles unbound states within the finite range approximation \cite{Sofianos97} for different values of the $\eta$ N scattering length (see text). }
\label{Fig:He4-real}
\end{figure}

\begin{figure}[h!]
\centering
\includegraphics[width=0.6\textwidth]{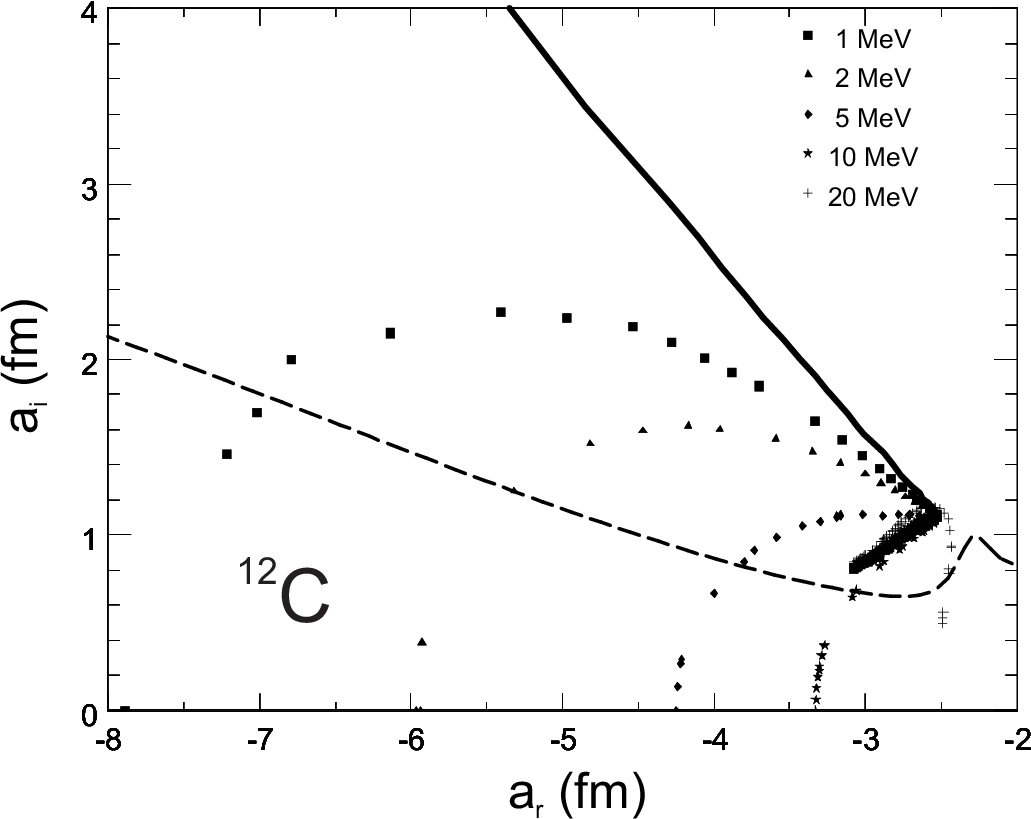}
\caption{Same as Fig. \ref{Fig:He4-real} but for {$^{12}$C}. }
\label{Fig:C12-real}
\end{figure}

\begin{figure}[h!]
\centering
\includegraphics[width=0.6\textwidth]{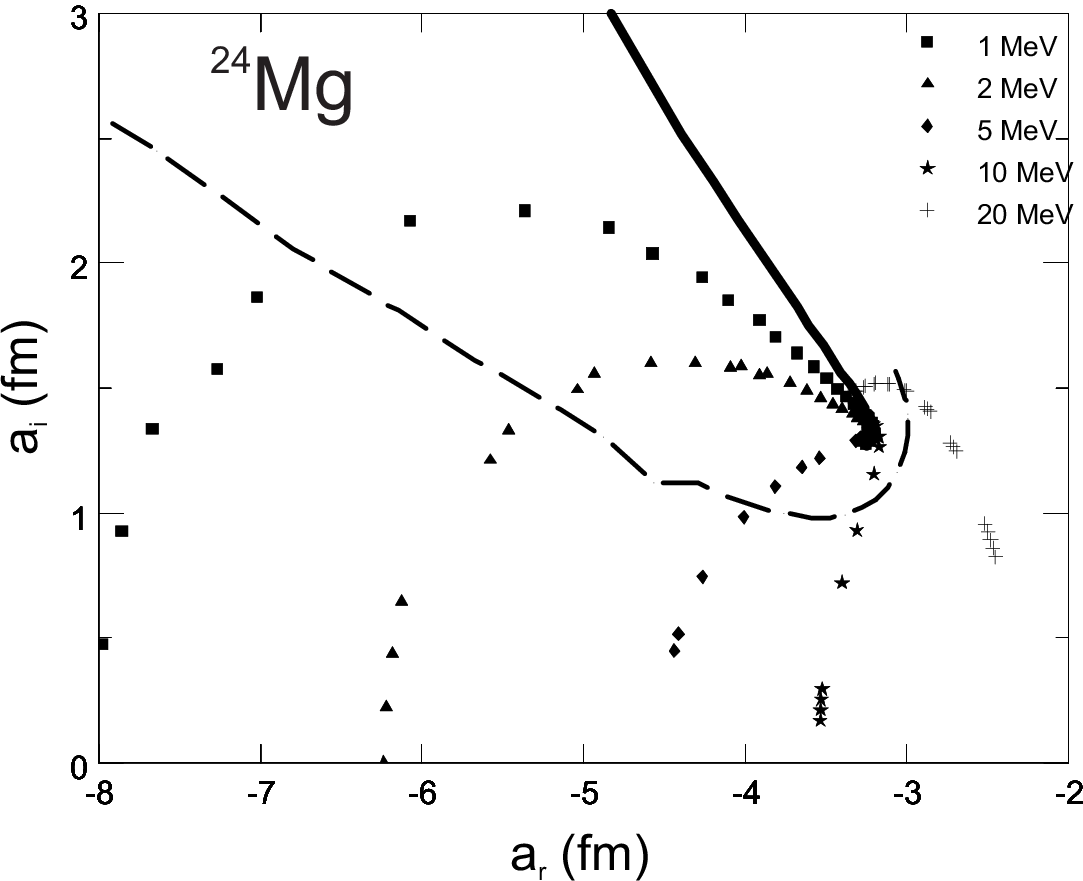}
\caption{Same as Fig. \ref{Fig:He4-real} but for {$^{24}$Mg}.}
\label{Fig:Mg24-real}
\end{figure}

In another publication  \cite{Sibirtsev04} they calculated the pole positions in the Lippman--Schwinger equation \eqref{equ:Lippman-Schwinger} in the case of $^3$He whereas Niskanen and Machner \cite{Niskanen13} extended these calculations to heavier nuclei. The real values of the poles, i.e. $B_\eta$ are shown in Figs. \ref{Fig:He3-real}, \ref{Fig:He4-real}, \ref{Fig:C12-real} and \ref{Fig:Mg24-real} as function of the real and imaginary parts of the $\eta$ nucleus scattering length.

The thick solid curves in the figures indicate the boundary for binding. Above this line $B_\eta>0$ and therefore no binding.

In Fig. \ref{Fig:He4-real} also calculations for the nucleus $^4$He are shown which were obtained within the framework of FRA \cite{Sofianos97} with scattering length $a_{\eta N}=0.2+0.1n+i0.3$ fm with $n$=1--8. The results are shown counter clockwise. Binding occurs for $\text{Re}(a_{\eta N})\geq 0.47$ fm. This is in agreement with the different calculational approach of \cite{Niskanen13}.

Results from Fix and Arenh\"{o}vel \cite{Fix02} for $^3$H, which should be almost the same as for $^3$He, are completely different than these results. For different values of the $\eta N$ scattering length they obtained within their model positive values for the real part of the $\eta ^3$H scattering length and even for a large value of the  scattering length $a_{\eta N}=0.75+i0.27$ fm they obtained $a_{\eta ^3\text{H}}=4.2+i5.7$ fm. This indicates the existence of a virtual state. Although the real part of the $\eta$ nucleus scattering length is negative for the different model parameters in the FRA approach \cite{Belyaev95} they violate the condition Eq. (\ref{equ:cond_1}) except for $^4$He. It is unclear to which extent the model assumption influences the extrapolation to negative energies.

As discussed above, one can only hope to find the bound state, if it exists, when the width or half width is smaller than the binding energy. This limit is indicated in the figures by dashed lines. Below these lines $|B_\eta|>|\Gamma_\eta/2|$.
\begin{figure}[h!]
\centering
\includegraphics[width=0.6\textwidth]{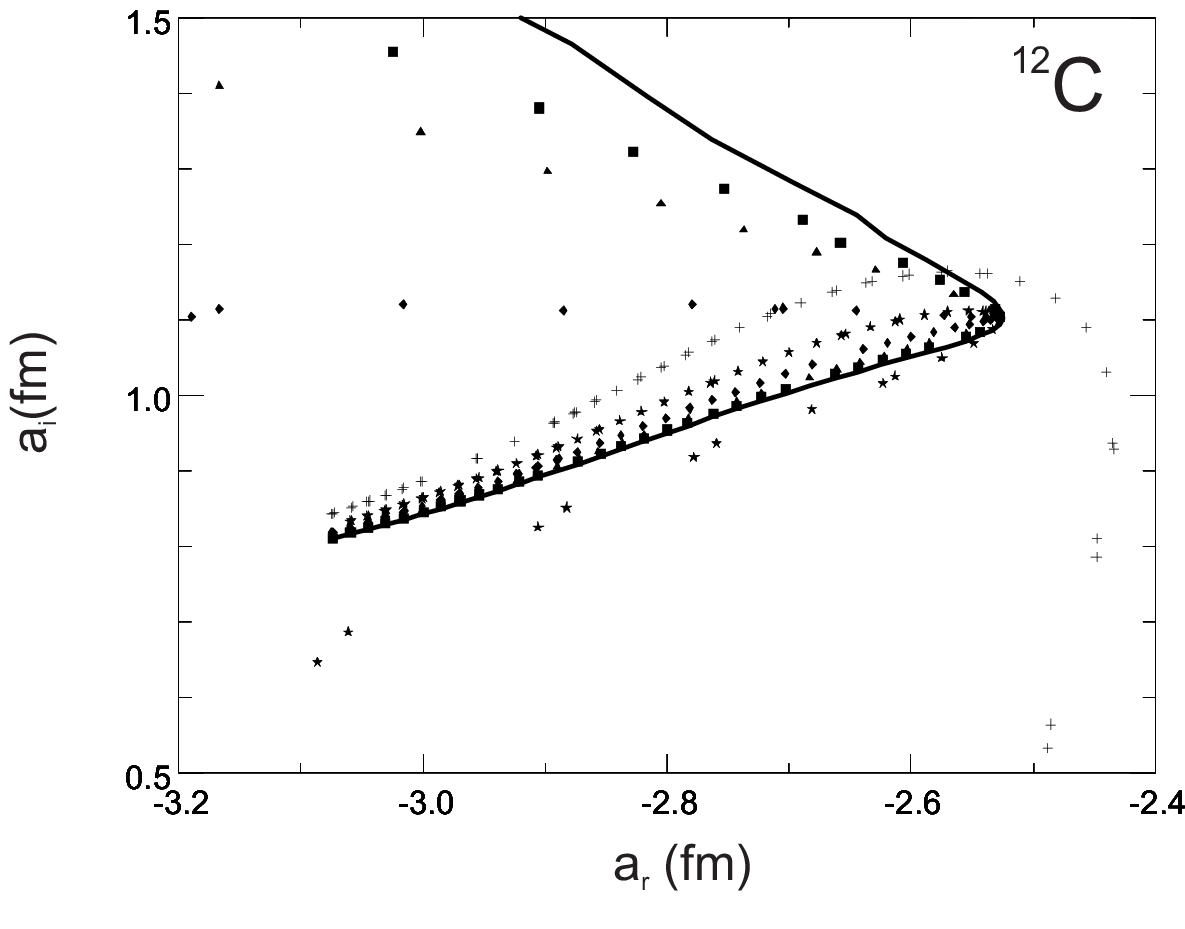}
\caption{Same as Fig. \ref{Fig:C12-real} except for an expanded scale. The expanded range shows the "back bending" region. }
\label{Fig:C12-expanded}
\end{figure}
In the case of carbon Fig. \ref{Fig:C12-real} a "back bending" like dependence is visible. This becomes more clear in Fig. \ref{Fig:C12-expanded} which is an expanded graph of the corresponding region. The upper branch corresponds to a rather weak potential while it is strong for the lower branch. It is just this region which is of interest for a bound system. The results are much more dense in this region than it is the case for the weak potential.
In Ref. \cite{Niskanen13} it was argued that a complex effective range has to be considered. They give a relation for this case between the effective range parameters and the binding energy \cite{Joachain79}, which reads when extended to the complex case as:
\begin{equation}
\frac{1}{a} = -\,\sqrt{-2\mu_{\eta A}E_\eta}
    - r_0\, \mu_{\eta A}\, E_\eta .
\label{relation}
\end{equation}
A bivariate expansion was found for the effective range:
\begin{equation}\label{equ:r_0}
r_{0,r}=c_r+d_ra_r+e_ra_i+f_ra_r^2+g_ra_i^2+h_ra_ra_i
\end{equation}
and
\begin{equation}\label{equ:r_1}
r_{0,i}=c_i+d_ia_r+e_ia_i+f_ia_r^2+g_ia_i^2+h_ia_ra_i.
\end{equation}
\begin{figure}[h!]
\begin{center}
\includegraphics[width=0.8\textwidth]{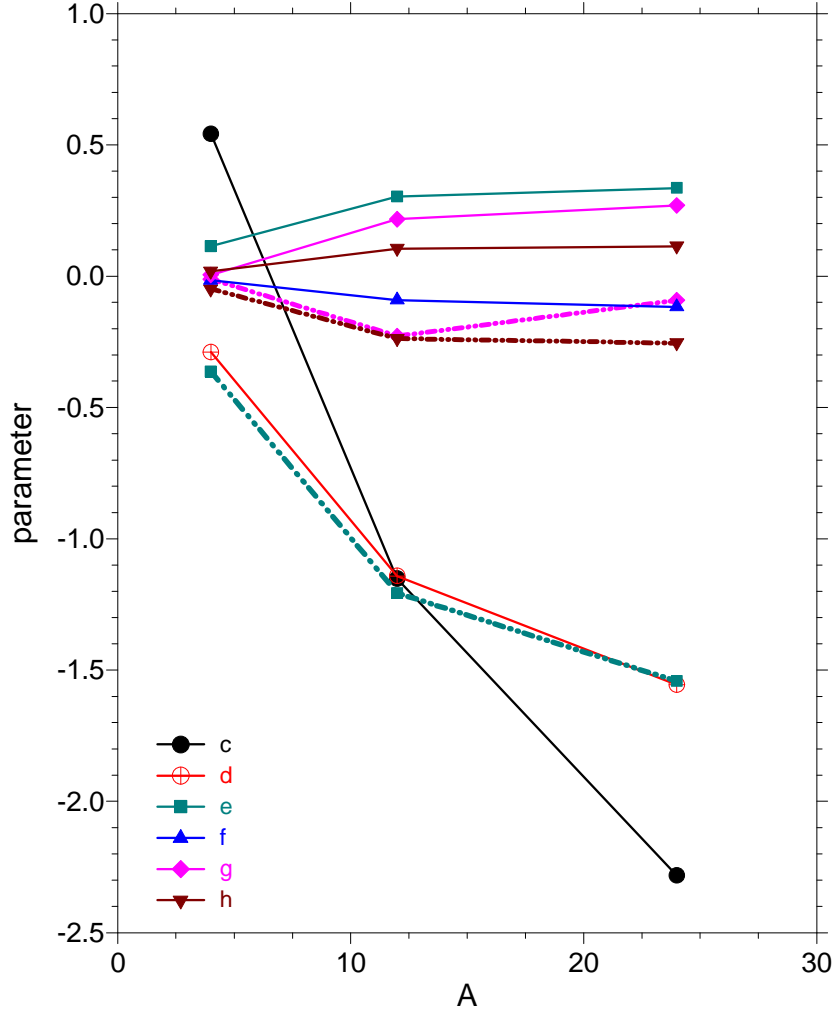}
\caption{The dependence of the parameters Eqs. (\ref{equ:r_0}) and (\ref{equ:r_1}) on the mass number $A$. The real parts are connected by solid lines, the imaginary parts by dotted lines. Parameters not included are compatible with zero. The dimensions of the parameters are such that effective range and scattering length are in fm.}
\label{Fig:R_0}
\end{center}
\end{figure}
The parameters $c-h$ are shown in Fig. \ref{Fig:R_0} for the real part as solid lines. Those for the imaginary part are shown as dotted lines. In this case $c$ and $d$ are zero. It is interesting to note that the parameters $c$ and $d$ show for the real part $r_{0,r}$ a strong mass dependence while it is weak for the other parameters. In case of the imaginary part $r_{0,i}$ only the parameter $e$ has the mass dependence. To summarise we have
\begin{gather}
r_{0,r}\approx c+da_r \\
r_{0,i}\approx ea_i\,.
\end{gather}

\subsubsection{Polarisation Observables}\label{sssec:Polaris-Observables}
The observables of reactions between particles with spin are  polarisation and analysing powers. Here we will concentrate on the case of a polarised beam and unpolarised target. We follow the Madison convention \cite{Madison70} in which a right handed coordinate system is used with the $z$ axis in beam direction. Polarisations are denoted by small letters, analysing powers by capital letters. In Cartesian coordinate system \cite{Goldfarb54} the polarisations are $p_i$ and $p_{i,j}$, the analysing powers $A_i$, $A_{i,j}$, $i,j=x,y,z$. In spherical coordinates \cite{Lakin55} $t_k$, $t_{k,q}$ and $T_k$, $T_{k,q}$, $|q|\leq k$.

For a large number of particles the polarisation is given by the statistical average of the individual polarisation whereas for a specific particle of spin $1$ the possible projection of the spin on the $z$ quantisation axis can be $-1,0,+1$. The vector and tensor polarisations of the beam can be defined respectively as
\begin{equation}
p_{z} = \frac{N_{+}-N_{-}}{N_{+}+N_{0}+N_{-}},
\label{Pzjeden}
\end{equation}
\begin{equation}
p_{zz} = \frac{N_{+}+N_{-}-2N_{0}}{N_{+}+N_{0}+N_{-}},
\label{Pzzjeden}
\end{equation}
where $N_{+}$, $N_{-}$ and $N_{0}$ denote the number of particles with spin projection $+1$, $0$ and $-1$, respectively. The allowed value of polarisation fulfills the following conditions $\vert p_{z} \vert \leq1$ and $-2 \leq p_{zz} \leq 1$.

An extensive discussion of reactions with polarised beams is given for instance in Ref. \cite{Ohlsen72}; we will review only the essentials here.

The reaction is determined by the unpolarised cross section and the analysing powers $T_{ij}$.  For the case of polarised particles in a cyclic accelerator with magnetic field perpendicular to the beam direction we have
{\setlength\arraycolsep{2pt}
\begin{eqnarray}
\Big( \frac{d\sigma}{d\Omega} (\theta_\eta, \phi_\eta) \Big)_{pol} & = &  \Big( \frac{d\sigma}{d\Omega} (\theta_\eta) \Big)_{unpol} \Big[1 + i\sqrt2 t_{10}T_{11}(\theta_\eta)cos\phi_\eta
\nonumber\\
& & {} -\frac{1}{2}t_{20}T_{20}(\theta_\eta) - \sqrt\frac{3}{2}t_{20}T_{22}(\theta_\eta)cos2\phi_\eta \Big]
\label{equ:dsigma_rho_tensors}
\end{eqnarray}
with $\theta_\eta$ the polar angle and $\phi_\eta$ the azimuth angle of the emitted $\eta$ meson.
Equivalently, in Cartesian notation:
\begin{gather}
\Big( \frac{d\sigma}{d\Omega} (\theta_\eta, \phi_\eta) \Big)_{pol}  = \Big( \frac{d\sigma}{d\Omega} (\theta_\eta) \Big)_{unpol} \Big[1 + \frac{3}{2}A_{y}(\theta_\eta)p_{z}cos\phi_\eta
\nonumber\\
  + \frac{1}{4}p_{zz}\Big( A_{yy}(\theta_\eta)(1+cos2\phi_\eta) + A_{xx}(\theta_\eta)(1 - cos2\phi_\eta) \Big) \Big].
\label{equ:dsigma_rho_cartesian}
\end{gather}
The relations between the analysing powers  in polar coordinates Eq. (\ref{equ:dsigma_rho_tensors}) and in Cartesian coordinates Eq.  (\ref{equ:dsigma_rho_cartesian}) are
\begin{eqnarray}
A_{y}(\theta_\eta) & = & \frac{2}{\sqrt3}iT_{11}(\theta_\eta),
\\ \label{Ay}
A_{yy}(\theta_\eta) & = & -\sqrt{2}\Bigg[ \frac{1}{2}T_{20}(\theta_\eta) + \sqrt{\frac{3}{2}}T_{22}(\theta_\eta) \Bigg],
\\ \label{Ayy}
A_{xx}(\theta_\eta) & = & -\sqrt{2}\Bigg[ \frac{1}{2}T_{20}(\theta_\eta) - \sqrt{\frac{3}{2}}T_{22}(\theta_\eta) \Bigg],
\label{Axx}
\end{eqnarray}
while the relation between vector and tensor polarisation in spherical and Cartesian representation is as follows
\begin{eqnarray}
t_{10}& = &\sqrt{\frac{3}{2}}p_{z}, \\
\label{t10_pz}
t_{20} & = & \frac{1}{\sqrt{2}}p_{zz}.
\label{t20_pzz}
\end{eqnarray}

\subsubsection{The Reaction $dd\to \eta\alpha$}\label{sub:dd2eta}

We will first discuss the reaction $dd\to \eta\alpha$. The two deuterons have positive parity like the $\alpha$ particle. The $\eta$ meson has spin and parity $0^-$. In deriving the possible transitions and corresponding amplitudes we apply the following symmetries or conservation laws:
\begin{align}
&L+S \text{ even} & \text{Bose Einstein symmetry},\\
&\vec{J}=\vec{L}+\vec{S}=\vec{J'}=\vec{L'}+\vec{S'}& \text{total angular momentum conservation},\\
&P=P'& \text{parity conservation},\\
&T=T'& \text{isospin conservation.}
\end{align}
Since in the entrance channel we are dealing with two identical particles the total wave function has to be symmetric and Bose Einstein statistics requires $S+L=$even. The two deuteron spins can couple to $S=2,\,1,\,0$. Bose Einstein symmetry requires then $L=2$ or $0$ for the $S=2$, $L=1$ or 3 for the $S=1$ and $L=2$ for $S=0$. In Table \ref{tab:dd-Amplitudes} we give these quantum numbers together with the parity $\pi$, and the total angular momentum $J$. Since both particles in the exit channel, i. e. the $\alpha$ particle and the $\eta$ have spin zero, the relation $L'=J$ holds.

\begin{table}[htb]
\caption{Allowed transitions and the corresponding partial wave amplitudes.}
\label{tab:dd-Amplitudes}
\begin{center}
\begin{tabular}{|c|c|c|c|c|c|c|c|}
\hline
$S$ & $L$ & $\pi$ & $J$ & $S'$& $L'$ & wave & amplitude \\
\hline
2 & 2 & +1 & 3 &0& 3 & $f$ & $a_4$ \\
2 & 2 & +1 & 1 &0& 1 & $p$ & $a_1$ \\
1 & 3 & -1 & 2 &0& 2 & $d$ & $a_3$ \\
1 & 1 & -1 & 2 &0& 2 & $d$ & $a_2$ \\
1 & 1 & -1 & 0 &0& 0 & $s$ & $a_0$ \\
\hline
\end{tabular}
\end{center}
\end{table}

In the range of angular momenta shown in the table only five partial waves amplitudes exist. For the energy range of interest we can safely ignore the $f$ wave amplitude $a_4$ so that $s$, $p$ and two $d$ waves remain. In the above discussion we have ignored isospin conservation so far. The initial state has $T=0$ as the final state $T'=0$ whereas in $\pi^0$ production $T'=1$ and hence the cross section in this case is strongly suppressed.

We now will derive the amplitudes given in the table. For that let us consider a general case of two-body scattering
\begin{equation}
a+b\to c+d,
\label{reaction}
\end{equation}
where $a$ denotes beam particle, $b$ target, $c$ and $d$ reaction products. With $s_i$ we denote the spins of the particles. In case of an unpolarised beam and target, the differential cross section can be expressed in the following way
\begin{equation}
\Bigg(\frac{d\sigma}{d\Omega}\Bigg)_{unpol} = \sum_{m_{a}m_{b}m_{c}m_{d}}\Big|F^{m_{a}m_{b}}_{m_{c}m_{d}}\Big |^{2},
\label{amplitude_2}
\end{equation}
where $m_{a}$, $m_{b}$, $m_{c}$, $m_{d}$ are the magnetic quantum numbers of the particles in a suitable frame. The scattering amplitude is written as
\begin{eqnarray}
F^{m_{a}m_{b}}_{m_{c}m_{d}} = \sum_{SLS'L'J} & &\langle s_{a},m_{a};s_{b},m_{b}|S,m_{a}+m_{b}\rangle \langle S,m_{a}+m_{b};L,0|J,m_{a}+m_{b} \rangle
\nonumber\\
 & & \langle s_{c},m_{c};s_{d},m_{d}|S',m_{c}+m_{d} \rangle
\nonumber\\
 & & \langle S',m_{c}+m_{d};L',m_{a}+m_{b}-m_{c}-m_{d}|J,m_{a}+m_{b}\rangle
\nonumber\\
& & {} a_iY^{m_{a}+m_{b}-m_{c}-m_{d}}_{L'}.
\label{Fmambmcmd}
\end{eqnarray}
The first Clebsch-Gordan coefficient is to couple the spins of the two particles in the entrance channel to a total spin $S$, the second to couple this spin with the angular momentum $L$ to the total angular momentum $J$. The next two account for the same couplings in the exit channel. $a_i$ with $i=\{S,L,S',L',J\}$ denotes the partial amplitude (see Table \ref{tab:dd-Amplitudes}) and the $Y^{m_{a}+m_{b}-m_{c}-m_{d}}_{L'}$ are the usual spherical harmonics.

In the following Tables \ref{tab:cross} and \ref{tab:T2} the decomposition of the observables $d\sigma/d\cos \theta_{unpol}$ and $T_{20}$ and $T_{22}$  for complex partial wave amplitudes $a_i$ is given. The tables should be read as $$d\sigma/d\cos\theta_\eta=\sum_{ij} CG_{ij}\text{Re}(a_ia_j^*| YY^*)$$ with $CG_{ij}$ the Clebsch-Gordan coefficients and $Y$ the spherical harmonics.

\begin{table}[!ht]
\caption{The decomposition of the unpolarised cross section into partial wave amplitudes. The values in the first column give the Clebsch Gordan coefficients, the last column the spherical function.}
\label{tab:cross}
\begin{center}
\begin{tabular}{|rcc|rcc|rcc|}
                    \hline
 & $\Big( \frac{d\sigma}{d\cos\theta_\eta} \Big)_{unpol}$ &  \\
\hline
$\frac{1}{27}$ & $|a_{0}|^{2}$ & Y$_0^{0}$Y$_0^{0*}$  \\

$\frac{1}{15}$ & $|a_{1}|^{2}$ & Y$_1^{1}$Y$_1^{1*}$  \\

$\frac{1}{9}$  & $|a_{2}|^{2}$ & Y$_2^{1}$Y$_2^{1*}$  \\

$\frac{2}{27}$ & $|a_{2}|^{2}$ & Y$_2^{0}$Y$_2^{0*}$  \\

$\frac{2}{63}$ & $|a_{3}|^{2}$ & Y$_2^{1}$Y$_2^{1*}$  \\

$\frac{1}{21}$ & $|a_{3}|^{2}$ & Y$_2^{0}$Y$_2^{0*}$  \\

$-\frac{2\sqrt{2}}{27}$  & $\text{Re}(a_0a_2^*)$ & Y$_0^{0}$Y$_2^{0*}$ \\

$\frac{2\sqrt{7}}{63}$   & $\text{Re}(a_0a_3^*)$ & Y$_0^{0}$Y$_2^{0*}$ \\

$\frac{2\sqrt{14}}{63}$   & $\text{Re}(a_2a_3^*)$ & Y$_2^{1}$Y$_2^{1*}$ \\

$-\frac{2\sqrt{14}}{63}$ & $\text{Re}(a_2a_3^*)$ & Y$_2^{0}$Y$_2^{0*}$ \\
\hline
\end{tabular}
\end{center}
\end{table}

\begin{table}[!ht]
\caption{Same as Table \ref{tab:cross} but for the analysing powers.}
\label{tab:T2}
\begin{center}
\begin{tabular}{|rcc|rcc|}
                    \hline
 & $\Big( \frac{d\sigma}{d\cos\theta_\eta} \Big)_{unpol}T_{20}$ & & &$\Big( \frac{d\sigma}{d\cos\theta_\eta} \Big)_{unpol} T_{22}$ & \\
\hline
 $\frac{\sqrt{2}}{54}$ & $|a_{0}|^{2}$ & Y$_0^{0}$Y$_0^{0*}$ &  $\frac{\sqrt{30}}{60}$ & $|a_{1}|^{2}$ & Y$_1^{1}$Y$_1^{1*}$    \\

 $-\frac{\sqrt{2}}{60}$ & $|a_{1}|^{2}$ & Y$_1^{1}$Y$_1^{1*}$ &  $\frac{\sqrt{3}}{36}$ & $|a_{2}|^{2}$ & Y$_2^{1}$Y$_2^{1*}$    \\

 $-\frac{\sqrt{2}}{36}$ & $|a_{2}|^{2}$ & Y$_2^{1}$Y$_2^{1*}$ & $\frac{\sqrt{3}}{126}$ & $|a_{3}|^{2}$ & Y$_2^{1}$Y$_2^{1*}$    \\

 $\frac{\sqrt{2}}{27}$ & $|a_{2}|^{2}$ & Y$_2^{0}$Y$_2^{0*}$  & $\frac{\sqrt{5}}{30}$ & $\text{Re}(a_1a_2^*)$ & Y$_1^{1}$Y$_2^{1*}$ \\

 $-\frac{\sqrt{2}}{126}$ & $|a_{3}|^{2}$ & Y$_2^{1}$Y$_2^{1*}$  & $\frac{\sqrt{70}}{210}$ & $\text{Re}(a_1a_3^*)$ & Y$_1^{1}$Y$_2^{1*}$ \\

 $\frac{\sqrt{2}}{42}$ & $|a_{3}|^{2}$ & Y$_2^{0}$Y$_2^{0*}$  & $\frac{\sqrt{42}}{126}$ & $\text{Re}(a_2a_3^*)$ & Y$_2^{1}$Y$_2^{1*}$ \\

$-\frac{2}{27}$ & $\text{Re}(a_0a_2^*)$ & Y$_0^{0}$Y$_2^{0*}$ & & & \\

$\frac{\sqrt{14}}{63}$ & $\text{Re}(a_0a_3^*)$ & Y$_0^{0}$Y$_2^{0*}$ & & & \\

$\frac{\sqrt{30}}{30}$ & $\text{Re}(a_1a_2^*)$ & Y$_1^{1}$Y$_2^{1*}$ & & & \\

$\frac{\sqrt{105}}{105}$ & $\text{Re}(a_1a_3^*)$ & Y$_1^{1}$Y$_2^{1*}$ & & & \\

$-\frac{\sqrt{7}}{63}$ & $\text{Re}(a_2a_3^*)$ & Y$_2^{1}$Y$_2^{1*}$ & & & \\

$-\frac{2\sqrt{7}}{63}$ & $\text{Re}(a_2a_3^*)$ & Y$_2^{0}$Y$_2^{0*}$ & & & \\
\hline
\end{tabular}
\end{center}
\end{table}

In order to establish the functional dependence of the amplitudes on emission angle we rewrite the cross section as
\begin{equation}
\frac{d\sigma}{d\Omega} =\frac{1}{(2s_a+1)(2s_b+1)}
 \frac{p_c}{p_a} |A|^2
 \label{equ:cross-section}
\end{equation}
with $a=d=$deuteron and $c=\eta$.
The elements of the scattering amplitude $A$ can be denoted by $b_i$ and we get the relation
\begin{equation}
b_i=\frac{a_i}{3\sqrt{4\pi(2L+1)}}\sqrt{ \frac{p_a}{p_c}}\,.
\end{equation}
The amplitudes ($a_i$ and $b_i$) should not be mixed with the particles $a$ and $b$ (no index) in the entrance channel.

The angle dependence of the scattering amplitude is
\begin{gather}
|A(\theta_\eta)|^2 = |b_0|^2+\frac{9}{2}|b_1|^2 + \frac{5}{2}|b_2|^2+  \frac{15}{4}|b_3|^2  \nonumber \\
-5\sqrt{\frac{3}{2}}\text{Re}(b_2b_3^*) + \sqrt(10)\text{Re}(b_0b_2^*) - \sqrt{15}\text{Re}(b_0b_3^*)\nonumber \\
+\cos^2(\theta_\eta)\left[30\sqrt{6} \text{Re}(b_2b_3^*) -3\sqrt{10}\text{Re}(b_2b_0^*) +3\sqrt{15}\text{Re}(b_3b_0^*) \right. \nonumber \\ \left. - \frac{9}{2}|b_1|^2 + \frac{15}{2}|b_2|^2 -\frac{15}{2}|b_3|^2
\right] \nonumber \\
+\cos^4(\theta_\eta) \left[\frac{75}{4}|b_3|^2 - 75\sqrt{\frac{3}{2}}\text{Re}(b_2b_3^*)\,\,  \right]\,.
\label{gat:ampldd}
\end{gather}

Although we have limited ourselves to only $s,\,p$ and $d$ waves we have in total four complex parameters. That is by far too many to extract them from the total cross section. This will be discussed further in section \ref{sub:Reaction_dd}. If data do not show a $\cos^4(\theta_\eta)$ dependence then the amplitude $b_3$ is either very small or zero. In that case we have
\begin{gather}
|A(\theta_\eta)|^2 = |b_0|^2+\frac{9}{2}|b_1|^2 + \frac{5}{2}|b_2|^2 \\ \nonumber
+\cos^2(\theta_\eta)\left[-3\sqrt{10}\text{Re}(b_2b_0^*) \right.  \left. - \frac{9}{2}|b_1|^2 + \frac{15}{2}|b_2|^2
\right] \,.
\label{gat:ampldd_red}
\end{gather}
The term proportional to  $\cos^2(\theta_\eta)$ depends on a $p$ wave and a $d$ wave and $s-d$ wave interference.

Another possibility to account for the observables is to apply  helicity amplitudes.
The cross section is given similar to Eq. (\ref{equ:cross-section}) by
\begin{equation}\label{equ:unpolarised}
\frac{d\sigma}{d\Omega} =\frac{(2s_c+1)(2s_d+1)}{(2s_a+1)(2s_b+1)}\frac{p_c}{p_a}|I|^2.
\end{equation}
with $a$, $b$, $c$, $d$ the projectile, target, and the two  ejectiles, respectively. For the present reaction $a=b=$deuteron, $c=\eta$, and $d=\alpha$.
Due to the identical nature of the incident deuterons only one spin factor is in the denominator and the spin degeneracy ratio first fraction bar is 2/3. Because of the same reason only three independent scalar amplitudes are necessary to describe the spin
dependence of the reaction. These three helicity amplitudes $A$, $B$
and $C$  correspond to helicities +1, 0, and $-1$ \cite{Conzett85}. If we let the incident deuteron cms
momentum be $\vec{p}_d$ and that of the $\eta$ be $\vec{p}_{\eta}$,
then one choice for the structure of the transition matrix \cite{Wronska05}
$\mathcal{M}$ is%
\begin{eqnarray}
\nonumber \mathcal{M}&=&A({\vec{\epsilon}}_1\times{\vec{\epsilon}}_2)
\cdot{\hat{p}_d} +B({\vec{\epsilon}}_1\times{\vec{\epsilon}}_2)\cdot
\left[\hat{p}_d\times(\hat{p}_{\eta}\times\hat{p}_d)\right]
(\hat{p}_{\eta}\cdot{\hat{p}_d})\\
&&\hspace{-2mm}+C\left[({\vec{\epsilon}}_1\cdot{\hat{p}_d})\,
{\vec{\epsilon}}_2\!\cdot\!(\hat{p}_{\eta}\times\hat{p}_d)
+({\vec{\epsilon}}_2\cdot{\hat{p}_d})\,
{\vec{\epsilon}}_1\!\cdot\!(\hat{p}_{\eta}\times\hat{p}_d) \right],
\label{A1}
\end{eqnarray}
where the ${\vec{\epsilon}}_i$ are the polarisation vectors of the
two deuterons and $\hat{p_i}=\vec{p_i}/p_i$ the unit vector. The matrix $|I|^2$ which is essentially the unpolarised cross section is then obtained by averaging over the spin directions of both deuterons
\begin{equation}
|I|^2 =\frac{1}{9} \sum_{m_1,m_2} \mathcal{M}\mathcal{M^\mathcal{y}}\,.
\end{equation}

If we restrict ourselves up to $d$ waves in the final system, then $B$ and $C$ do not depend implicitly on the emission angle. This leads to the expressions\footnote{We use here capital letters for the analysing powers according to the Madison convention \cite{Madison70} although the authors of the original publication used small letters.}:
\begin{eqnarray}
|I(\theta_\eta)|^2&=&\left(|A|^2+|B|^2\sin^2\theta_\eta\cos^2\theta_\eta
+|C|^2\sin^2\theta_\eta\right),\label{INT}\\
|I(\theta_\eta)|^2\,T_{20}(\theta_\eta)&=&\frac{1}{2\sqrt{2}}\left(2|A|^2-|B|\sin^2\theta_\eta
\cos^2\theta_\eta\right. \nonumber \\
&&\left.-|C|^2\sin^2\theta_\eta-6\,\text{Re}\{B^*C\}\sin^2\theta_\eta
\cos\theta_\eta\right), \label{t20}\\
|I(\theta_\eta)|^2\,T_{21}(\theta_\eta)&=&-\frac{\sqrt{3}}{2}\,\text{Re}\left\{A^*(B\cos\theta_\eta
+C)\right\}\sin\theta_\eta, \label{t21}\\
|I(\theta_\eta)|^2\,T_{22}(\theta_\eta)&=&\frac{\sqrt{3}}{4}\,|B\cos\theta_\eta
-C|^2\sin^2\theta_\eta,\label{t22}\\
|I(\theta_\eta)|^2\,iT_{11}(\theta_\eta)&=&\frac{\sqrt{3}}{2}\,Im\left\{A^*(B
\cos\theta_\eta+C)\right\}\sin\theta_\eta,\\
|I(\theta_\eta)|^2\,T_{10}(\theta_\eta)&=&0\,.\label{t11}
\end{eqnarray}

The helicity amplitude $A$ contains an angular dependence which can be accounted for by
\begin{equation}
A(\theta_\eta) = A_0 + A_2\,P_2(\cos\theta_\eta)
\label{equ:A0-A2}
\end{equation}
if we still restrict ourselves up to $d$ waves. Then $|A|^2$ in Eqs. [\ref{INT}]-[\ref{t11}] has to be replaced by $|A_0|^2+2\,\text{Re}\{A_0A_2^*\}P_2+|A_2|^2P_2^2$. So far only one data set exists. If one wants to study $\eta$ production, it may be useful to treat the energy dependence of the amplitudes explicitly: $A_2\to p_\eta A_2$, $B\to p_\eta^2 B$ and $C\to p_\eta C$.

Thus, measurements of the angular distributions of the unpolarised
cross section $d\sigma(\theta_\eta)/d\Omega$, of $T_{20}(\theta_\eta)$, and of $iT_{11}(\theta_\eta)$, would allow one to extract
the values of $|A_0|^2$, $\text{Re}(A_0^*A_2)$, $|C|^2$, $\text{Im}(A_0^*B)$,
and $\text{Im}(A_0^*C)$. This would then lead to two two--fold
ambiguities that could only be resolved by the measurement of
$T_{21}$.

Inspection of Eq. (\ref{INT}) yields two terms proportional to $\cos^4\theta_\eta$ (when $\sin^2\theta =1-\cos^2\theta$ is applied): one proportional to $|B|^2$ and one proportional to $|A_2|^2$. Hence, these two amplitudes are related to the two $d$ waves. If we restrict ourselves to only $s$ and $p$ waves we have the following relations between partial wave amplitudes and helicity amplitudes:
\begin{eqnarray}
A&=& \sqrt{\frac{3}{2}}b_0 \\
C&=& \frac{\sqrt{27}}{4}b_1.
\end{eqnarray}

In Ref. \cite{Wronska05} the relations between analysing powers and invariant amplitudes are given in polar coordinates. They can be converted into Cartesian coordinates employed here by the transformation Eq. (\ref{Axx}). We can then write
\begin{eqnarray}
\nonumber \left(1 - A_{xx}\right)\frac{d\sigma}{d\Omega} &=&
\frac{p_{\eta}}{p_d}\left[|A_0|^2
+2\text{Re}(A_0A_2^*)P_2(\cos\theta_\eta)+|A_2|^2\left(P_2(\cos\theta_\eta)
\right)^2\right]\,,\\
\label{comb1}\\
\left(1 + 2 A_{xx}\right)\frac{d\sigma}{d\Omega}
&=&2\frac{p_{\eta}}{p_d}\left(|B|^2\sin^2\theta_\eta\cos^2\theta_\eta+|C|^2\sin^2\theta_\eta\right)\,,
\label{comb2}
\end{eqnarray}
where the results have been expressed in terms of convenient linear combinations. From these it is seen that both the cross section and $A_{xx}$ are even functions of $\cos\theta_\eta$.

\subsubsection{$pd\to \eta^3$He Reaction}\label{sub:pd_theory}
The reaction $pd\to \eta^3$He is studied quite extensively. Here we follow the prescription discussed in the previous section.
In order to dig out possible transitions from initial states, indicated by unprimed quantities, to final states, indicated by primed quantities, we make use of conservation of total spin $J$, parity $\pi$ and isospin $T$. The isospin in the entrance channel is $T(p)+T(d)=1/2+0=1/2$ while in the final channel we have $T(\eta)+T(^3\text{He})=0+1/2=1/2$. The particles in the incident channel can couple to total spin $3/2$ and $1/2$. In the exit channel the total spin is $S=0+1/2=1/2$. The particles in the entrance channel have positive parity as has the $^3$He. The $\eta$ has parity $\pi_\eta = -1$. Hence we have
\begin{equation}
\pi = (-1)^L=(-1)^{L'\pm 1}
\end{equation}
or $\Delta L=L'-L=\text{odd}$. All allowed transitions are given in Table \ref{tab:pd-Amplitudes}.
\begin{table}[ht]
\caption{Allowed transitions and the corresponding partial wave amplitudes for the reaction $pd\to \eta^3$He.}
\label{tab:pd-Amplitudes}
\begin{center}
\begin{tabular}{|c|c|c|c|c|c|c|c|}
\hline
$S$ & $L$ & $\pi$ & $J$ & $S'$ & $L'$ & wave & amplitude \\
\hline
3/2 & 0 & +1 & 3/2 & 1/2 &1& $p$  & $a_6$ \\
3/2 & 1 & -1 & 5/2 & 1/2 &2& $d$  & $a_8$ \\
1/2 & 0 & +1 & 1/2 & 1/2 &1& $p$  & $a_3$ \\
1/2 & 1 & -1 & 3/2 & 1/2 &2& $d$  & $a_7$ \\
3/2 & 1 & -1 & 1/2 & 1/2 &0& $s$  & $a_1$ \\
1/2 & 1 & -1 & 1/2 & 1/2 &0& $s$  & $a_0$ \\
3/2 & 2 & +1 & 3/2 & 1/2 &1& $p$  & $a_2$ \\
1/2 & 2 & +1 & 3/2 & 1/2 &1& $p$  & $a_4$ \\
3/2 & 2 & +1 & 1/2 & 1/2 &1& $p$  & $a_5$ \\
\hline
\end{tabular}
\end{center}
\end{table}
In summary we have two independent $s$ waves and five $p$ waves in the exit channnel. This is much more than in the case of the $dd\to\eta\alpha$ reaction. Similar to Eq. [\ref{gat:ampldd}] we can write down the scattering amplitude squared:
\begin{gather}
|A|^2 =  2|b_0|^2 +2|b_1|^2 +4|b_2|^2 +2|b_3|^2 +2|b_4|^2 +2|b_5|^2 +4|b_6|^2 \nonumber \\
+\cos(\theta_\eta)\left[-4\sqrt{2}\text{Re}(b_1b_6^*) +4\sqrt{2}\text{Re}(b_1b_2^*) +4\text{Re}(b_0b_3^*) +8\text{Re}(b_0b_4^*)\right. \nonumber  \\  \left.  +4\text{Re}(b_2b_6^*) +4\text{Re}(b_1b_5^*) +2\sqrt{2}\text{Re}(b_5b_6^*) - 2\sqrt{2}\text{Re}(b_2b_5^*) -4\text{Re}(b_3b_4^*)\right] \nonumber  \\
+\cos^2(\theta_\eta) \left[ -12\text{Re}(b_2b_6^*) -6\sqrt{2}\text{Re}(b_5b_6^*) +6\sqrt{2}\text{Re}(b_0b_5^*) +12\text{Re}(b_3b_4^*)\right. \nonumber \\ \left. +6|b_4|^2\right]\,.
\label{gat:amppd}
\end{gather}

This corresponds to thirteen observables. It is clear that so many parameters together with their energy dependencies can not be extracted from experiments. In order to solve this complicated problem one has to apply approximations which reduce the number of parameters. The simplest approximation is to reduce the number of waves considered. If the data do not show a $\cos^2(\theta_\eta)$ dependence, the amplitudes $b_4$, $b_5$ and $b_6$ are either very small or zero. Then one is left with
\begin{gather}
|A|^2 =  2|b_0|^2 +2|b_1|^2 +4|b_2|^2 +2b_3^2
+\cos(\theta_\eta)\left[4\sqrt{2}\text{Re}(b_1b_2^*) +4\text{Re}(b_0b_3^*)\right]
\label{gat:amppd-red}
\end{gather}
The angle dependence is then only due to interference between the $s$ and $p$ waves.

Instead of the partial waves in Table \ref{tab:pd-Amplitudes} one can again make use of the related helicity amplitudes. The transition operator in the non-orthogonal basis is given by \cite{Germond89},\cite{Wilkin07}
\begin{gather}
\label{equ:T_op}
\mathcal{M}=\vec{\varepsilon}\,\vec{T}=
 A{ \varepsilon}\cdot {\hat { p}}_p+
iB[{\vec {\varepsilon}} \times \vec { \sigma}]\cdot {\hat {p}}_p+
 C{\vec{\varepsilon}}\cdot \vec{{{p}}_\eta}+
iD[{\vec{\varepsilon}}\times \vec{ \sigma}]\cdot\vec{{ { p}}_\eta} \nonumber \\
+iE({ \vec{\varepsilon}}\cdot \vec{ n})(\vec{\sigma}\cdot{\hat {  p}}_p)
+iF({\vec{\varepsilon}}\cdot \vec{ n})(\vec{ \sigma}\cdot\vec{  p}_\eta),
\end{gather}
where $\vec{\sigma}$ is the Pauli matrix, $\vec{\varepsilon}$ is the
polarisation vector of the deuteron, ${\hat {p}_p}=\vec{p}_p/|{p}_p|$ is the unit vector
along the proton beam direction, $\vec{{ p}_p}$ and
 $\vec{{ p}_\eta}$ are the cms momenta of the proton and
the $\eta-$meson, respectively, and
$\vec{ n}=[\vec{ p}_\eta \times \hat{p}_p]$. These vectors define a coordinate system with $\vec{n}\times {\hat { p}}_p$ the x-axis, $\vec{n}$ the y-axis and $\vec{{ p}_p}$ the z-axis. The helicity amplitudes are named by $A$ to $F$. The $s$ waves are contained in $A$ and $B$ only. Close to threshold  the terms  $A$ and $B$ could contain mainly  an admixture of p-waves,
like $A'(\vec{ p}_p\cdot \vec{ p}_\eta)(\vec{\varepsilon} \cdot \vec{p}_p)$ and
 $iB'(\vec{p}_p\cdot \vec{ p}_\eta)[\vec{\varepsilon}\times
 \vec{\sigma}]\cdot {\hat {\bf p}}_p$. These two $p$ waves and the $E$ amplitude correspond to the transitions with $L=2$. The amplitudes $C$ and $D$ correspond to the two transitions with $L=0$, $F$ to the $d$ wave and higher partial waves (see Table \ref{tab:pd-Amplitudes}).

The unpolarised cms cross section is given  by
\begin{equation}\label{equ:unpol}
|I|^2=\frac{1}{3}\sum_\alpha T_\alpha T^\mathcal{y}_\alpha
\end{equation}
with the spin degeneracy factor 1/3.
This leads to the angular dependence of the unpolarised cross section
\begin{eqnarray}
\label{eqn:unpol-cs}
|I|^2= \left|A|^2+2|B|^2+p_\eta^2(|C|^2+2|D|^2)\right.\nonumber \\
+2p_\eta \text{Re}(AC^*+2BD^*)\cos{\theta_\eta}+
p_\eta ^2 (p_\eta ^2|F|^2+|E|^2)\sin^2{\theta_\eta}+\nonumber \\
 +2p_\eta ^2 \text{Re}(DE^* -BF^*+ p_\eta EF^*\cos{\theta_\eta})\sin^2{\theta_\eta}
\end{eqnarray}
with $\theta_\eta$ being the  angle between the vectors $\vec{ p}_p$ and $\vec{ p}_\eta$. In the following we ignore the amplitudes $E$ and $F$. Then the remaining $p$ waves depend linearly on $\cos{\theta_\eta}$. Nine unknowns remain to be extracted from experiments. The measurement of the unpolarised cross section is not sufficient. In principle one can make use of vector polarised protons and/or deuterons and/or tensor polarised deuterons and measure spin dependent observables in order to access information on the the amplitudes. In practice only polarised deuteron beams have been applied so far. The reason is that in cyclic accelerators deuterons have only very small depolarising resonances due to their small magnetic moment. We will therefore concentrate on mainly tensor polarised deuterium beams.

Now we express the spin dependent observables in terms of the four amplitudes $A$, $B$, $C$, and $D$ discussed above. The analysing powers of the deuteron are given by \cite{Wilkin07}
\begin{gather}
\sqrt{2}\,|I|^2\,T_{20}=2\left(|B|^2-|A|^2\right)
+\left(|D|^2-|C|^2\right)(3\cos^2\theta_{\eta}-1)
\nonumber \\
 +\cos\theta_{\eta}\,Re\left(B^*D-A^*C\right)\,,
\label{equ:t20}
 \end{gather}
\begin{equation}\label{equ:T21}
|I|^2\,T_{21}=\sqrt{3}\,\left[Re\left(A^*C-BD^*\right)
\sin\theta_{\eta}
+\:(|C|^2-|D|^2)\sin\theta_{\eta}\cos\theta_{\eta}\right]
\,,
\end{equation}
\begin{equation}\label{equ:T22}
 2|I|^2\,T_{22}=\sqrt{3}\,\left(|D|^2-|C|^2\right)
 \sin^2\theta_{\eta}\,,
 \end{equation}
\begin{equation}\label{equ:T11}
|I|^2\,iT_{11}=\sqrt{3}\,\text{Im}\left(A^*C-BD^*\right)
\sin\theta_{\eta}\,,
\end{equation}
\begin{equation}\label{equ:T10}
|I|^2\,T_{10}=0\,
\end{equation}
and the one of the proton by
\begin{equation}\label{equ:Ay}
|I|^2A_y^{\,p}=2\,\text{Im}\left(A^*D-B^*D+CB^*\right)\, \sin\theta_{\eta}\,.
\end{equation}
It is interesting to note that $T_{21}$ and $T_{11}$ are sensitive, respectively, to the real and imaginary parts of an $s-p$ interference. Another combination is to be found in the forward/backward asymmetry of $T_{20}$. The measurement of this interference would allow one to deduce the sign of the scattering length and thus prove or disprove the existence of a bound $\eta-^3$He state. There are more observables esp. when proton and deuteron are polarised. In addition to polarisations one can measure spin transfer coefficients from the polarised proton to the $^3$He, the vector polarised deuteron to the $^3$He, and the tensor polarised deuteron to the $^3$He. Furthermore there are proton-deuteron spin-tensor correlation parameters as well as proton-deuteron vector correlation parameters. Not all observables are independent of each other. In order to determine four moduli of the amplitudes and their relative phases one has to measure ten observables. The number is even larger for all $s$ and $p$ waves. General formulae are given by Hanhart \cite{Hanhart04}, specific ones for the present reaction by  Uzikov \cite{Uzikov08}.

For completeness we give the relations between partial wave amplitudes and helicity amplitudes similar to the previous section. In the limit of only four partial waves without any admixtures we have
\begin{eqnarray}
A &=& \sqrt{2}b_0\\
B &=& b_1 \\
C &=& \sqrt{2}b_3\\
D &=& \sqrt{2}b_2\,.
\end{eqnarray}

\subsubsection{The Reaction $p+{^6\text{Li}}\to \eta+{^7\text{Be}}$}\label{sub:Reaction_p-Li6}

Finally we will discuss the $S$ wave dependence of the cross section for the $p+{^6\text{Li}}\to \eta+{^7\text{Be}}$ reaction. From the conservation laws similar to the two reactions discussed above we obtain the transitions given in Table \ref{tab:p-Li-Amplitudes}.
\begin{table}[htb]
\caption{Allowed transitions and the corresponding partial wave amplitudes.}
\label{tab:p-Li-Amplitudes}
\begin{center}
\begin{tabular}{|c|c|c|c|c|c|c|c|}
\hline
$S$ & $L$ & $\pi$ & $J$ & $S'$ & $L'$ & wave & amplitude \\
\hline
2/2 & 0 & +1 & 2/2 & 2/2 &0& $s$  & $a_0$ \\
3/2 & 2 & +1 & 3/2 & 3/2 &0& $s$  & $a_1$ \\
3/2 & 1 & -1 & 3/2 & 3/2 &1& $p$  & $a_3$ \\
3/2 & 1 & -1 & 5/2 & 3/2 &1& $p$  & $a_4$ \\
3/2 & 3 & -1 & 3/2 & 3/2 &1& $d$  & $a_5$ \\
3/2 & 3 & -1 & 5/2 & 3/2 &1& $p$  & $a_6$ \\
1/2 & 1 & -1 & 1/2 & 3/2 &1& $p$  & $a_7$ \\
1/2 & 1 & -1 & 3/2 & 3/2 &1& $p$  & $a_8$ \\
1/2 & 3 & -1 & 5/2 & 3/2 &1& $p$  & $a_9$ \\
3/2 & 1 & -1 & 1/2 & 3/2 &1& $p$  & $a_{10}$ \\
1/2 & 2 & +1 & 3/2 & 3/2 &0& $s$  & $a_2$ \\
\hline
\end{tabular}
\end{center}
\end{table}
Here we restrict ourselves to only $s$ and $p$ waves. There are three $s$ waves and eight $p$ waves. This means that there are 21 independent observables. In other words: it will be impossible to deduce them all from data. The $p$ waves give rise to terms proportional to $\cos\theta_\eta$ and $\cos^2\theta_\eta$. The terms proportional to $\cos\theta_\eta$ are all $s-p$ wave interferences.

\section{Experiments}\label{sec:Experiment}
\subsection{Production Experiments}\label{sec:production}
\subsubsection{Searches with $(\pi, p) $ reactions}\label{sub:Pi-p}

The number of experimental searches is small compared to the number of theoretical papers. The most favourable method produces the meson with almost zero momentum $q$ relative to the nucleus. This can be seen from the probability distribution which is typically
\begin{equation}\label{equ:Form-factor}
f(q)=\exp[-(q^2/2\mu B_\eta)]
\end{equation}
with $\mu$ the reduced mass and $B_\eta$ the binding energy.  At first we will mention some inconclusive searches.

Chrien et al. \cite{Chrien88} applied the $(\pi,\eta)$ reaction employing the pion beam at a momentum of 800 MeV/c at BNL. They made use of different targets: $^{7}$Li, $^{12}$C, $^{16}$O, and $^{27}$Al. Let us concentrate for the moment on the carbon case. The experiment is thought to proceed
\begin{equation}\label{equ:reaction}
\pi ^ +   + \left( {\begin{array}{*{20}c}
   n  \\
   {{}^{11}C}  \\
\end{array}} \right) \to \left( {\begin{array}{*{20}c}
   \eta   \\
   {{}^{11}C}  \\
\end{array}} \right) + p
\end{equation}
with $^{11}$C acting as a spectator. The emerging proton was detected with a magnetic spectrometer at a laboratory angle of less than 15\degr . However, at this angle the $\eta$ had a momentum of $q=$200 MeV/c relative to the nucleus and therefore the form factor $f(q)$ is small. Hence, no effect was seen.
\begin{figure}[h!]
\begin{center}
\includegraphics[width=0.60\textwidth]{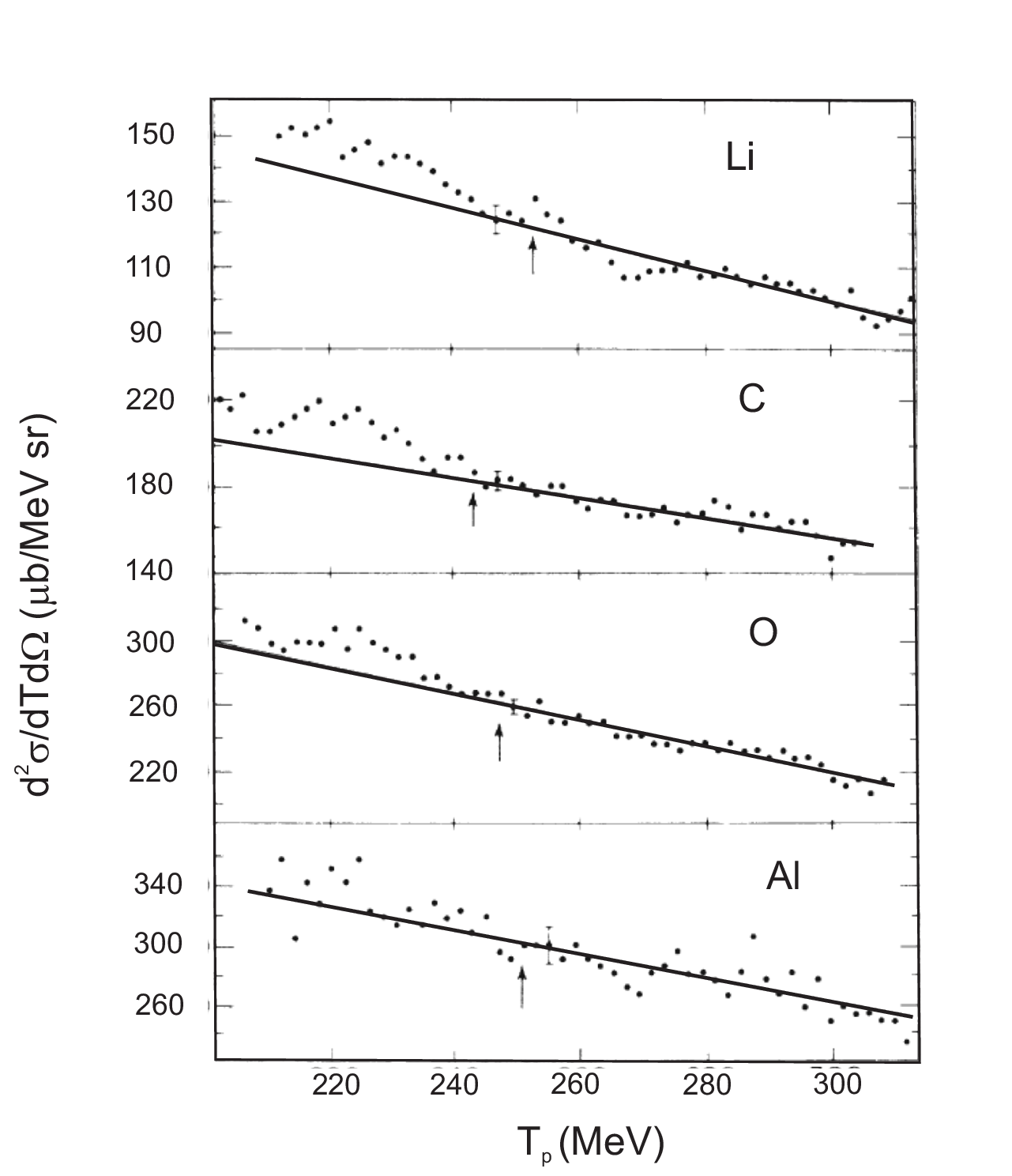}
\caption{Differential cross sections for the $\pi^++A\to p+X$ reaction with $A$ indicated in the figure next to the appropriate data set. The detection angle was 15\degr in the laboratory system. The straight lines indicate the non resonant yield, the arrows the expected position of binding (from Ref. \cite{Chrien88}). }
\label{Fig:Chrien}
\end{center}
\end{figure}
\begin{figure}[h!]
\begin{center}
\includegraphics[width=0.60\textwidth]{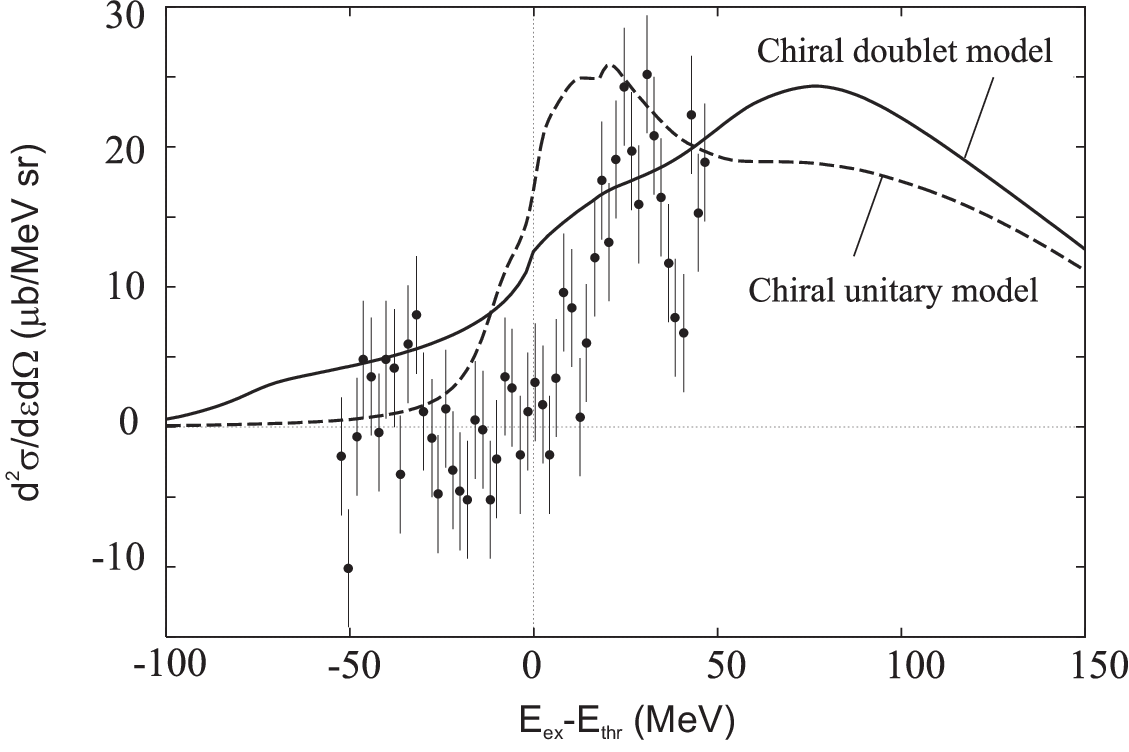}
\caption{The data for the case of carbon from Fig. \ref{Fig:Chrien} after subtracting the linear cross section. The curves are calculations within the indicated models \cite{Nagahiro09}.}
\label{Fig:Chrien_model}
\end{center}
\end{figure}
The data are shown in Fig. \ref{Fig:Chrien}. Exponential slopes of the spectra are shown. Possible bound states should show up as enhancement near the arrows.  The enhancement at low proton energies was attributed to quasi free $\eta$ production. Data for carbon are shown after subtraction of the exponential in the Fig. \ref{Fig:Chrien_model}. There is indeed no sign of a bound state, which would show up as some strength below the threshold energy. Also shown is a detailed study \cite{Nagahiro09} within the two models discussed in the theory section. Neither of these models shows an enhancement expected for a bound state although these models do so for more favourable kinematical conditions. From this study one can learn that recoil free production of the $\eta$ is mandatory for the production of a bound state.

\begin{figure}[h!]
\begin{center}
\includegraphics[width=0.6\textwidth]{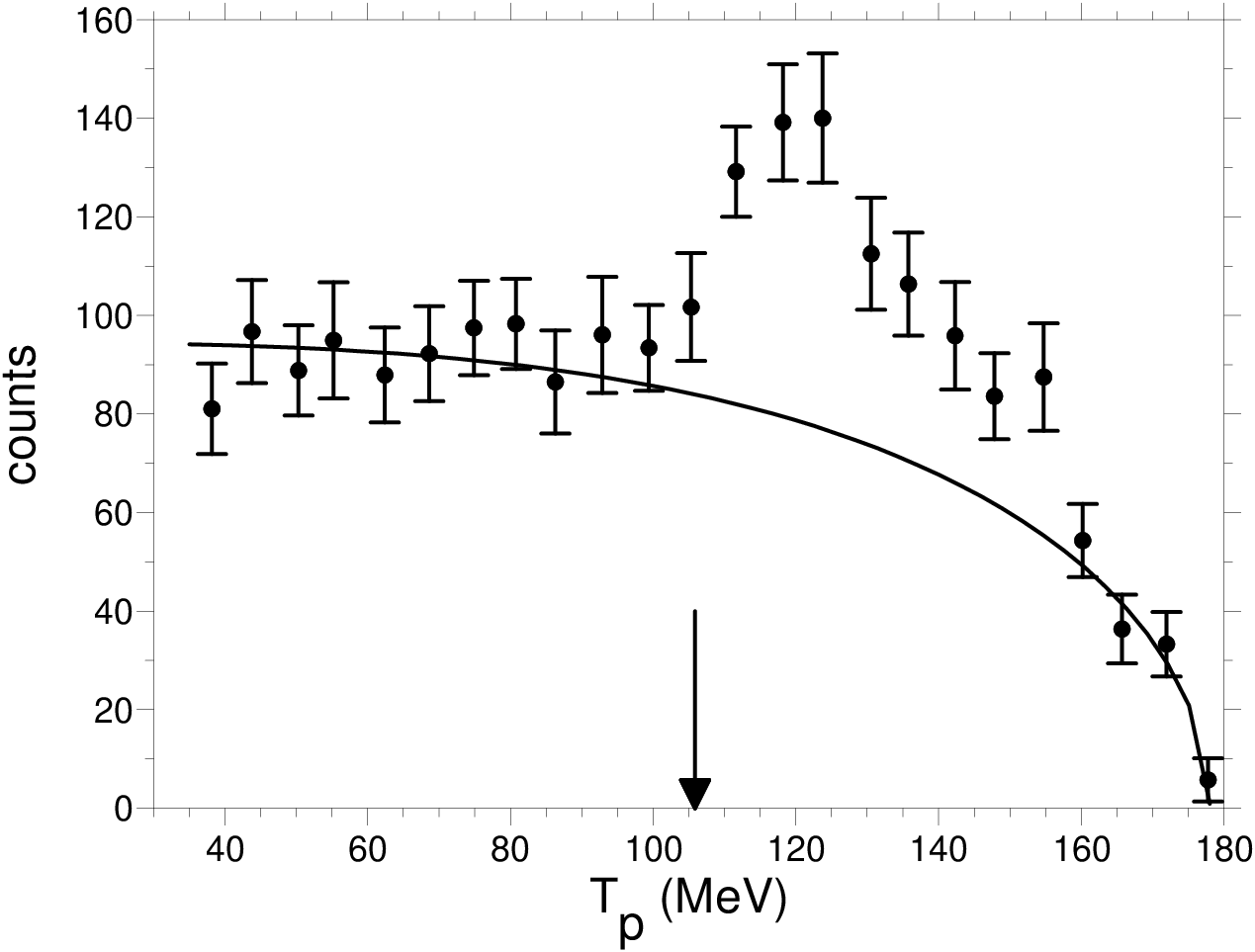}
\caption{Combined spectrum of protons from the reaction $\pi^++{^{l6}\text{O}}\to p+(\pi+p)+X$ \cite{Lieb88}. The solid curve indicates a possible non resonant cross section. The area to the right of the arrow should contain the bound state.}
\label{Fig:LANL}
\end{center}
\end{figure}
Another experiment employing the same reaction on the same targets as in the BNL experiment was performed at the Los Alamos Meson Physics Facility (LAMPF) accelerator at the Los Alamos National Laboratory (LANL) \cite{Lieb88}. The pion beam momentum was 657 MeV/c. The detector used was the Los-Alamos BGO ball which covers almost 4$\pi$ of the solid angle. The ball consists of 30 crystals with scintillators in front as $\Delta E$ detectors to identify protons. The most forward five elements detected the proton from the reaction (\ref{equ:reaction}). The reaction is thought to then have a second step
\begin{equation}\label{equ:second_step}
\eta+N\to N^*\to \pi+N\,.
\end{equation}
If the nucleon in this step is a neutron the decay of the resonance can be $\pi^-+p$. This is the only channel with two charged particles in the final state. Fig. \ref{Fig:LANL} shows events with a threefold coincidence, i. e. a proton in four of the five forward crystals, another proton and a pion detected at conjugate directions. The fifth forward crystal showed a spectrum not in accord with the other four. The arrow indicates the nuclear threshold for free $\eta$ production. The region to the right of the arrow is the region where a bound state is expected. Indeed a peak is visible. However, it is not clear whether the decrease of the count rate above $\approx$120 MeV is due to detector efficiency \cite{Lieb88b} for which the data are not corrected.

\subsubsection{Pion-nucleon back to back experiments}\label{sub:Back_to_Back}

A Lebedev Institute group \cite{Sokol99} performed an experiment in a similar spirit to the LAMPF experiment. Instead of incident pions they used bremsstrahlung photons with the maximum energies below and above the $\eta$ production threshold. The first step is assumed to be either
\begin{equation}\label{equ:reaction_1a}
\gamma  + \left( {\begin{array}{*{20}c}
   p  \\
   {{}^{11}B}  \\
\end{array}} \right) \to \left( {\begin{array}{*{20}c}
   \eta   \\
   {{}^{11}B}  \\
\end{array}} \right) + p
\end{equation}
or
\begin{equation}\label{equ:reaction_1b}
\gamma   + \left( {\begin{array}{*{20}c}
   n  \\
   {{}^{11}C}  \\
\end{array}} \right) \to \left( {\begin{array}{*{20}c}
   \eta   \\
   {{}^{11}C}  \\
\end{array}} \right) + n\,.
\end{equation}
In both cases the second step can lead to
\begin{equation}\label{equ:decay}
\eta+p\to N^{+*}\to\pi^++n
\end{equation}
with an intermediate nucleon resonance excited similar as Eq. \eqref{equ:second_step}. In their experiment the mass of the bound state and its width is not measured by the energy of the emerging nucleon in step (\ref{equ:reaction_1a}) or (\ref{equ:reaction_1b}), but from the decay of such a tentative state into particles number 1 $(=\pi^+)$ and number 2 $(=n)$ (see reaction (\ref{equ:decay})). In such an experiment the mass of the decaying system is given by
\begin{equation}
m_{12} = \sqrt{m_1^2+m_2^2+ 2\left[E_1E_2-2p_1p_2 \cos\left(\theta_1+\theta_2\right)\right]}
\end{equation}
with $E_i$ the total energy of particle $i$. Still some theory is needed to deduce from the mass $m_{12}$ the binding energy of the $\eta$ in the $A\otimes\eta$ system. If $m({^{11}A\otimes\eta})$ is a broad peak on top of a huge background, excellent measurement of momenta or velocity and particle identification is required. For a particle decaying in the laboratory system at rest into two particles, the total angle between the two has to be $\theta_1+\theta_2=180$\degr.

In the experiment \cite{Sokol99} this case was biased: the detection system consisted of two time of flight (TOF) arms: a neutron arm in one direction and a pion arm in the opposite direction. The neutron arm consisted of several plastic scintillators behind a veto counter, suppressing charged particles. Both arms had a flight path of 1.4 m. Although they can distinguish between pions and protons in the pion arm they do not measure the charge. Therefore the pions can be $\pi^+$ or $\pi^-$ stemming from quite different reactions. In reality they measured
\begin{equation}\label{equ:Lebedev}
\gamma +^{12}\text{C}\to \pi^\pm +n+X\,.
\end{equation}

In Fig. \ref{Fig:Lebedev} the measured velocities $\beta=v/c$ in the two arms are shown. This is for the case of $E_{\gamma,max }=850$ MeV which has a portion extending above the $\eta$ threshold at 708 MeV. The strongest yield is observed for $\beta_{neutron}\approx 0.42$ and $\beta_{charged}\approx 0.50$.  This can not be attributed to a decay of an $\eta$ bound state. For both velocities $\beta\approx$1 a second enhancement is visible. This will be most probably an $e^+e^-$ event and to a slightly smaller velocity two pion events. The crossing of the two dashed lines indicated the locus of the decay $\eta+n\to \pi^++p$. No enhancement of the count rate is visible there.
\begin{figure}[h!]
\begin{center}
\includegraphics[width=0.60\textwidth]{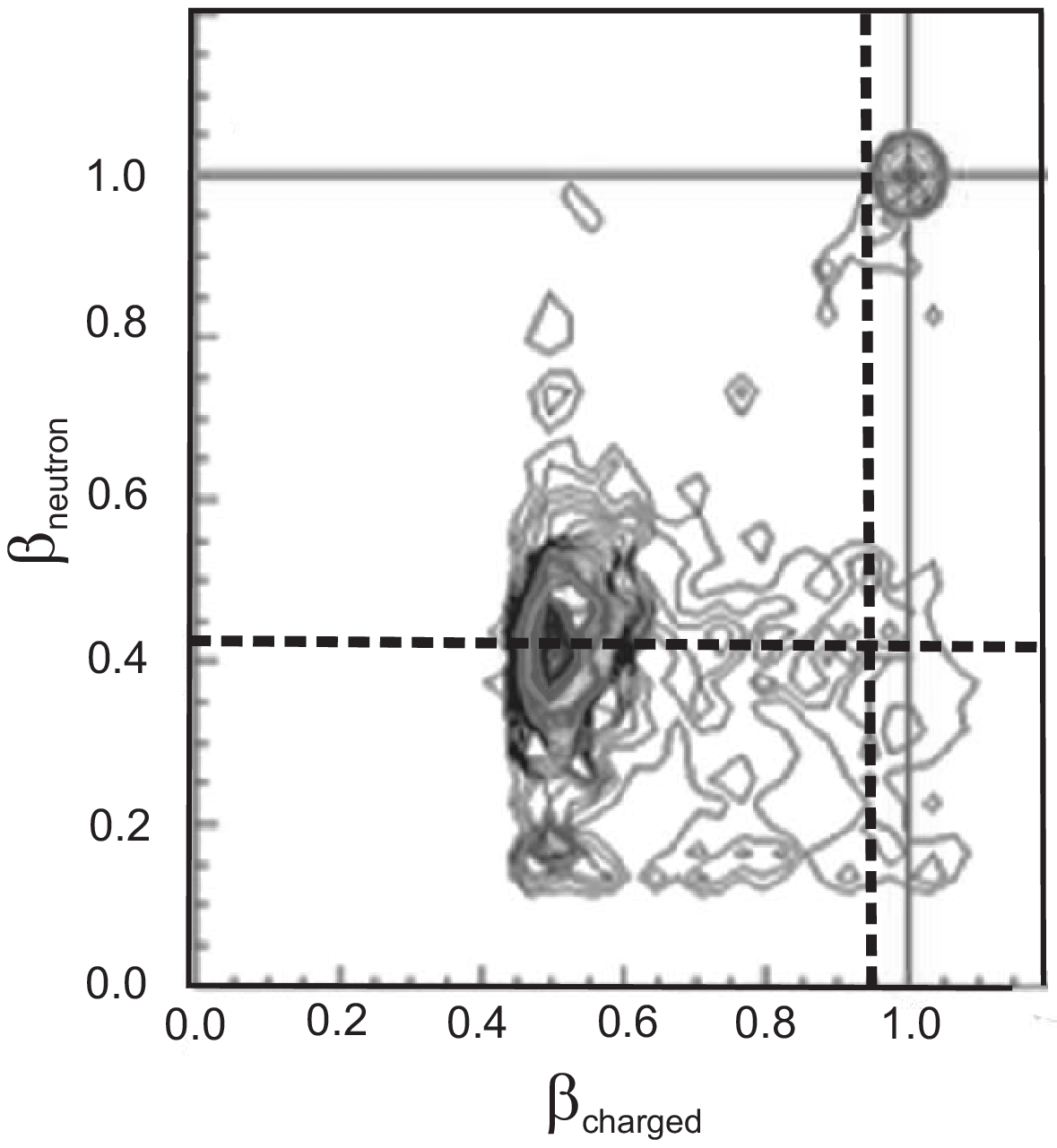}
\caption{Contour plot of the neutron velocity versus the velocity of the charged particle. The solid lines indicate the velocities of fully relativistic particles, the dashed lines possible decay of $\eta$ bound state. }
\label{Fig:Lebedev}
\end{center}
\end{figure}
\begin{figure}[h!]
\begin{center}
\includegraphics[width=0.60\textwidth]{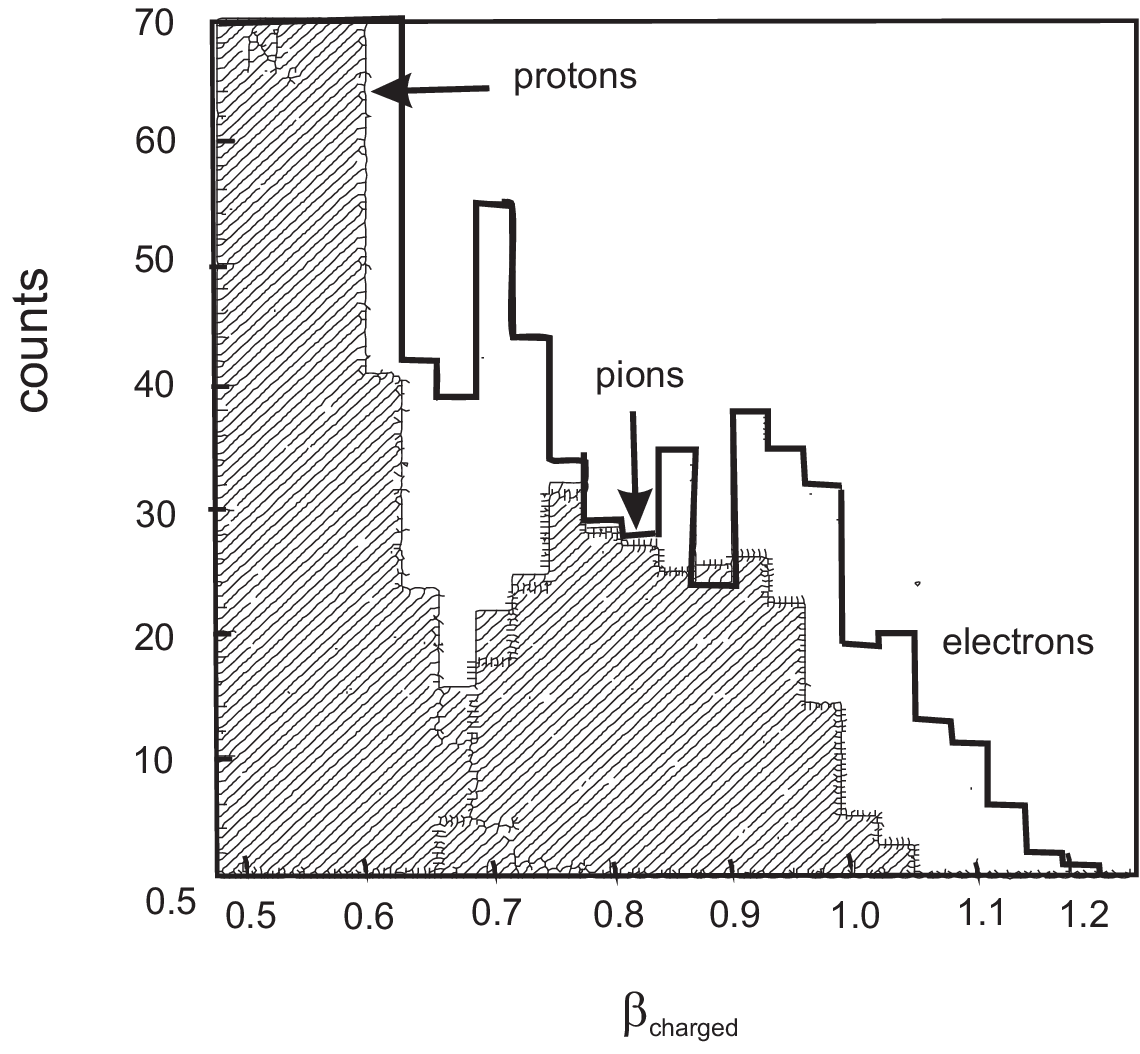}
\caption{The projection of the events from Fig. \ref{Fig:Lebedev} onto the charged particle axis. The shaded areas are from Monte Carlo simulations, the histogram is the measured velocity distribution \cite{Baskov12}.}
\label{Fig:Lebedev_1}
\end{center}
\end{figure}
In Fig. \ref{Fig:Lebedev_1} the velocity spectrum measured in the pion arm is shown.  Also shown are simulations for proton and pion emission. At the high velocity end relativistic particles show up, which can be electrons or positrons. From this part of the spectrum we can see that the leptons have a distribution with $FWHM\approx 0.2$. This indicates that the momentum resolution is only 20$\%$. In Ref. \cite{Baskov12} it is claimed that the enhancement of measured pions compared to the simulation at $\beta\approx 0.95\pm 0.05$ is just the indication of the decay of the bound state. However, a momentum resolution of 20$\%$ does not allow one to distinguish between quasi-free $\eta$ mesons and a state bound by only a few MeV. This resolution transforms into $\Delta m({^{11}A\otimes\eta})\approx 0.15*1486$ MeV $\approx$ 225 MeV. The velocity $\beta\approx 0.95$ for a pion corresponds to a momentum of 431 MeV/c which is expected. At a velocity of $\beta\approx$0.7 there is a gap between the calculated pion distribution and the proton distribution. Also at this velocity an enhancement is seen which is without explanation. So far no spectrum of the invariant mass is shown. In summary the resolution of the experiment is not sufficient to confirm the position and width of a tentative $\eta$ bound state. We therefore state that this experiment is not conclusive. The choice of the angle between the two detectors selects decay of a system at rest, assuming a two body decay. Such a system can only be formed if one or a group of several particles has carried away the beam momentum. The remaining system without linear momentum has an excitation energy of up to $\approx 600$ MeV.

A group with large overlap with the Lebedev group repeated the experiment at the NUCLOTRON accelerator at Dubna \cite{Afanasiev13}. They used a deuteron beam of 2.1 GeV/nucleon. This time one arm detected protons and the other pions. Again no charge selection was made nor was the particle emitted in the first step of the reaction measured. But the masses of the two particles were measured. From the TOF measurement the velocity $\beta$ of the particles were determined and then the mass of the decaying system is
\begin{equation}\label{equ:effective_mass}
m(\pi p)= \frac{m_\pi}{\sqrt{1-\beta_\pi^2}}+\frac{m_p}{\sqrt{1-\beta_p^2}}.
\end{equation}
in the effective mass spectrum a peak was found on an exponential background. The count rate remaining after subtracting the background is shown in Fig. \ref{Fig:Sokol}.
\begin{figure}[h]
\begin{center}
\includegraphics[width=0.6\textwidth]{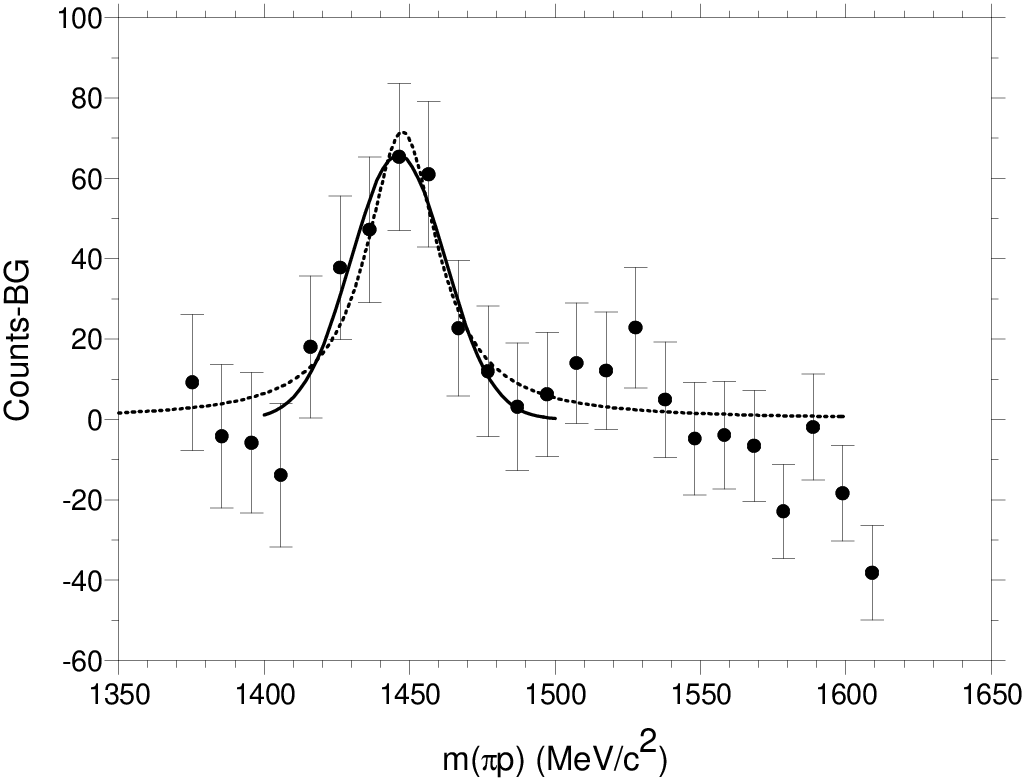}
\caption{Count rate minus background for the reaction $d+^{12}\text{C}\to \pi+p+X$ \cite{Afanasiev13}. The solid curve is a Gaussian fitted to the data, the dashed curve a Breit--Wigner distribution. }
\label{Fig:Sokol}
\end{center}
\end{figure}
A fit of a Gaussian plus exponential to the data yielded a peak position at 1447.8$\pm$3.6 MeV/c$^2$ and a width (FWHM) of 38.8$\pm$10.4 MeV/c$^2$. In the experiment the angle between the two arms has been 180\degr. This  implies that the decaying system was at rest. From momentum and energy conservation we can then calculate that the struck nucleon in the target must have had a Fermi momentum of 393 MeV/c which is not impossible. However, a total cross section for the peak of $11\pm 8\,\mu$b was reported, which is quite large with respect  to the strong constrains applied the experiment. From the reported uncertainties and repeating the fit to the data we can extract the uncertainty in the background, which is of course given by the fit. Assuming a statistical dependence of this uncertainty and the background counts we can extract the significance of the peak which is then only 2.5$\sigma$. This indicated a quite low significance of the peak.

\subsubsection{Formation experiments with photons}
\label{sub:Formation_photons}

Another photoproduction experiment was performed at the MAMI accelerator in Mainz making use of the TAPS spectrometer \cite{Pfeiffer04}. The experiment differs from the one at the Lebedev Institute in the that a tagged photon beam with 800 MeV maximum energy, a $^3$He target were used. Furthermore it differs in the possibility of measuring charged particles and photons, thus allowing to reconstruct neutral pions.
\begin{figure}[h!]
\begin{center}
\includegraphics[width=0.47\textwidth]{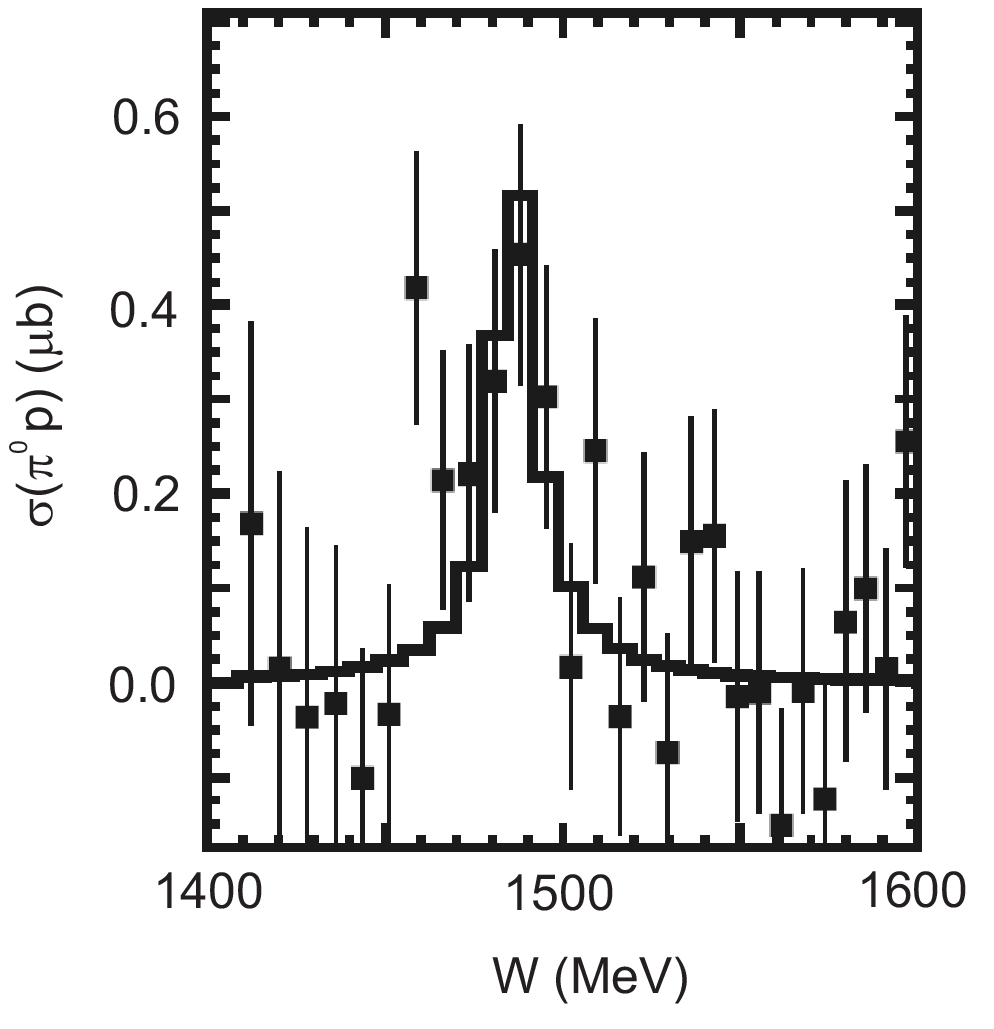}
\includegraphics[width=0.47\textwidth]{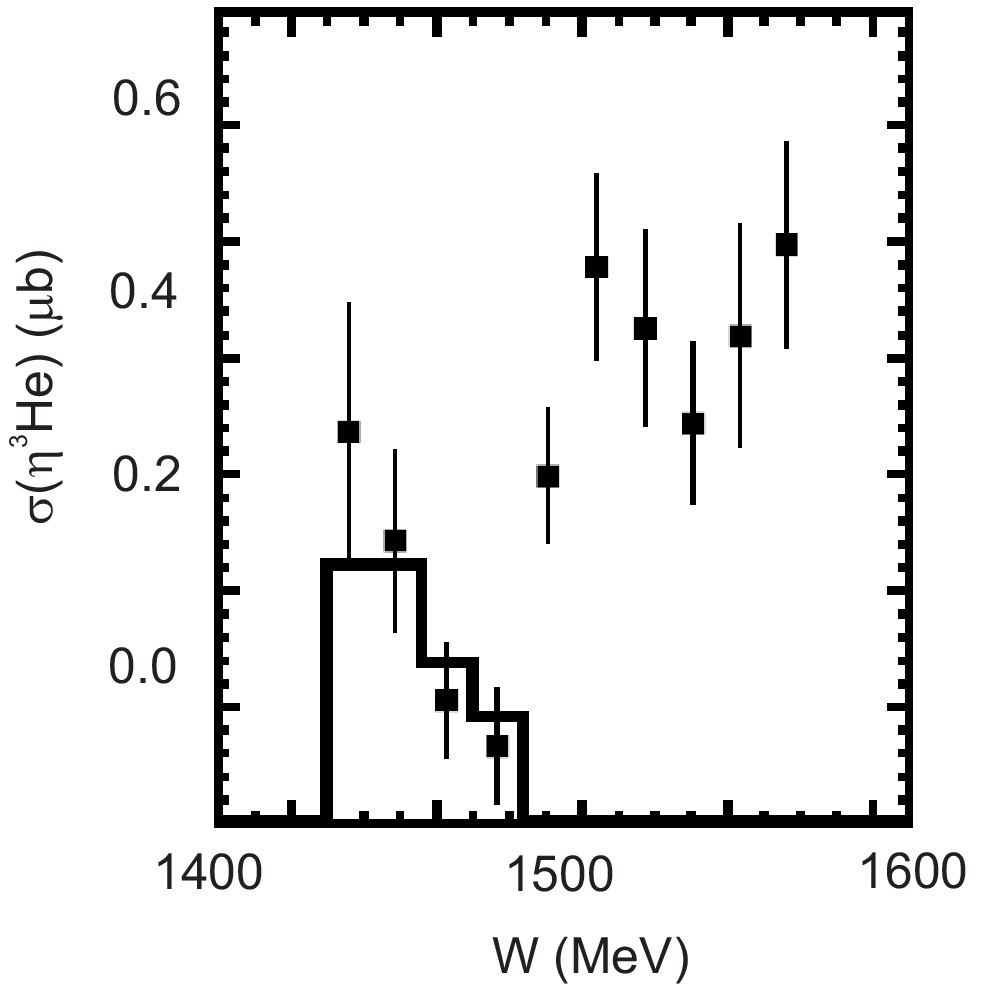}
\caption{Left panel: The cross sections for the inelastic $\gamma+{^3\text{He}}\to\pi^0+p+X$ decay channel as function of $W$ Eq. \eqref{equ:W}. Right panel: Cross sections for the elastic $\gamma+{^3\text{He}}\to\eta+{^3\text{He}}$ decay channel. The histogram is a Flatt\'{e} fit to both data sets  \cite{Pfeiffer05}.}
\label{Fig:Pfeiffer}
\end{center}
\end{figure}

\noindent
The reaction studied  was
\begin{equation}\label{equ:Taps_Reaction}
\gamma+{^3\text{He}}\to \pi^0+p+X
\end{equation}
as a function of $W$, which is the CM energy reduced by the deuteron mass and the {$^3$He} binding energy
\begin{equation}\label{equ:W}
W=\sqrt{2E_\gamma m_{^3He}+m^2_{^3He}}-m_d-B_{^3He}\,.
\end{equation}

Contrary to the Lebedev experiment the emerging $\pi^0$ and proton were not measured to reconstruct the missing mass, but to capture the signature from the decay of an $\eta$-mesic nucleus.
An enhancement was found in the difference spectrum between the angular range 180\degr to 170\degr and 170\degr to 150\degr. This difference spectrum after subtraction of the background of quasi-free $\pi^0$ production is shown in Fig. \ref{Fig:Pfeiffer}. The authors \cite{Pfeiffer04} claim to have seen a bound $\eta$ state which implies that the first step
\begin{equation}
\gamma+^3\text{He}\to \eta\otimes ^3\text{He}
\end{equation}
occurred followed by
\begin{equation}\label{equ:TAPS_step2}
\eta+p\to N^{+*}\to \pi^++p.
\end{equation}
A resonance position $W_R=1483\pm 3$ MeV was obtained from a fit with a Breit--Wigner form times phase space. This corresponds to a binding energy of the $\eta$ of   $B_\eta=W_R-(m_\eta+m_p)=4.4\pm 4.2$ MeV. A full width $\Gamma_\eta =25.6\pm 6.1$ MeV was fitted. The significance of the peak is $\approx 3.5\sigma$.

In a comment Hanhart \cite{Hanhart05} pointed out that  also the non-resonant channel
\begin{equation}\label{equ:Taps_Non_resonant}
\gamma+{^3\text{He}}\to \eta+{^3\text{He}},
\end{equation}
which will be strongly affected by the bound state, is open. The width of the Breit--Wigner will be changed to $\Gamma = \Gamma_{inel.}+\Gamma_{elast.}$ While $\Gamma_{inel.}$ is constant in the momentum range of interest, $\Gamma_{elast.}$ is momentum dependent $\Gamma_{inel.}=gp$ with $p$ the momentum of the $\eta$ relative to the $^3$He and $g$ the coupling constant. Pfeiffer et al. \cite{Pfeiffer05} repeated their analysis by fitting simultaneously both channels with a Flatt\'{e} distribution. They found the fit insensitive to the coupling constant $g$. The results for position and width did not change very much from the previous analysis. Both spectra together with the fits are shown in Fig. \ref{Fig:Pfeiffer}.

So far it seems that a bound state $^3\text{He}\otimes\eta$  had been experimentally seen. However, almost the same group repeated the experiment with again the TAPS spectrometer plus the Crystal Ball detector \cite{Pheron12}. The experiment benefitted not only from the now almost 4$\pi$ acceptance but also from much higher statistics. The photon energies ranged from 0.45 GeV to 1.4 GeV. The results of this measurement are shown in Figs. \ref{Fig:Pheron} and \ref{Fig:Pheron_a}. The strong rise of the $\pi^0 p$ cross section above the $\eta$ production threshold is similar to the previous experiment and supports the possibility of a resonance in the threshold region. However, the structures visible at higher energies have not been seen in \cite{Pfeiffer04}. They are in the so called second and third resonance region and their walk with angle is purely kinematical. Fig. \ref{Fig:Pheron_a} compares the elastic channels measured in the two MAMI experiments. They have the same binning of 8 MeV. The minimum in the earlier data around 620 MeV is not visible in the newer data.
\begin{figure}[ht]
\begin{center}
\includegraphics[width=0.50\textwidth]{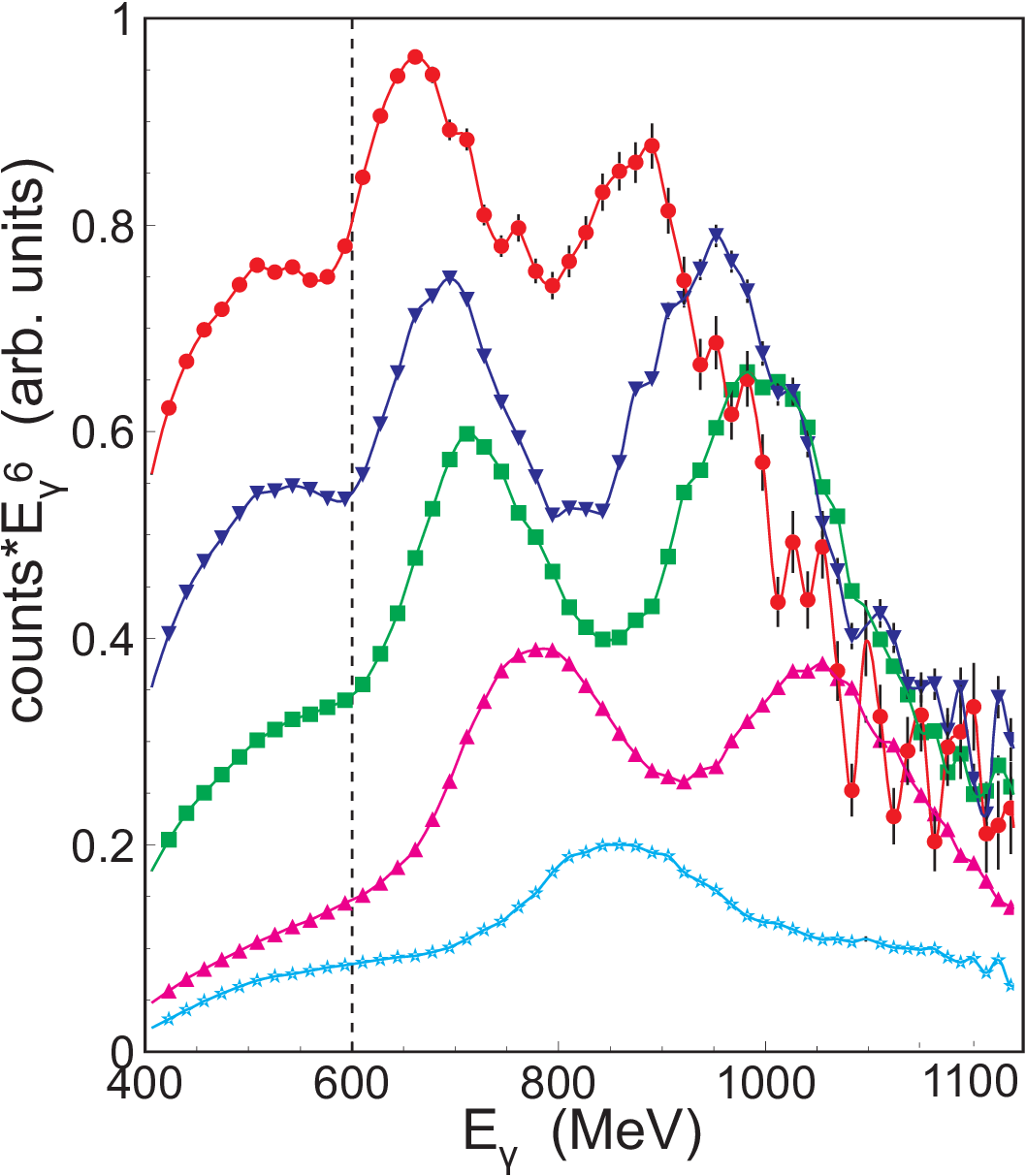}
\caption{Similar as Fig. \ref{Fig:Pfeiffer} but for the experiment Ref. \cite{Pheron12}. Note that $E_\gamma\propto W^2$ (see Eq. (\ref{equ:W})). Excitation functions of $\pi^0p$ back-to-back pairs for different ranges of the opening angle $\theta_{\pi^0}+\theta_p$ after removal of the overall energy dependence $\propto E_\gamma^{-6}$. From top to bottom opening angle ranges of: 165\degr - 180\degr, 150\degr - 165\degr, 140\degr - 150\degr, 130\degr - 140\degr, and 120\degr - 130\degr. The vertical line indicates the
$\eta$-production threshold. }
\label{Fig:Pheron}
\end{center}
\end{figure}
\begin{figure}[ht]
\begin{center}
\includegraphics[width=0.50\textwidth]{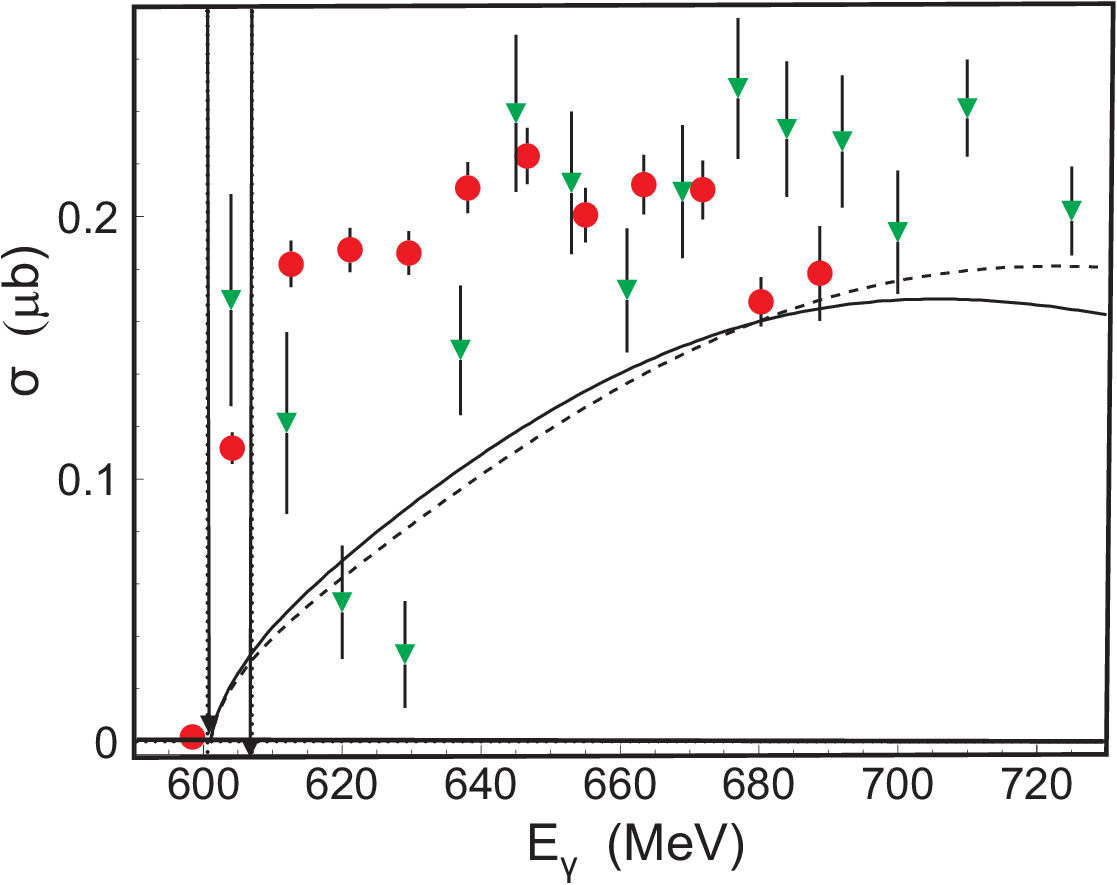}
\caption{Same as Fig. \ref{Fig:Pfeiffer} but for the experiment Ref. \cite{Pheron12}. Note that $E_\gamma\propto W^2$ (see Eq. (\ref{equ:W})). Excitation function of the total cross section for the $\gamma+{^3\text{He}}\to \eta+{^3\text{He}}$ reaction. The dots are data from Ref. \cite{Pheron12}, the triangles are data from \cite{Pfeiffer04}. The two vertical lines indicate the coherent and the break up thresholds. The solid curve is a PWIA calculation with realistic angular distribution for the elementary reaction, the dashed curve with an isotropic angular distribution. }
\label{Fig:Pheron_a}
\end{center}
\end{figure}
A direct comparison between the two experiments is done in Fig. \ref{Fig:Pheron_a}. The structure seen in the first experiment was also seen in the second experiment with much better statistics. But it could be shown that it is an artefact of the complicated behaviour of the background. So again there is no convincing result for a bound state.

\subsubsection{Formation experiments with hadrons}
\label{sub:Formation_hadrons}

The WASA collaboration \cite{Adlarson13} studied the reaction $d+d\to \pi^-+p+{^3\text{He}}$ reaction. The idea is that an intermediate $\eta\,\alpha$ bound state might exist. The whole reaction chain is then
\begin{equation}\label{equ:WASA-dd}
d+d\to \eta\otimes\alpha\to N^*(1535)+{^3\text{He}}\to (\pi^- + p)+{^3\text{He}}\,.
\end{equation}
The deuteron beam momentum varied between 2.185 GeV/c and \- 2.400 GeV/c. This interval covers the $\eta$ production threshold at 2.336~GeV/c. Thus the variation is in terms of $\eta$ production from -51.3 MeV$\leq Q\leq 22$ MeV with $Q$ the excess energy. The angle of the decay $N^{*}\to \pi^-p$ has to be $\approx$180\degr in the $N^*$ rest frame. Furthermore one expects more bound events at small relative momenta than at large relative momenta.  The WASA collaboration could not find an anomaly in the excitation function for beam momenta below and above threshold (see Fig. \ref{Fig:Wasa}).
\begin{figure}[ht]
\begin{center}
\includegraphics[width=0.6\textwidth]{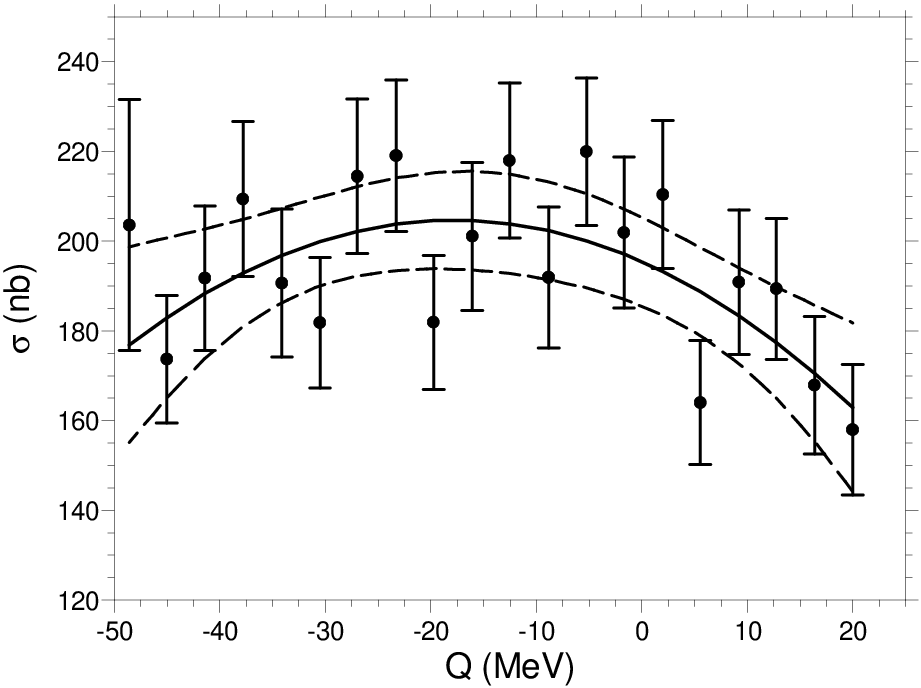}
\caption{The excitation function for the reaction $dd\to ^3$He$p\pi^-$. The data \cite{Adlarson13} are shown with error bars. A polynomial fit up to second order is shown as solid curve. The dashed curves indicate the 95$\%$ confidence interval. }
\label{Fig:Wasa}
\end{center}
\end{figure}
A bound state would result in a peak like structure on top of a smooth continuum. This was approximated by us with a polynomial $\sigma = a_0+a_1Q+a_2Q^2$ which is also shown in the figure together with a 95$\%$ confidence interval. There is surely no peak above that limit. From a detailed analysis the authors conclude an upper limit for the decay of a possible intermediate $\eta$ mesic system into this particular channel of 20~nb. However, this experiment suffered from unfavourable boundary conditions \cite{Moskal14a}. The experiment was repeated with much higher statistics. In addition another isospin channel was also studied: $d+d\to \alpha\otimes\eta\to N^*(1535)+{^3\text{He}}\to (\pi^0 + n)+{^3\text{He}}\,$. This new experiment has taken much more data and will be sensitive to a total cross section for a peak of a few nb.

Is this upper limit excluding the existence of a bound state? It should be mentioned that at threshold the relative velocity between the $\eta$ and the $\alpha$ particle is zero and hence the formation of a bound state is favoured. On the other hand at just this energy the cross section for producing a free $\eta$ is zero. We can therefore conclude that the cross section in the vicinity of the threshold is very small and may be undetectable on top of a larger cross section for the same final state without a bound state in between. Recently Wilkin \cite{Wilkin14} presented a model where an iformed guess for cross sections for the formation of bound $\eta$ mesons on light nuclei is given. The relevant numbers were extracted for the excitation functions of real $\eta$ production in hadron and photon induced reactions. The guess for the present case is above the upper limit cited here. However, a form factor for the $^3$He in the final state has still to be considered which might bring the numbers to better agreement.

The same group \cite{Moskal14a} has recently performed a similar experiment on the reaction
\begin{equation}
d+p\to ^3\text{He}\otimes\eta\to N^*(1535)+{pp}\to (\pi^- + p)+pp\,.
\end{equation}
The upper limit for this reaction so far is 270 nb \cite{Moskal10} and about 70 nb for the reaction $d+p\to {^3\text{He}}\otimes\eta\to \pi^0+^3$He. As pointed out above the cross section for a bound $\eta-^3$He system might be too small to be detected. The higher statistics experiments will allow a smaller upper limit for this process to be determined.

\subsubsection{Charge exchange experiments}
\label{sub:Charge-exchange}

Also in a pion double charge exchange (DCX) experiment
\begin{equation}
\label{equ:charge_exchange}
\pi^++{^{A}\text{Z}}\to \pi^-+{^{A}\text{(Z+2)}}
\end{equation}
close to the $\eta$ threshold the $\eta$ has small momenta relative to the nucleus. Haider and Liu \cite{Haider_Liu87} had predicted the occurrence of an $\eta$ bound state in DCX. This process is thought to proceed via $\pi^++N\to\pi^0+N\to\pi^-+N$. However, the process $\pi^++N\to\eta+N\to\pi^-+N$ is also possible. In addition to an unbound $\eta$ a bound one may exist. More specific they predicted a resonance in the reaction $\pi^++^{14}$C$\to \pi^-+^{14}$O at a momentum transfer of $q=210$ MeV/c.
\begin{figure}[h!]
\begin{center}
\includegraphics[width=0.9\textwidth]{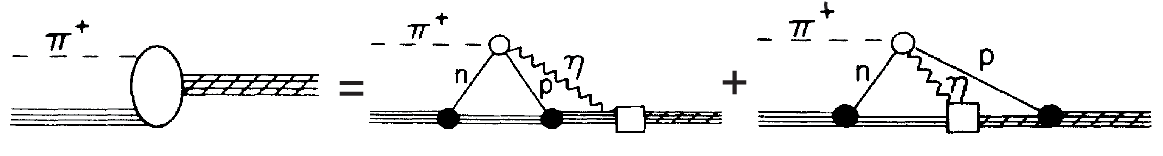}
\caption{Diagrammatic representation of the formation of $\eta$ bound states in DCX. The bound state may be the (A,Z+2)$\otimes\eta$ (first graph) or  (A,Z+1)$\otimes\eta$ (second graph) with a final capture of the proton. }
\label{Fig:DCX}
\end{center}
\end{figure}
The possible formation of such a process is sketched in Fig. \ref{Fig:DCX}. Two possible reactions are shown which lead to a final double isobar analog state (DIAS) with the following steps:
\[
\begin{gathered}
  \pi ^ +   + \left( {\begin{array}{*{20}c}
   n  \\
   {{}^{17}O}  \\

 \end{array} } \right) \to \left( {\begin{array}{*{20}c}
   {\eta  + p}  \\
   {{}^{17}O}  \\

 \end{array} } \right) \to \left( {\begin{array}{*{20}c}
   \eta   \\
   {{}^{18}F}  \\

 \end{array} } \right) \to \left( {\begin{array}{*{20}c}
   {\eta  + n}  \\
   {{}^{17}F}  \\

 \end{array} } \right) \to \pi ^ -   + {}^{18}Ne \hfill \\
  \pi ^ +   + \left( {\begin{array}{*{20}c}
   n  \\
   {{}^{17}O}  \\

 \end{array} } \right) \to \left( {\begin{array}{*{20}c}
   {\pi ^0  + p}  \\
   {{}^{17}O}  \\

 \end{array} } \right) \to \left( {\begin{array}{*{20}c}
   {\pi ^0  + n}  \\
   {{}^{17}F}  \\

 \end{array} } \right) \to \pi ^ -   + {}^{18}Ne \hfill \\
\end{gathered}
\]

Such an experiment has been performed at LAMPF in Los Alamos \cite{Johnson93} at an incident $\pi^+$ beam of 350 MeV to 440 MeV kinetic energy.
\begin{figure}[h!]
\begin{center}
\includegraphics[width=0.6\textwidth]{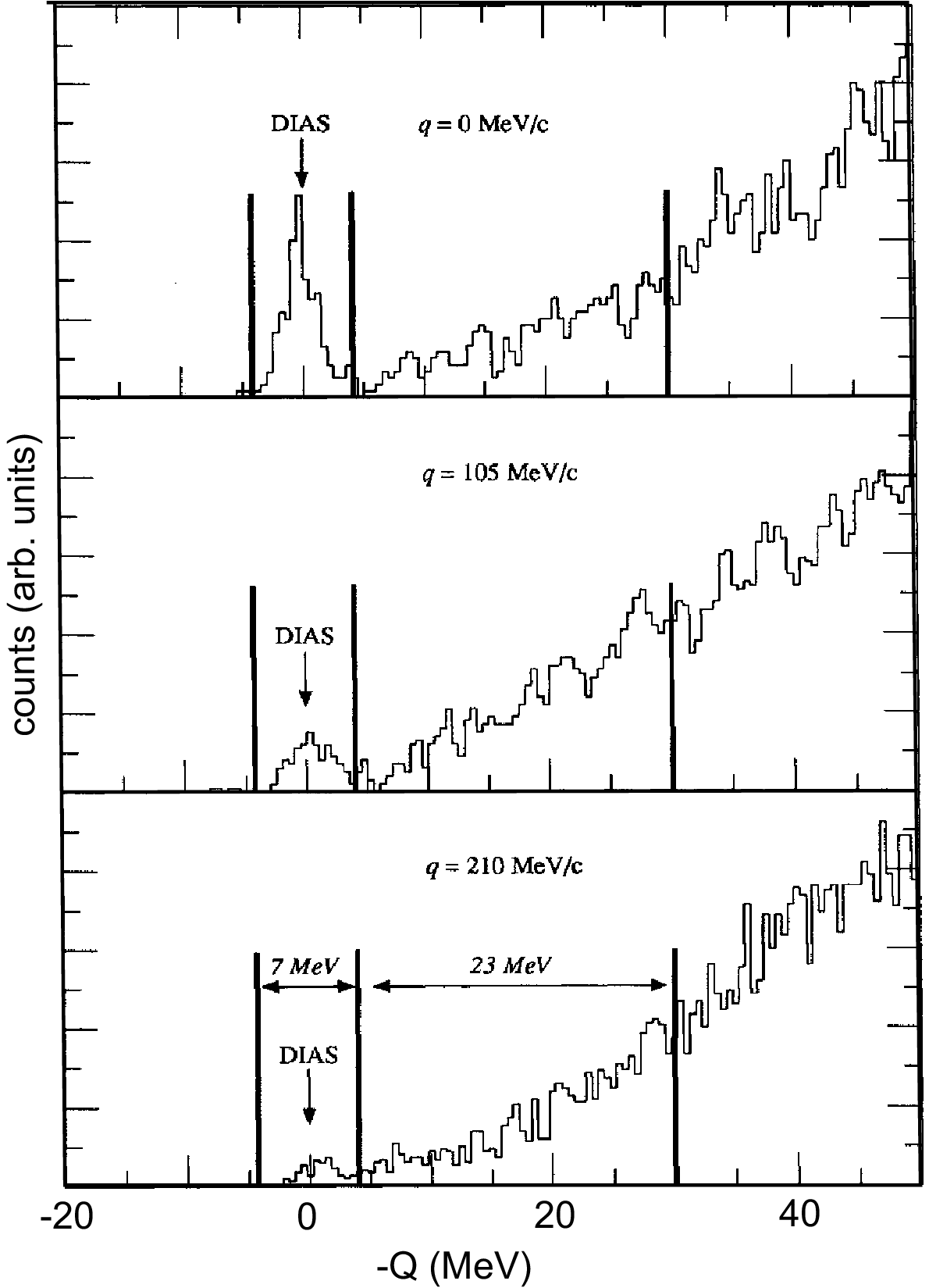}
\caption{Missing mass spectra  for fixed momentum transfers $q$ of 0, 105, and 201 MeV/c for the
reaction ${^{18}\text{O}}+\pi^+\to \pi^-+{^{18}\text{Ne}}$ at 420 MeV kinetic beam energy. The 7 MeV wide interval is the double isobaric analog state DIAS. The 23 MeV wide bin contains continuum states \cite{Johnson93}.}
\label{Fig:Johnson}
\end{center}
\end{figure}
\begin{figure}[h!]
\begin{center}
\includegraphics[width=0.5\textwidth]{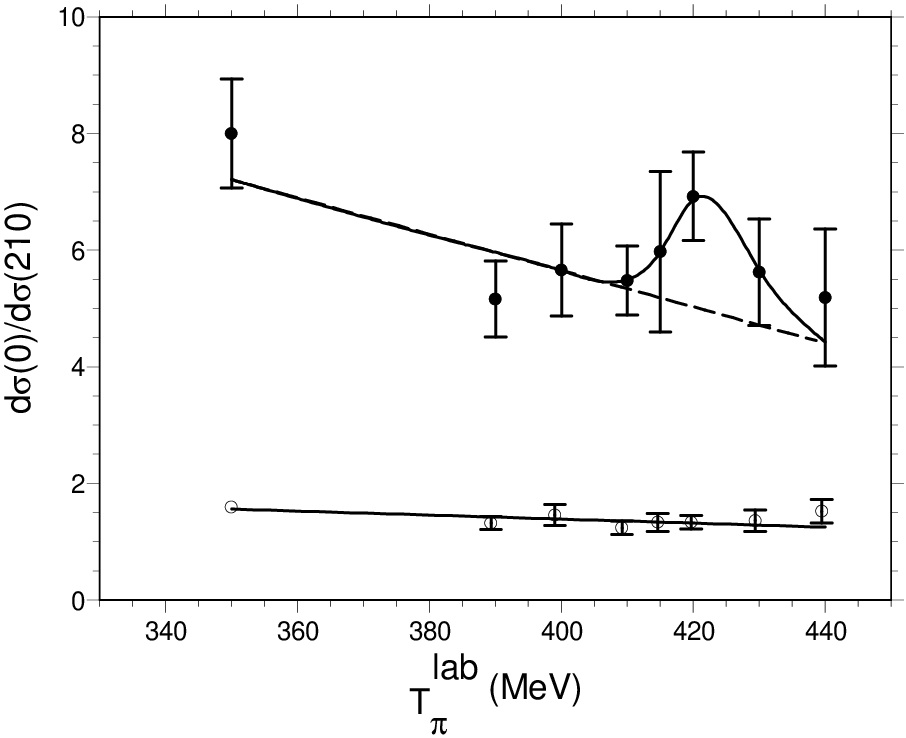}
\caption{The ratios of spectral ranges for different momentum transfer measurements. The upper part is for the DIAS, the lower part for the continuum states. The dashed line is a straight line fit, the solid line a Gaussian fit on top of the straight line. }
\label{Fig:Joh_ratio}
\end{center}
\end{figure}
Examples of missing mass spectra are shown in Fig. \ref{Fig:Johnson} for fixed momentum transfers $q$ of 0, 105, and 201 MeV/c. Then the ratios between the lowest spectrum ($q=0$ MeV/c) and the highest spectrum ($q=210$ MeV/c) for the indicated areas are plotted as a function of the pion beam energy $T_\pi$ in Fig. \ref{Fig:Joh_ratio}. While for the continuum region the ratio shows an almost linear dependence on the pion energy, it shows some enhancement for the DIAS region, which might be an indication of an $\eta$ bound state. We can investigate the significance of the enhancement. For this purpose we fit a linear curve to the data excluding those in the enhancement. This dependence is shown in the figure. In a second step we extract the number of  counts above the background by fitting a Gaussian to the remaining part. From this procedure we find a significance of only 1.6$\sigma$, so the peak is not significant. This is supported by the finding that the mean of the data is $<d\sigma(0)/d\sigma(210)>=5.90\pm 0.33$ with a $\chi^2$/free=1.3. So the ''peak'' is most probably a statistical fluctuation of the background. Unfortunately this experiment was never repeated with better statistics and other nuclei.

\subsubsection{Transfer reactions}

Experiments employing transfer reactions are favourable; in such experiments the whole beam momentum can be transferred to a nucleon or a cluster of nucleons. The remaining system then does not carry linear momentum and thus favours the probability that a produced $\eta$ is bound to the residual nucleus. This method, originally developed in the production of hypernuclei  \cite{Bruckner76}, was successfully applied in the study of pionic atoms \cite{Yamazaki96}. In order to transfer the beam momentum almost completely to the emerging particle it has to be emitted in the forward direction close at zero degrees. The momentum transferred from the beam to $^3$He being emitted at zero degrees for two different reactions
\begin{equation}\label{equ:reac_dHe}
d+{^{12}\text{C}}\to {^3\text{He}}+{^{11}\text{B}}\otimes\eta
\end{equation}
and
\begin{equation}\label{equ:reac_pHe}
p+{^{27}\text{Al}}\to {^3\text{He}}+{^{25}\text{Mg}}\otimes\eta
\end{equation}
are shown in Fig. \ref{Fig:Magic_Kine} and for different assumptions of the $\eta$ binding energy.
\begin{figure}[h!]
\parbox[b]{0.480\textwidth}{ \centering
\includegraphics[width=0.45\textwidth]{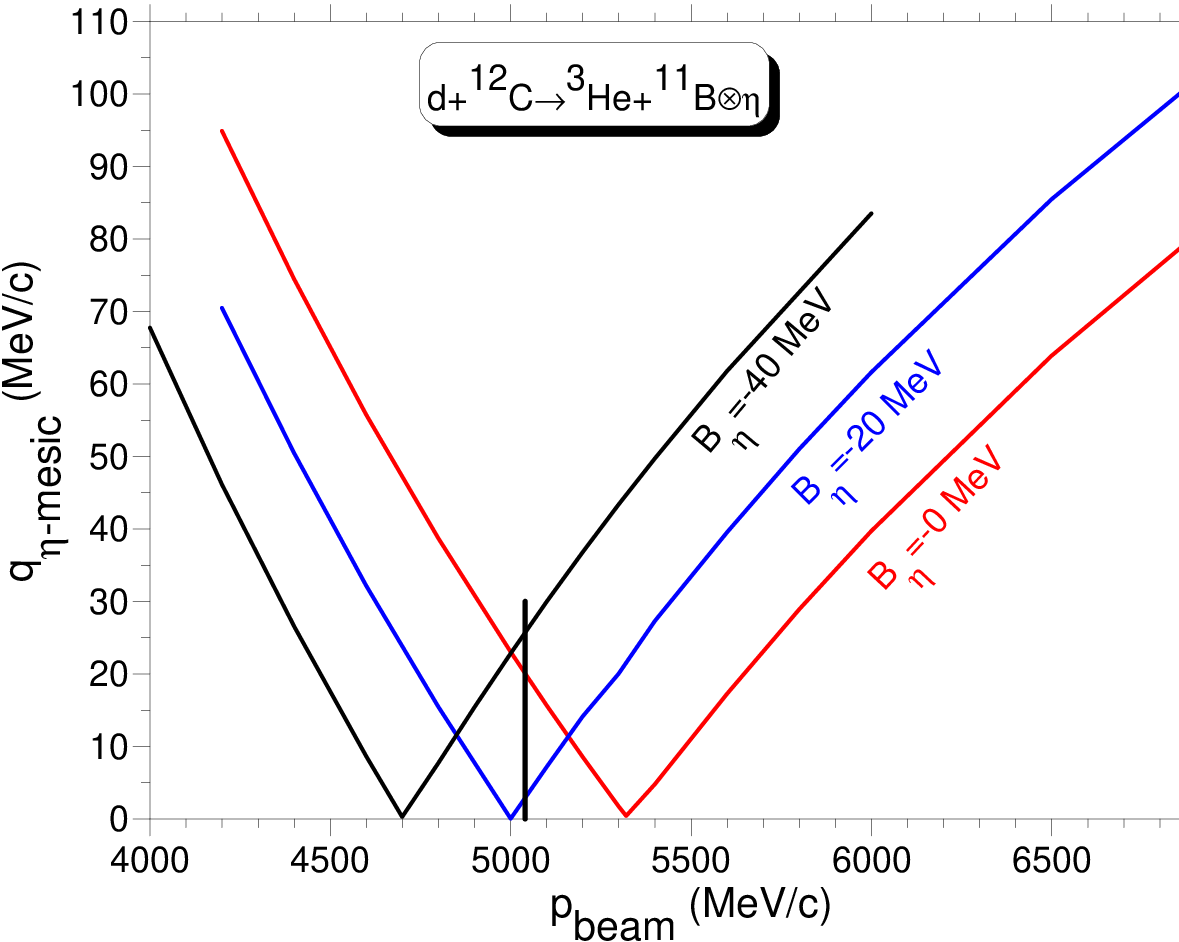}}
\parbox[b]{0.480\textwidth}{ \centering
\includegraphics[width=0.45\textwidth]{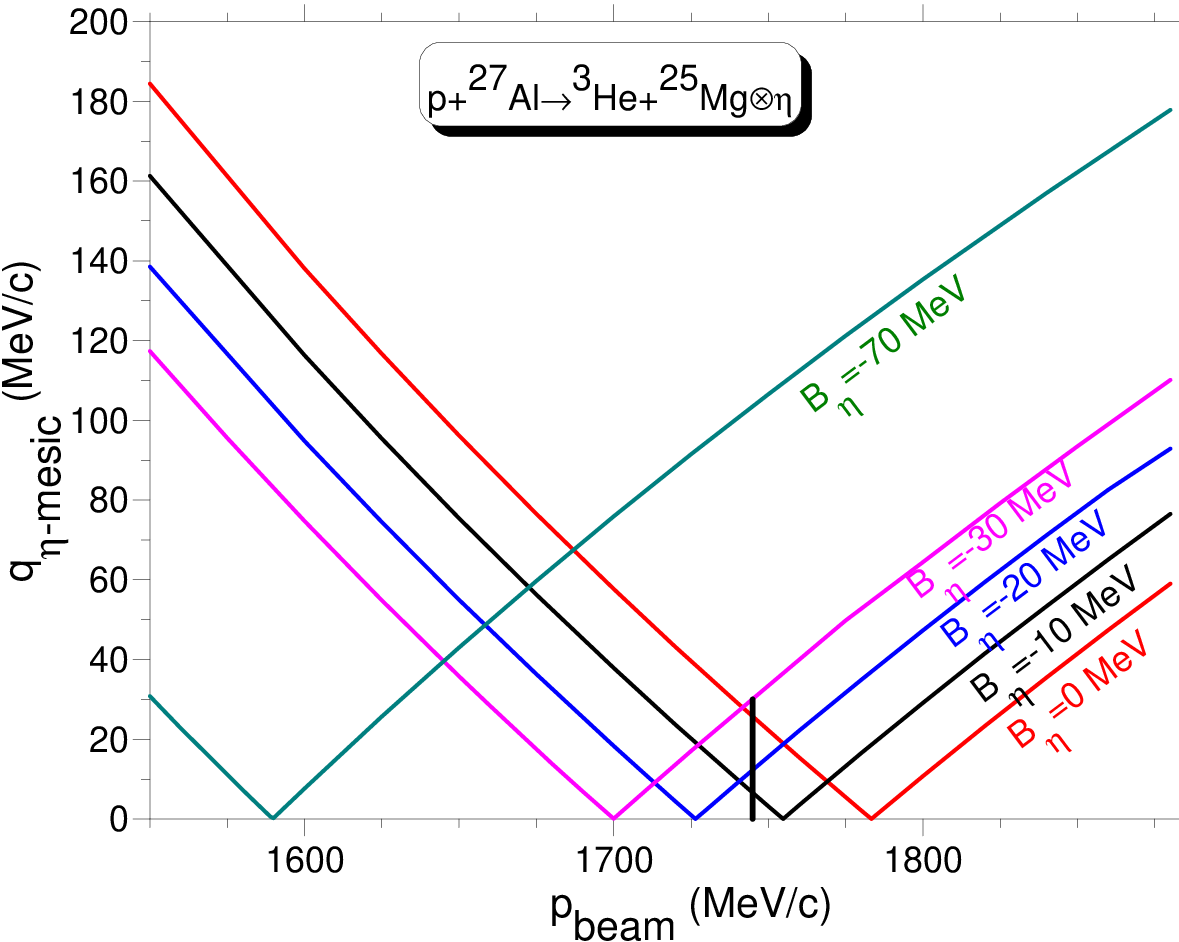}}
\caption{Linear momentum transfer from the incoming deuteron (left frame) or proton (right frame) onto the $\eta$ mesic system. The vertical marks indicate the beam momenta in the corresponding experiments \cite{Gillitzer06} and \cite{Budzanowski09}. }
\label{Fig:Magic_Kine}
\end{figure}
The proton transfer reaction at recoil free kinematics was already successfully applied in the production of pionic atoms \cite{Yamazaki96}. There, the detector employed was the GSI fragment separator \cite{Geissel92}.
\begin{figure}[h!]
\begin{center}
\includegraphics[width=0.9\textwidth]{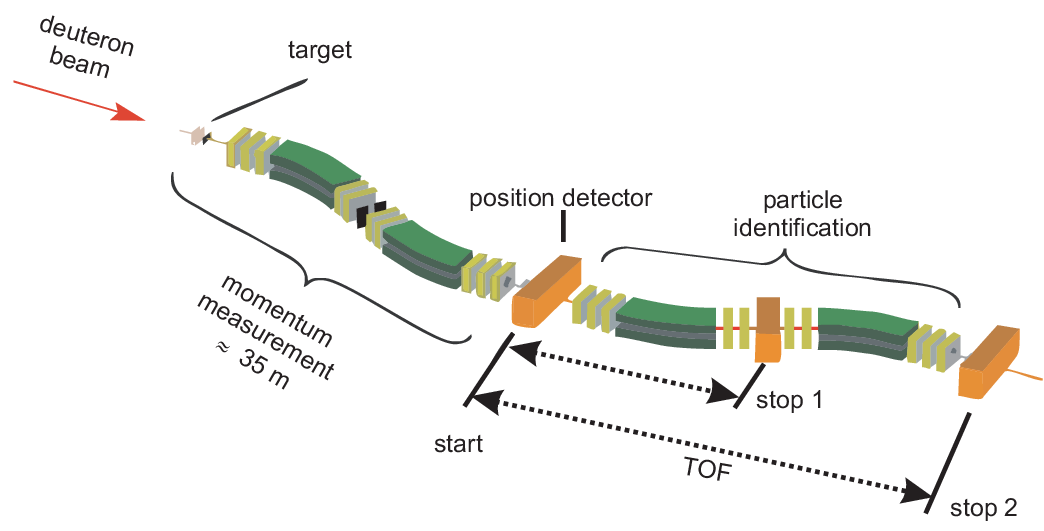}
\caption{Typical layout of the GSI fragment separator. Each half consists of two dipole magnets and several quadrupole magnets. A typical length scale is indicated.  }
\label{Fig:GSI_separator}
\end{center}
\end{figure}
The experiment \cite{Hayano97}, \cite{Gillitzer06} to search for a bound $\eta$ mesic state used the same apparatus. It is schematically shown in Fig. \ref{Fig:GSI_separator}. The first half of the 76 m long apparatus consists of two dipole magnets, allowing a high resolution momentum analysis in a dispersive mode. The second half allows particle identification via two different TOF and $\Delta E$ measurements. The spectrometer is flooded by break up protons having beam velocity and therefore the same magnetic rigidity $p/Z$ as the $^3$He particles of interest and are thus undistinguishable. The protons and $^3$He particles were differentiated in part by an additional \'{C}erenkov counter. A deuteron beam of 5040 MeV/c was used. This momentum is indicated in Fig. \ref{Fig:Magic_Kine}. At this momentum $\eta$ mesic states with binding energies $-40\text{ MeV}\leq B_\eta\leq 0\text{ MeV}$ can be produced with a momentum transfer $q\leq$ 30 MeV/c. So far no final result is published \cite{Gillitzer14}.

One nucleon transfer guarantees a rather large cross section. This is not the case for two nucleon transfer reactions (\ref{equ:reac_pHe}). However, it is just this experiment by the GEM collaboration \cite{Budzanowski09} which claims to have observed an $\eta$ mesic bound state with sufficient significance. We will therefore discuss this experiment in more detail. The experiment made use of both techniques discussed above simultaneously: transfer reaction with recoil free kinematics and back to back emission of a pion and a nucleon. The magic kinematics for the present reaction is shown in the right panel of Fig. \ref{Fig:Magic_Kine}. A proton beam from the COSY accelerator with momentum of 1745 MeV/c was used. This momentum is also indicated in Fig. \ref{Fig:Magic_Kine}. At this momentum $\eta$ mesic states with binding energies $-30\text{ MeV}\leq B_\eta\leq 0\text{ MeV}$ can be produced with a momentum transfer $q\leq$ 30 MeV/c.
\begin{figure}[ht]
\begin{center}
\includegraphics[width=0.75\textwidth]{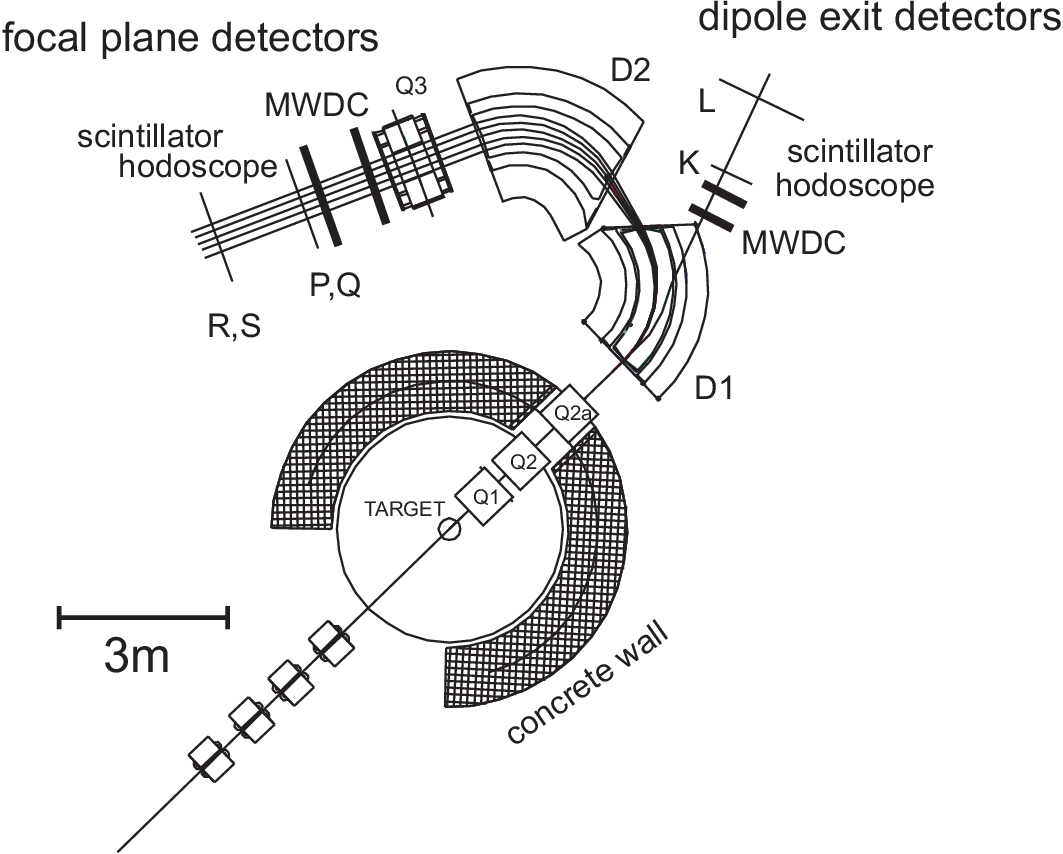}
\caption{Layout of the magnetic spectrometer BIG KARL Refs. \cite{Drochner98} and \cite{Bojowald02}. D and Q denote magnetic dipole and quadrupole magnets, MWDC multi wire drift chambers and four layers of scintillator paddles with a TOF measurement between the PQ and RS layers in the standard focal plane (used in the discussed experiment) and between K and L in the dipole exit. The former bends protons up to 1080 MeV/c while the dipole exit bends protons up to 3240 MeV/c, although with a smaller acceptance. A typical length scale is indicated.}
\label{Fig:BK}
\end{center}
\end{figure}
\begin{figure}[ht]
\begin{center}
\includegraphics[width=0.50\textwidth]{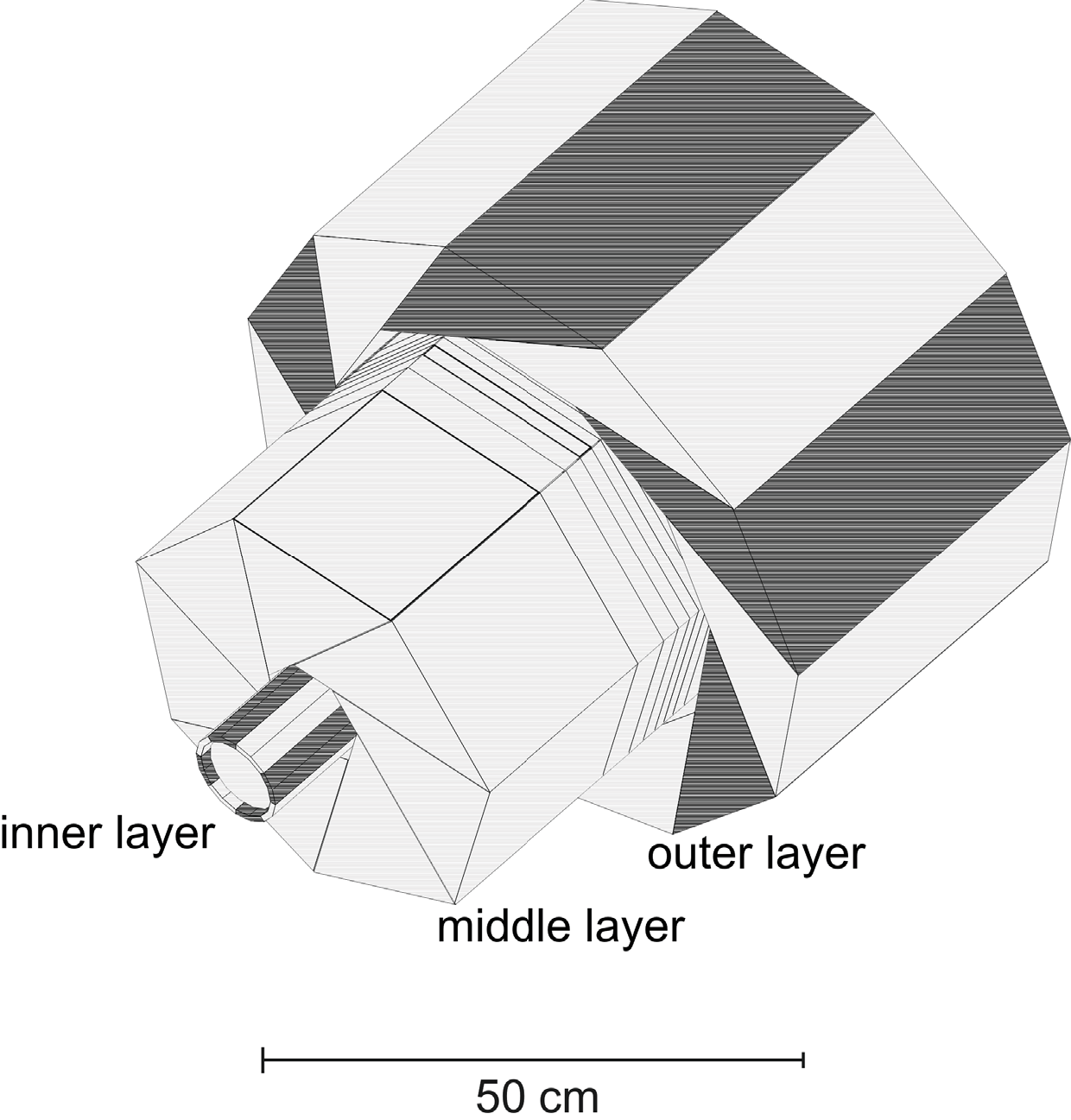}
\caption{The ENSTAR detector \cite{Betigeri07} surrounding the target. It consists of wedges from scintillating material. Read out is performed by scintillating fibres collecting the light in grooves milled in the wedges and transporting it to photo tubes.  One half of this detector is shown. The inner two layers are extruded for clarity. A typical length scale is indicated}
\label{Fig:ENSTAR}
\end{center}
\end{figure}
Again a high resolution magnetic spectrograph was utilised.
The {$^3$He} ions were  identified by the focal plane detectors of the BIG KARL spectrometer (see Fig. \ref{Fig:BK}) and their momenta measured with the same device \cite{Drochner98}. The decay $\eta n\to N^{0*}(1535)\to \pi^- p$ with the two final particles emitted almost back to back to each other was measured with a dedicated detector ENSTAR \cite{Betigeri07}. It surrounds the target and one half of it is shown in Fig. \ref{Fig:ENSTAR}. By construction it is capable of determining azimuth and polar angle.

\begin{figure}[ht]
\begin{center}
\includegraphics[width=0.48\textwidth]{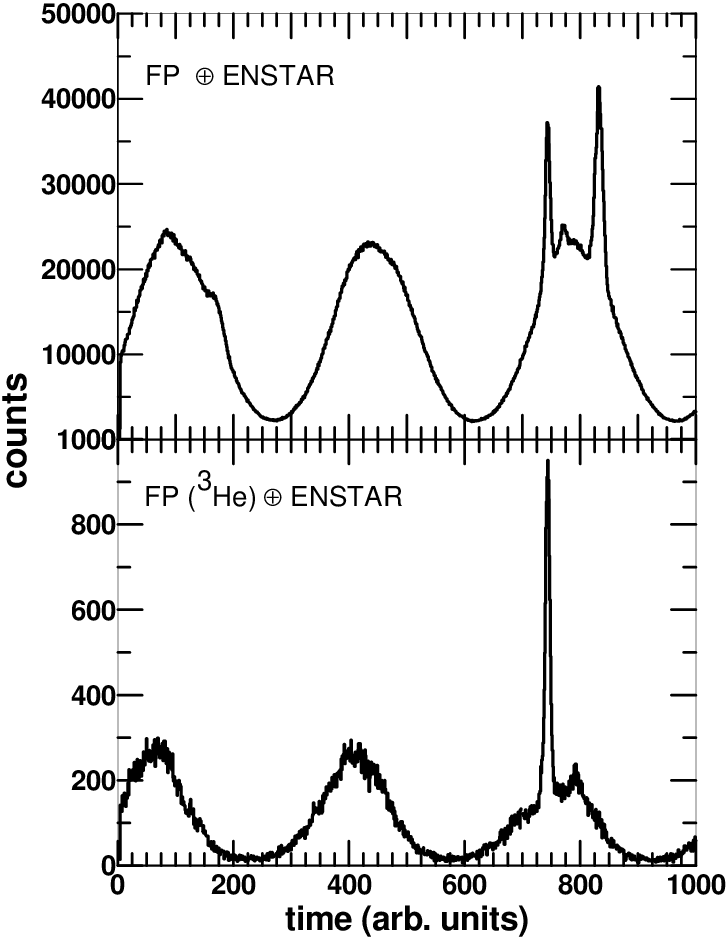}
\includegraphics[width=0.48\textwidth]{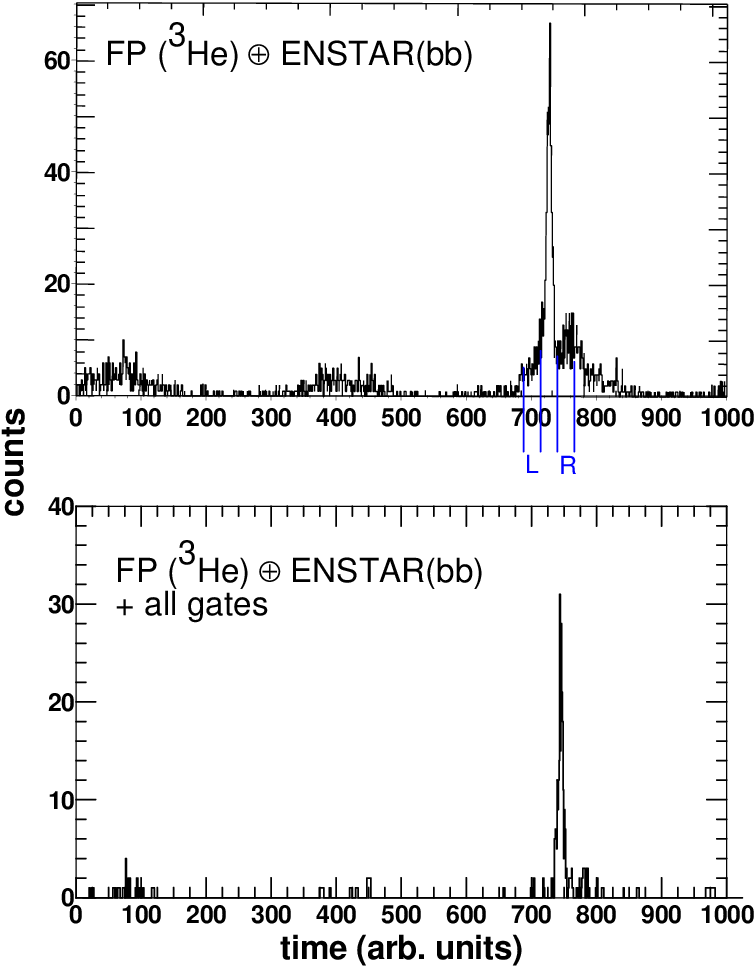}
\caption{Time measurement between the focal plane detectors (FP) and the ENSTAR detector under various constraints \cite{Machner10}. Note the reduction in count rate with increasing boundary conditions.}
\label{Fig:Time_spectrum}
\end{center}
\end{figure}

\begin{figure}[ht]
\begin{center}
\includegraphics[width=0.45\textwidth]{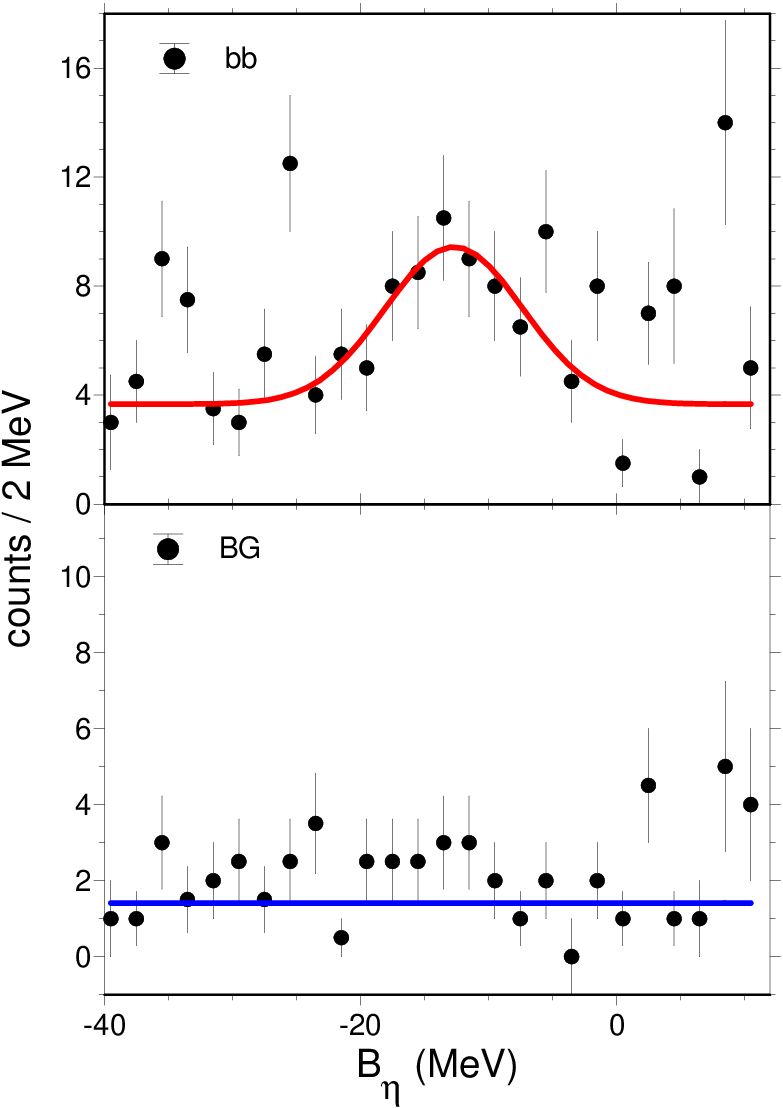}
\caption{Binding energy spectra. Upper panel: data with the requirement of a coincidence (a gate on the range between L and R indicated in the time spectrum \ref{Fig:Time_spectrum}) between a $^3$He in the focal plane and a $\pi^-$ and proton being back to back (bb) emitted recorded in the ENSTAR detector. The solid curve is a fitted Gaussian together with a constant. Lower panel: Same as upper panel but for non resonant events in L and R (see Fig. \ref{Fig:BB_Bg}). The solid line is a fitted constant.}
\label{Fig:BB_Bg}
\end{center}
\end{figure}
In Fig. \ref{Fig:Time_spectrum} time spectra between the scintillation layers in the focal plane (FP) and the ENSTAR detector are shown measured under different conditions. The spectrum upper left is obtained if only a coincidence between FP and ENSTAR is required. The broad structures correspond to different extracted beam particles usually separated by three revolutions in the synchrotron (slow or resonance extraction).  On top of one structure are needles corresponding to physical coincidences with particles registered in the focal plane. The TOF is given by $t=lmZ/p$ with $l$ the flight path, $m$ and $Z$ the mass and charge number of the particle and $p$ its momentum. If out of identified particles only $^3$He is selected, the lower left spectrum emerges. So the remaining needle is due to $^3$He coincidence. The counting rate for the accidental coincidences is reduced.  Requiring now a $\pi^-p$ close to back to back in the ENSTAR detector gives the time spectrum shown in the upper right part. We now define two background intervals left (L) and right (R) of the needle and produce a background energy spectrum. This spectrum is subtracted from the upper right one and the remaining spectrum is shown in the lower right part of Fig. \ref{Fig:Time_spectrum}. The latter is now almost free of background.

We want now to discuss the effect of the conditions applied not only to the time spectrum but also to missing mass spectrum or binding energy spectrum.  The momenta measured in the  FP are shown in Fig. \ref{Fig:BB_Bg}, converted to binding energy. The upper part corresponds to the upper right time spectrum. The background spectrum due to the two intervals L and R in the time spectrum yields the binding energy spectrum shown in the lower part of Fig. \ref{Fig:BB_Bg}. Finally this background spectrum is subtracted from the upper spectrum and the result is the spectrum shown in Fig. \ref{Fig:Final} which is the final spectrum of Ref. \cite{Budzanowski09}.
\begin{figure}[h!]
\begin{center}
\includegraphics[width=0.60\textwidth]{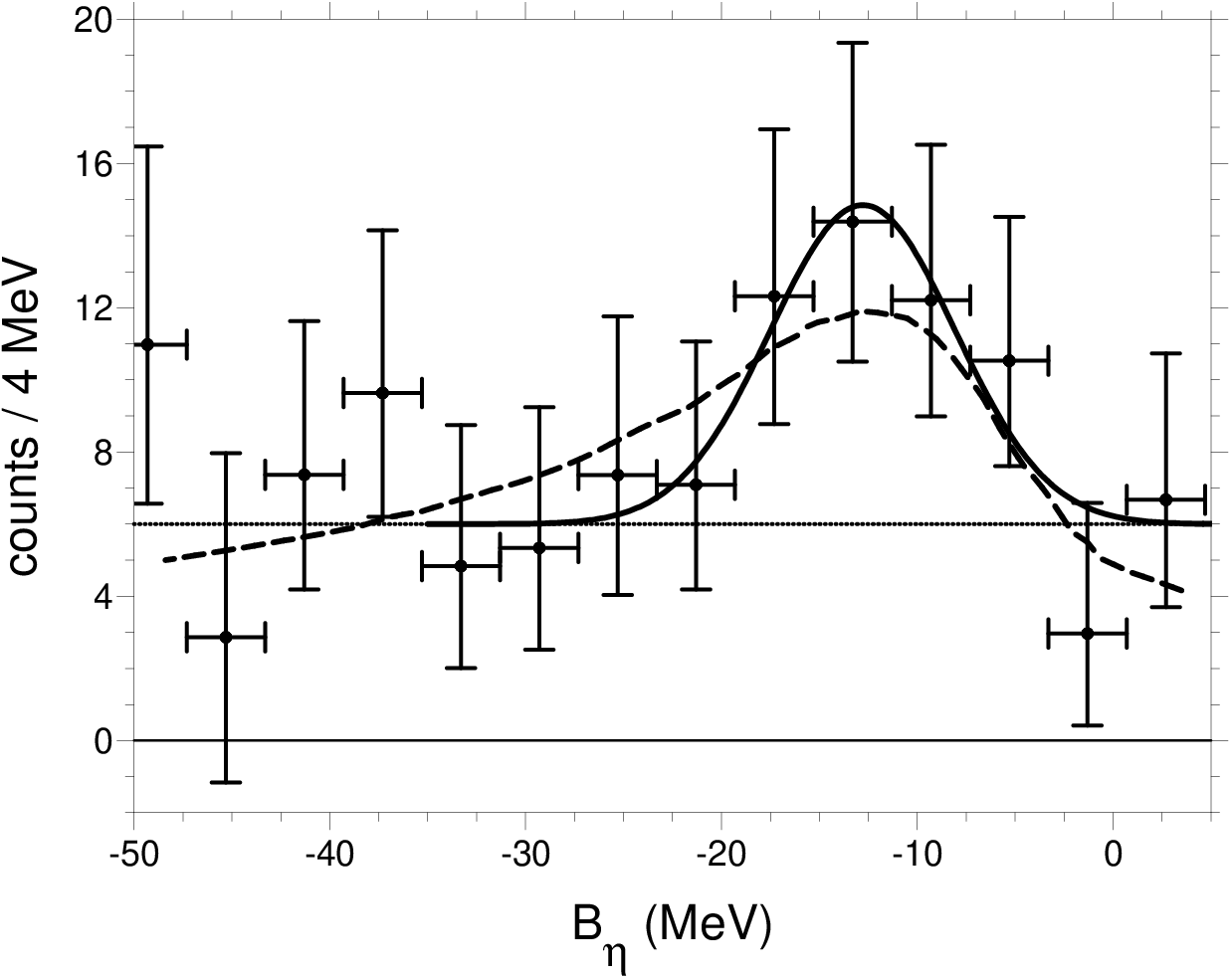}
\caption{The final binding energy spectrum. Note the expanded view and the different binning compared to Fig. \ref{Fig:BB_Bg}.  The data are shown with Poisson error bars. A fit to the data with a Gaussian and a constant background are shown (solid curve). A fit with a Breit-Wigner form with a coherent non resonant fraction is shown as dashed curve (from \cite{Haider-Liu10}).}
\label{Fig:Final}
\end{center}
\end{figure}
This spectrum shows a peak on a continuum. This continuum was parameterised by a constant as well as polynomials while for the peak a Gaussian was assumed. In addition fits were performed applying Poisson statistics. The significance of the peak is around 5$\sigma$ \cite{Budzanowski09}. The centroid $E_B$ and Gaussian width $\sigma$ were found to be -12.0$\pm$2.2 MeV and 4.7$\pm$1.7 MeV.

The procedure applied by the GEM collaboration \cite{Budzanowski09} to assume a further background below the peak was questioned by Haider and Liu \cite{Haider-Liu10}. The final state can also be reached by a non resonant reaction for which they used a microscopic-theory based nearly energy-independent amplitude. The need of adding non-resonant amplitude is further discussed in \cite{Liu14}. Then there will be an interference between this amplitude and the one for the resonant production. They fitted the corresponding amplitudes to the experimental data and found indeed a serious interference effect which shifts the calculated Breit-Wigner maximum towards the experimental maximum. The same is true for the width. One such fit is also shown in Fig. \ref{Fig:Final}. For this curve the $\eta N$ scattering length is (0.250 + 0.123i) fm, corresponding to an imaginary-to-real ratio $R$ of 0.49. This is in agreement with the values given in Table \ref{Tab:eta-N} except for the Green Wycech (GW) model \cite{Green05}. If the interference is neglected a scattering length $a=(0.292\pm0.077i)$ fm is necessary to get the experimental peak parameters. The ratio is then $R= 0.24$. This is in accord with the Green Wycech result although then the real part of the scattering length is much larger, almost 1 fm. Such a large value was found necessary by Friedman et al. \cite{Friedman13}. It should be mentioned that both values obtained for the imaginary part of the scattering length violate the condition Eq. \eqref{equ:optical}.

\subsection{Final State Interactions}\label{sub:Final-State-Interaction}
Another method proposed to search for $\eta$ bound states is to
study the final state interaction (\fsi) between the $\eta$ meson
and a nucleus.

\subsubsection{The reaction $d+p\to ^3$He$+\eta$}
The system  most intensively studied is the $d+p\to ^3$He$+\eta$ reaction. Data are from Refs.  \cite{Berger88},  \cite{Betigeri00}, \cite{Adam07}, \cite{Rausmann09}, \cite{Mayer96}, \cite{Smyrski07}, \cite{Mersmann07}. Those close to threshold are shown in Fig. \ref{Fig:pd2eta_old}.
\begin{figure}[ht]
\begin{center}
\includegraphics[width=0.75\textwidth]{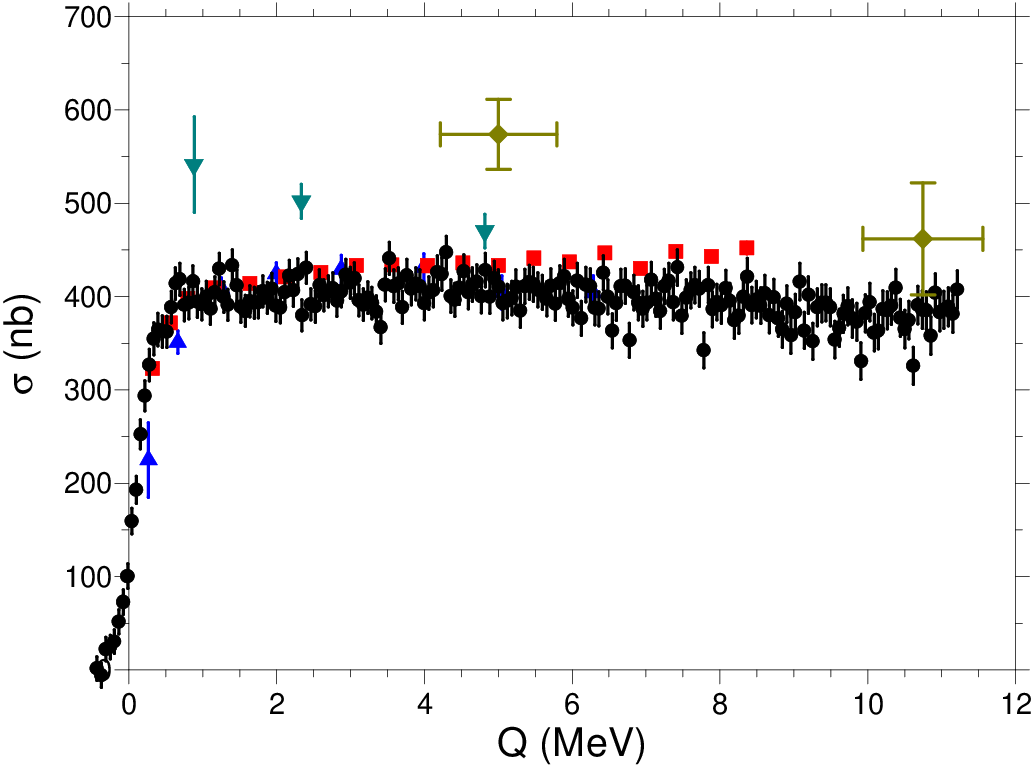}
\caption{Published total cross sections for the reaction $p+d\to \eta+^3$He. The fat triangles down are from \cite{Berger88}, triangles up from \cite{Mayer96}, diamonds with error bars in both directions from \cite{Bilger02} and \cite{Bilger04}, squares from \cite{Smyrski07}, and those with small dots from \cite{Mersmann07}.  }
\label{Fig:pd2eta_old}
\end{center}
\end{figure}
In this figure the published data are compiled. Obviously there are differences between the different data sets. This leads of course to different results for the final state parameters. Here we will ignore the data sets which are completely different to the bulk of data.

COSY~11 \cite{Smyrski07}, quoted also as Smyrski, and ANKE \cite{Mersmann07}, quoted also as Mersmann,  measured at COSY applying the internal deuteron beam. The momentum of the beam increased linearly with time. Data were taken continuously and later put into bins with widths $\Delta Q$. The strong nonlinearity of the cross section may lead to a deviation of the measured cross section, assumed to be the one in the middle of the momentum bin, relative to the real one. The relation between both is
\begin{equation}
\sigma^\text{meas}(Q) = \frac{1}{\Delta Q}\int_{Q-\Delta Q/2}^{Q+\Delta Q/2}dQ_1 \int_{Q-\delta Q/2}^{Q+\delta Q/2} dQ_2\, w(Q_1-Q_2)\, \sigma^\text{real}(Q_2)\,,
\end{equation}
with $\delta Q$ the width of the beam distribution and $w$ the distribution function normalised to one. COSY~11 used an inverted parabolic distribution while ANKE made use of a Gaussian. Such a smearing was found sufficient by the ANKE collaboration to bring the points measured below the threshold to above. COSY~11 assumed in addition a downward shift of the mean beam momentum by 3 MeV/c. This is in agreement with a precision determination of the beam momentum  at 1930 MeV/c, where the beam momentum was found to be 2 MeV/c lower than its nominal value \cite{Betigeri99}.

COSY~11 \cite{Smyrski07} applied Eq. (\ref{equ:scattering_length}) to their data and obtained
\begin{equation}
a_{\eta {^3He} } = {\pm}(2.9 {\pm} 2.7)+i(3.2 {\pm} 1.8) \text{ fm}.
\end{equation}
This corresponds to a possible bound state at $B_\eta=-0.2\pm 0.8$~MeV. So this result points more to a virtual than to a bound state. The half width is $\Gamma/2=1.9\pm 0.4$~MeV. This result is close to the one of earlier data \cite{Mayer96}:
\begin{equation}\label{equ:Mayer}
a_{\eta {^3He}}= \pm (3.8\pm 0.6) +i(1.6\pm 1.1)\text{ fm}\,.
\end{equation}
However, when we repeated the fit for the data from Ref. \cite{Smyrski07} we found different values and moreover they depend on the fit interval.
\begin{figure}[ht]
\begin{center}
\includegraphics[width=0.6\textwidth]{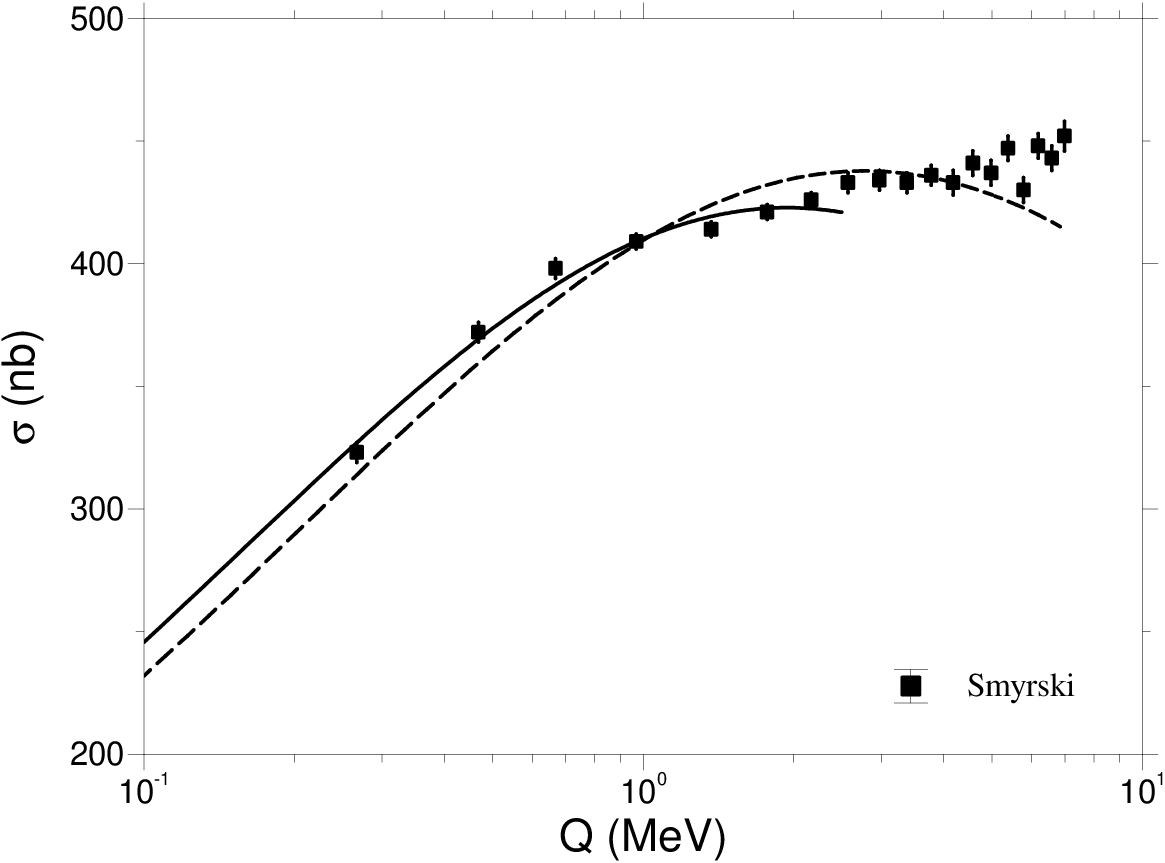}
\caption{Cross sections for the reaction $d+p\to \eta+^3$He from Ref. \cite{Smyrski07} (COSY 11 collaboration, squares with error bars). The solid curve is a fit to the data assuming \fsi to be represented by a scattering length only in the range up to $Q\approx 2.2$ MeV while the dashed curve is a fit to all data points, which results into useless numbers (see text).  }
\label{Fig:Smyrski}
\end{center}
\end{figure}
While the value for $a_i$ is quite stable, $a_r$ varied from $0.0\pm
6000$ fm, when the full data set is included in the fit, to $2.1\pm
2.7$ fm, when the range is limited to 2.2~MeV. The large error is an
indication that the option of fitting the full range is useless, because the assumption of pure $s$ wave is wrong. The
imaginary part is $3.6 \pm 1.2$ fm. These numbers are in agreement
with the published values. The corresponding curves are compared to
the data in Fig. \ref{Fig:Smyrski}. This finding is an indication
that already for excess energies above 2.2~MeV Eq.
(\ref{equ:scattering_length}) is no more applicable and the
effective range has to be considered in addition to the scattering
length as has been stressed in Ref. \cite{Niskanen13}. We have
therefore tried to fit the full expansion Eq.
(\ref{equ:effective_range}) to the data. However, even in this case
the fit curve drops similarly to the results in Fig.
\ref{Fig:Smyrski}, while the data increase slightly with energy.
Therefore no satisfactory result could be obtained.

On the other hand the ANKE data show after the rapid rise a gentle
decrease with increasing energy. Mersmann \cite{Mersmann07a} has
performed  a corresponding fit to the ANKE data including the
smearing as discussed above. This fit yielded
\begin{gather}\label{equ:Anke_fsi}
a_{\eta ^3He }  = \left[ { \pm \left( {0.000 \pm 2.416 } \right) + i \cdot \left( {6.572 \pm 0.501} \right)} \right]{\text{fm}} \\
r_{0,\eta ^3He }  = \left[ {  \left( {0.000 \pm 2.416 } \right) + i \cdot \left( {1.268 \pm 0.212} \right)} \right]{\text{fm}}\,.
\end{gather}
The scattering length and the effective range are thus determined by
the imaginary parts alone. The data and the fit curve are shown in
Fig. \ref{Fig:pd2he3_eta}. In order to make the curve visible we
have not shown the data as in Fig. \ref{Fig:pd2eta_old}, but added
five consecutive channels together. The bin width thus comes to the
same order of magnitude as the beam width: $\Delta Q/2=150$ keV and
170 keV, respectively. In addition the error bars become compatible
with those from the COSY~11 experiment. The fit results do neither
fulfil the criterion $|a_r|>|a_i|$ nor the more general relation Eq.
(\ref{equ:cond_full})

The ANKE collaboration \cite{Mersmann07} applied in addition another
fitting form. They assumed a two pole representation of the final
state interaction
\begin{equation}
f_s(p)=\frac{f_B}{(1-\frac{p}{p_1})(1-\frac{p}{p_2})}\,
\label{equ:poles}
\end{equation}
with $p_1$ and $p_2$ two complex pole positions. By comparing with
Eq. (\ref{equ:effective_range}) one gets
\begin{equation}\label{equ:a_pole}
a=-i\left(\frac{1}{p_1}+\frac{1}{p_2}\right)
\end{equation}
and
\begin{equation}\label{equ:r_pole}
r_0=\frac{2i}{p_1+p_2}\,.
\end{equation}
The second pole is assumed to have no physical meaning. It will just
represent the additional energy dependence. For $p_2\to \infty$ the
relation (\ref{equ:pole_position}) is regained. The fit results for
the poles were \cite{Mersmann07}
\begin{gather}
p_1 =  \left[ {\left( {-5 \pm 7_{ -1}^{ + 2} } \right) \pm i \cdot \left( {19 \pm 2 \pm 1 } \right)} \right] \text{MeV/c} \\
p_2 =  \left[ {\left( {106 \pm 5 } \right) \pm i \cdot \left( {76 \pm 13 _{ -2}^{ + 1} } \right)} \right] \text{MeV/c}\,.
\end{gather}
This result yields the low energy parameters
\begin{equation}
a_{\eta ^3He }  = \left[ { \pm \left( {10.7 \pm 0.8_{ - 0.5}^{ + 0.1} } \right) + i \cdot \left( {1.5 \pm 2.6_{ - 0.9}^{ + 1.0} } \right)} \right]{\text{fm}}
\end{equation}
and
\begin{equation}
r_{0,\eta ^3He }  = \left[ {\left( {1.9 \pm 0.1} \right) + i \cdot \left( {2.1 \pm 0.2_{ - 0.0}^{ + 0.2} } \right)} \right]{\text{fm}}.
\end{equation}
In obtaining these values a smearing of the energy scale due to a
finite beam momentum distribution was applied. This results in a
pole (if exists) at $B_\eta=0.30\pm 0.15 \pm 0.04$ MeV and
$\Gamma_\eta/2 = 0.21 \pm 0.29\pm 0.6$ MeV.
\begin{figure}[ht]
\begin{center}
\includegraphics[width=0.6\textwidth]{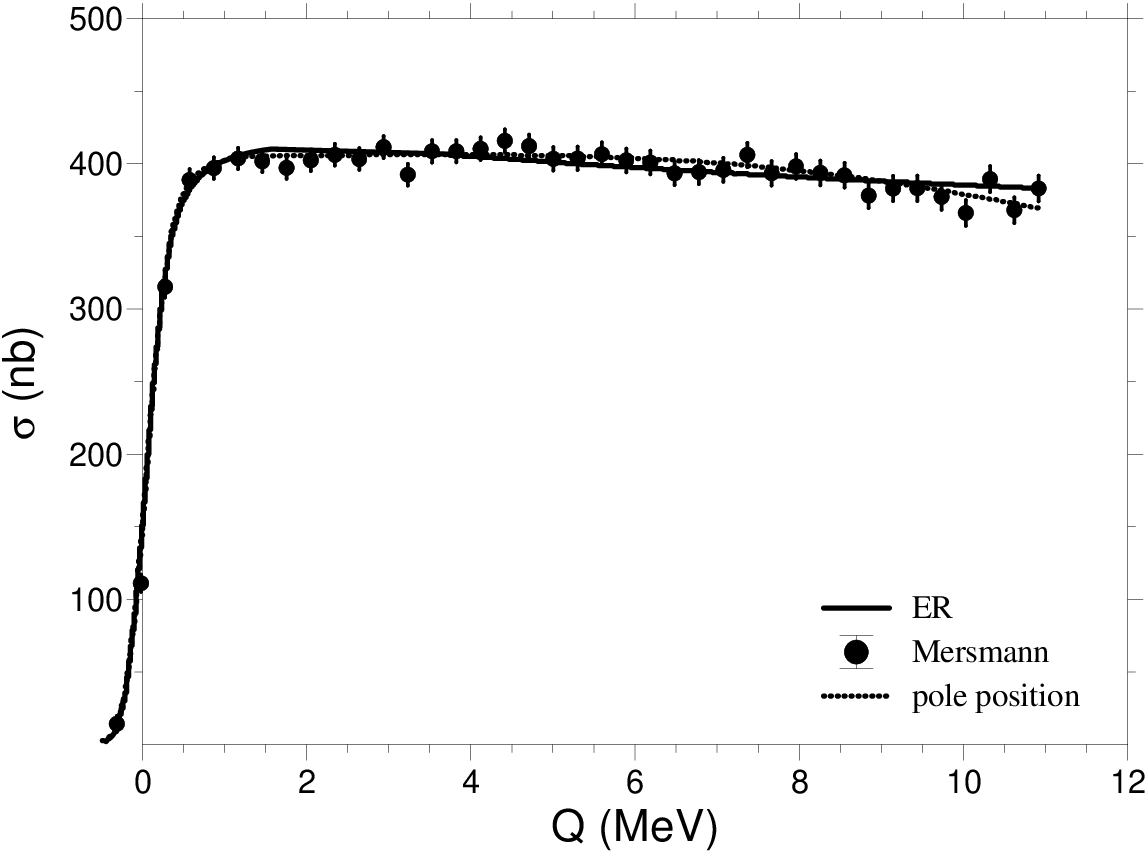}
\caption{Excitation function for the $d+p\to ^3$He$+\eta$ reaction.
The total cross sections are from \cite{Mersmann07} (ANKE
collaboration, full dots with error bars). The uncertainty of the
beam energy leads to an uncertainty in Q of 9 keV.  For visibility
of the curves we have combined five data points together.  The solid
curve is a fit with scattering length and effective range, while the
dashed curve is one applying the two pole expansion
(\ref{equ:poles}). }
\label{Fig:pd2he3_eta}
\end{center}
\end{figure}
The fit curve is also shown in Fig. \ref{Fig:pd2he3_eta}. Although
the \fsi parameters differ drastically from those of the fit
(\ref{equ:Anke_fsi}) the two fit curves are practically
indistinguishable especially in the strong rising  part which is
decisive for the scattering length. It is somewhat surprising that
two fits with five parameters each and a one to one correspondence
give so different results. As stated above, the data from COSY~11
have the tendency to rise while the ANKE yields drop (see Fig.
\ref{Fig:pd2eta_old}). This might be due to different acceptance
corrections in the two experiments. Another difference between the
two experiments is that COSY~11 corrected the beam energy with thus
almost no need to apply the smearing, while ANKE get rid of data
below threshold by the smearing method.

The extraction of \fsi parameters is valid only if the total cross sections are all due to $s$ wave. This can be proven by checking if the emission is isotropic.
\begin{figure}[ht!]
\begin{center}
\includegraphics[width=0.6\textwidth]{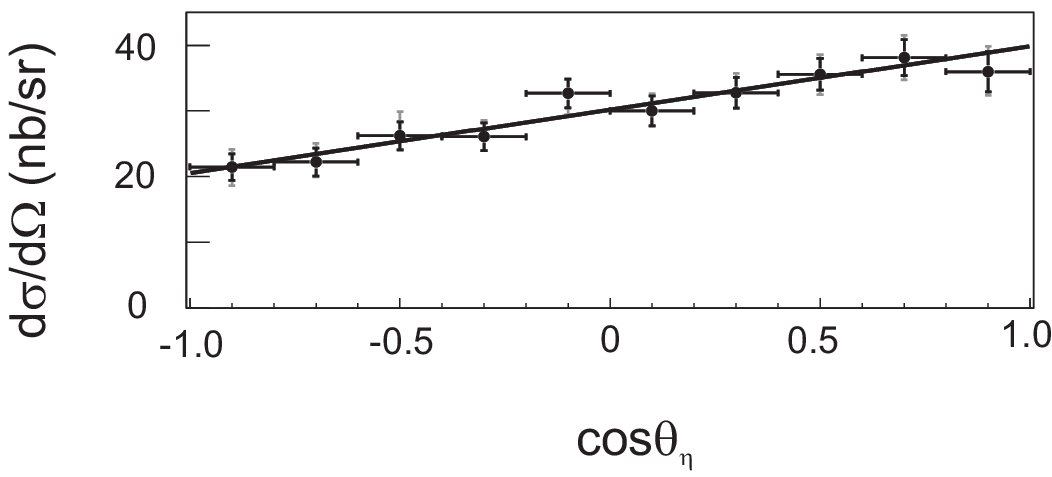}
\caption{The angular distribution of $\eta$ mesons in the $dp\to \eta ^3$He reaction at  $p_\eta = 97.0\pm 0.6$ MeV/c, adapted from Ref. \cite{Mersmann07}.}
\label{Fig:angdist_dp}
\end{center}
\end{figure}
Such a test case is shown in Fig. \ref{Fig:angdist_dp}. The angular distribution is clearly not isotropic and hence $p$ waves contribute to the cross section already close to threshold. We can define an asymmetry as
\begin{equation}
\label{equ:asymmetry}
\frac{d\sigma(\theta_\eta)}{d\Omega} =\frac{\sigma_\text{tot}}{4\pi} (1+\alpha \cos\theta_\eta).
\end{equation}
This ansatz yields a linear dependence on $\cos\theta_\eta$ and a fit of (\ref{equ:asymmetry}) is also shown in the figure. The fit does not require partial waves higher than $p$ waves. Fig. \ref{Fig:asymmetry} shows the asymmetry $\alpha$ as function of the $\eta$ momentum.
\begin{figure}[h!]
\begin{center}
\includegraphics[width=0.6\textwidth]{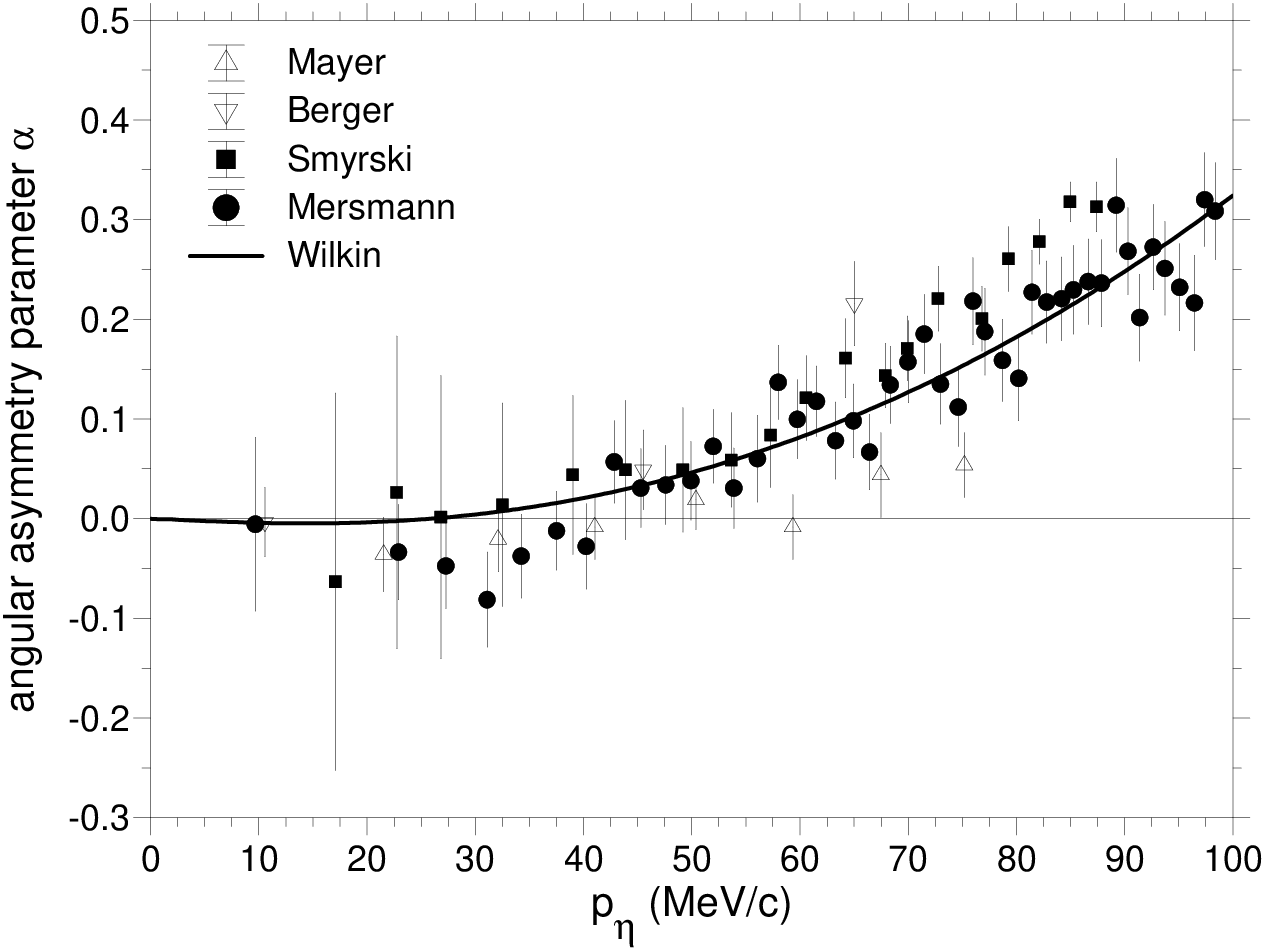}
\caption{The asymmetry parameter $\alpha$ (Eq. (\ref{equ:asymmetry})) as a function of the $\eta$ momentum. The data are from \cite{Mayer96} (open triangles up), \cite{Berger88} (open triangles down), \cite{Smyrski07} (full squares) and \cite{Mersmann07} (full dots). Shown are the statistical  errors only. The solid curve is a fit with the model from Ref.  \cite{Wilkin07} (see text). }
\label{Fig:asymmetry}
\end{center}
\end{figure}

The asymmetry $\alpha$ can be deduced from the data via
\begin{equation}
\alpha =\frac{d\sigma(\cos\theta_\eta=+1)-d\sigma(\cos\theta_\eta=-1)} {d\sigma(\cos\theta_\eta=+1)+d\sigma(\cos\theta_\eta=-1)}\,.
\end{equation}
Insertion of Eq. (\ref{eqn:unpol-cs}) into this equation and limiting ourselves to only the two $s$ wave amplitudes $A$ and $B$ and the two $p$ waves with amplitudes $C$ and $D$ which have a linear dependence on $\cos\theta_\eta$ yields
\begin{equation}
\alpha=2p_{\eta}\,\frac{\text{Re}(A^*C+2B^*D)}
{|A|^2+2|B|^2+p_{\eta}^{\:2}|C|^2+2p_{\eta}^{\:2}|D|^2}\,.
\label{alpha2}
\end{equation}
We are now left with only two observables for four complex amplitudes. It should be remembered that $A$ and $B$ contain admixtures of $p$ waves (see \ref{sub:pd_theory}). In Ref. \cite{Wilkin07} it was argued that the rapid change in size of the cross section is due to \fsi in the $s$ wave and not to fast changes in the $p$ waves which were believed to vary smoothly. With the assumptions $A=B=f_s$ and $C=D$ the cross section and the asymmetry parameter $\alpha$ reduce to
\begin{equation}\label{equ:ABfs}
\sigma=\frac{4\pi
p_{\eta}}{p_p}\left[|f_s|^2+p_{\eta}^{\,2}|C|^2\right],
\end{equation}
\begin{equation}
\alpha=2p_{\eta}\,\frac{\text{Re}(f_s^*C)}
{|f_s|^2+p_{\eta}^{\:2}|C|^2}\:\cdot \label{eqn:alpha3}
\end{equation}
A fit of Eq. (\ref{eqn:alpha3}) to the ANKE data as performed by Wilkin et al. \cite{Wilkin07} is also shown in Fig. \ref{Fig:asymmetry}. It reproduces the data quite nicely. Also an excellent fit was obtained with negative $\alpha$'s for small momenta $p_\eta$, when a phase variation caused by the $s$ wave pole was considered. At this point we want to stress the impact of high quality data. In Ref. \cite{Sibirtsev04a} it was stated, based on the then existing data, that the asymmetry parameter $\alpha$ is practically zero although the data showed an increase. However, the statement was valid at that time because of the huge error bars.

If the assumptions made above are reasonable then the tensor analysing power $T_{20}$ should be very small (see Eq. (\ref{equ:t20})).
\begin{figure}[ht!]
\begin{center}
\includegraphics[width=0.6\textwidth]{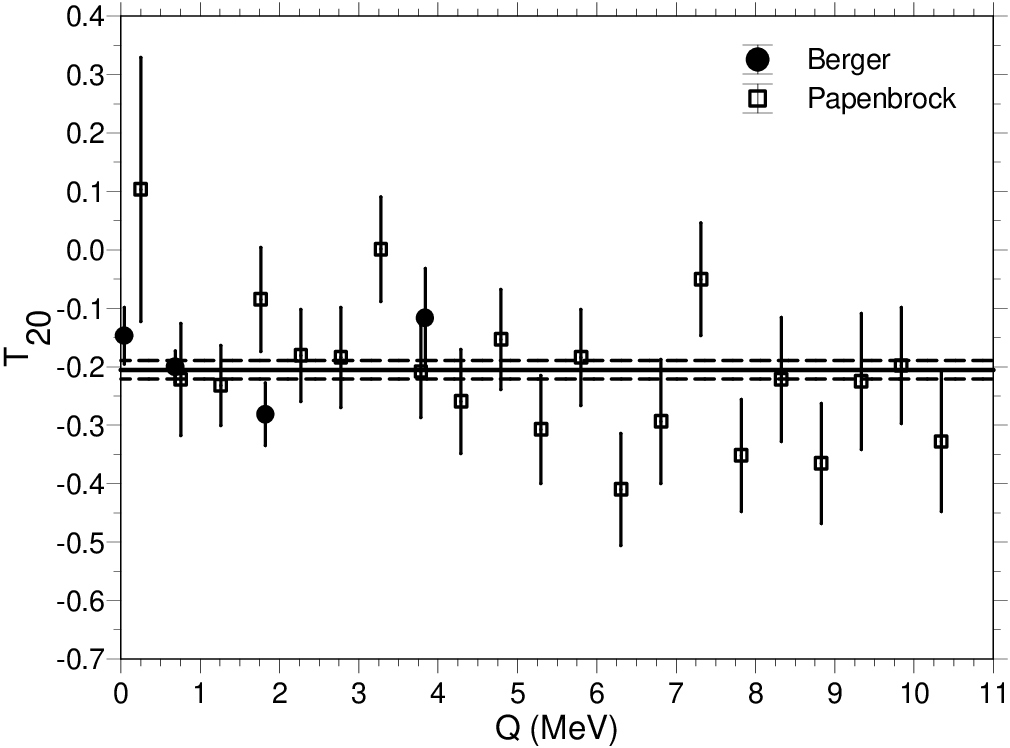}
\caption{The angle averaged tensor analysing power $T_{20}$ for the reaction $d+p\to \eta+^3$He as function of the excess energy $Q$. The data are from \cite{Berger88} (full dots) and \cite{Papenbrock14} (open squares).}
\label{Fig:Bergert20}
\end{center}
\end{figure}
Indeed Berger et al. \cite{Berger88} measured this quantity and the
results are shown in Fig. \ref{Fig:Bergert20}. The authors found
that the variation with angle is smaller than the error bar of the
mean. So only the mean (angle average) is shown in Fig.
\ref{Fig:Bergert20}. Recent measurement of the tensor analysing
power by the ANKE collaboration \cite{Papenbrock14}, which will be
discussed below, shows a similar behaviour and therefore also the
values averaged over $\cos(\theta_\eta)$ are shown in Fig.
\ref{Fig:Bergert20}. These values vary around $T_{20}\approx -0.2$.
This implies $C\approx D$ and $|A|^2\gtrapprox |B|^2$.

The question is, how well are the assumptions made? If this is the
case we have from the fit the $s$ wave amplitude $f_s$ and one can
deduce the \fsi. However, one cannot from the data base discussed so
far deduce that the $p$ waves in the amplitudes $A$ and $B$ are
negligible.  Recently Papenbrock et al. \cite{Papenbrock14} (ANKE
collaboration)  have measured the differential cross section using a
vector and tensor polarised deuteron beam again in the acceleration
ramp. They made use of Eq. \eqref{equ:dsigma_rho_tensors} to extract
the vector and tensor analysing power $T_{11}$ and $T_{20}$. They
find $T_{11}\leq 0.04$ for excess energies below 10 MeV. This
implies that the amplitudes $C$ and $D$ must be very small (see Eq.
\eqref{equ:T11}). This means that it will be impossible to extract
from the interference between $s$ and $p$ waves the sign of the
scattering length (see section \ref{sub:pd_theory}).  On the other
hand the asymmetry parameter $\alpha$ has then to be zero which is
not the case. The only possibility is then that the $p$ wave
fractions in $A$ and $B$ are not necessary small and the assumption
$A=B$ is not quite right.

In the following we then treat only the amplitudes $A$ and $B$. In this limit the tensor analysing power is
\begin{equation}\label{equ:T20_red}
T_{20}(\theta_\eta) = \sqrt{2}\frac{|B(\theta_\eta)|^2-|A(\theta_\eta)|^2} {|A(\theta_\eta)|^2+2|B(\theta_\eta)|^2}\,.
\end{equation}
Papenbrock et al. \cite{Papenbrock14} found that $T_{20}$ does not depend on angle. Hence the angular dependencies of $|A|^2$ and $|B|^2$ have to be the same. Furthermore $T_{20}$ is fairly constant over the measured range. Fitting a constant to all data shown in Fig. \ref{Fig:Bergert20} yields $T_{20}=-0.205\pm 0.016$ with a $\chi^2$/free=1.2. It seems therefore natural to assume $A\propto B$ which then yields from inserting into Eq. \eqref{equ:T20_red} $|A|^2\approx 3/2|B|^2$.

The present data body is not sufficient to extract the $s$ wave part on an amplitude base. However, the asymmetry of the differential cross section is practically zero for momenta $p_\eta$ up to 40 MeV/c. Therefore, it is safe to extract \fsi parameters within this near threshold interval.

\subsubsection{The Reaction $d+d\to \eta+^4$He}\label{sub:Reaction_dd}

Measurements of reaction $d+d\to \eta+^4$He were reported in \cite{Frascaria94}, \cite{Willis97}, \cite{Wronska05} and more recently in \cite{Budzanowski09b}. The last one is here  sometimes also called GEM data. The cross section is much smaller than for the previously discussed reaction $d+p\to \eta+^3$He (see Fig. \ref{Fig:Compare_he3-He4}).

In a simultaneous analysis of the $d+p\to ^3$He$+\eta$ reaction and the $d+d\to ^4$He$+\eta$ reaction in terms of a simple optical model approach \cite{Willis97} it was found that for the possible binding energies the relation $B_\eta(^3\text{He}+\eta)<B_\eta(^4\text{He}+\eta)$ holds. However, in the measurements with a polarised beam on which the analysis was based, the full polar angle could not be measured. The $s$-wave cross section was extracted by assuming isotropic emission. This isn't true for the data measured at the higher momenta as can be seen by comparing to the data from Ref. \cite{Wronska05}. The anisotropy could be either due to $s$ waves plus $p$ waves or to a $s-d$ wave interference. The problem could be only solved by applying polarised deuterons. Such an experiment was performed by the GEM collaboration at COSY J\"{u}lich \cite{Budzanowski09b} which will be discussed now in some detail.

The experiment was performed at a deuteron beam momentum of 2385.5 MeV/c corresponding to an excess energy of 16.6 MeV. Recoiling $\alpha$ particles were identified and their four momentum vector measured with the magnetic spectrograph Big Karl (see Fig. \ref{Fig:BK}). The experiment made use of polarised as well as unpolarised deuteron beams. The tensor polarisation $p_{zz}$ of the tensor polarised beam was obtained by measuring the polarised cross section of the elastic $d+p$ scattering and comparing to the known analysing power of this reaction. Later in the experiment it was monitored by detecting reaction particles in a scintillator disc consisting of 16 wedges.  The angular distribution of the unpolarised cross section is shown in Fig. \ref{Fig:Ma-Wro}.
\begin{figure}[h]
\begin{center}
\includegraphics[width=0.45\textwidth]{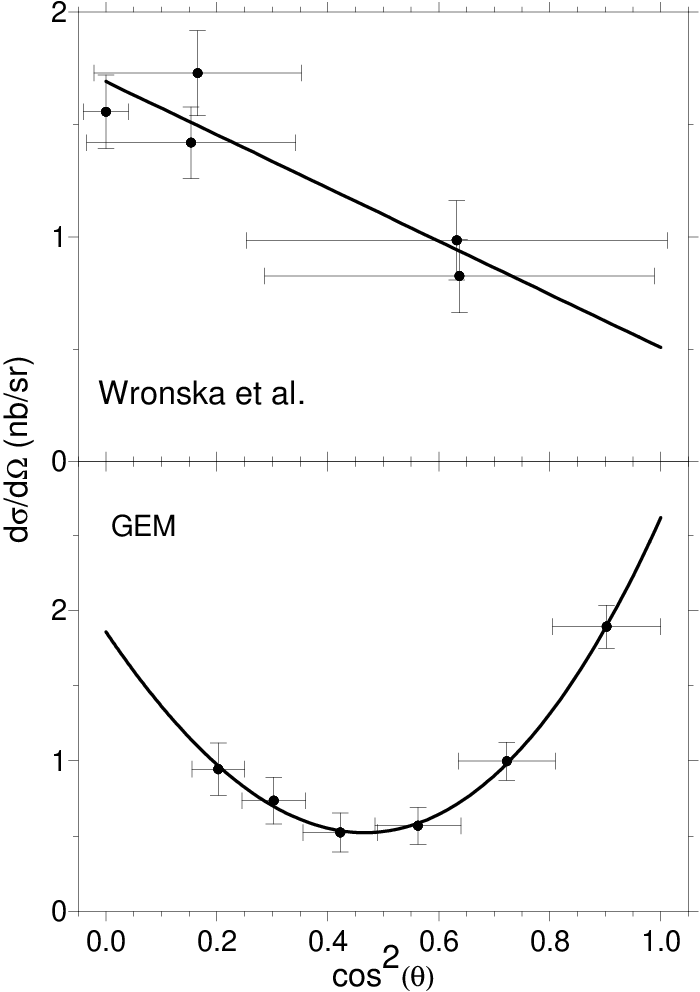}
\caption{Comparison of the angular distributions of the unpolarised differential cross sections from Ref. \cite{Wronska05} (ANKE collaboration, upper part) and Ref. \cite{Budzanowski09b} (GEM collaboration, lower part).  The data in the upper part were taken at a deuteron beam momentum of 2.358 GeV/c corresponding to an excess energy of $Q = 7.7\pm 0.8$ MeV, the data in the lower part at 2.385 GeV/c corresponding to $Q = 16.6$ MeV. The solid curves are Legendre polynomial fits up to $l=1$ and $l=2$, respectively.}
\label{Fig:Ma-Wro}
\end{center}
\end{figure}
From the fit with Legendre polynomials it becomes clear that partial waves up to $l=2$ contribute to the cross sections. From this fit, which is also shwon in Fig. \ref{Fig:Ma-Wro}, we obtain a total cross section which can be compared with the world data (see Fig. \ref{Fig:exfu_dd_ae}).
\begin{figure}[h]
\begin{center}
\includegraphics[width=0.6\textwidth]{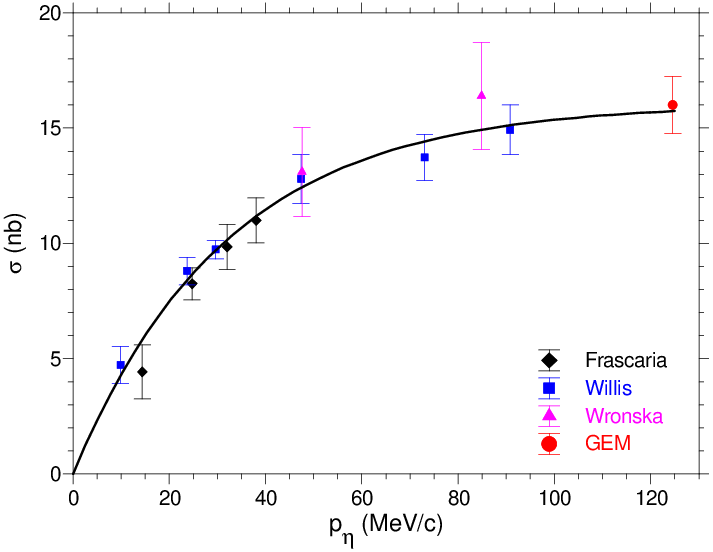}
\caption{Excitation function of the total cross section for the reaction $dd\to \eta \alpha$ close to threshold. The data points are from Frascaria et al. \cite{Frascaria94}, Willis et al. \cite{Willis97}, Wronska et al. \cite{Wronska05}, and the GEM collaboration Budzanowski et al. \cite{Budzanowski09b}. The solid line is to guide the eye.}
\label{Fig:exfu_dd_ae}
\end{center}
\end{figure}
The cross section rises almost linearly with momentum from threshold and seems to saturate above 100 MeV/c.

The GEM experiment Ref. \cite{Budzanowski09b} had full polar angle $\theta_\eta$ acceptance only in the range $(\pi-1)/2$ to $(\pi+1)/2$. Integration over this range leads to  $\int\cos \phi_\eta d\phi_\eta=-0.03$, i.e. vanishing of the term proportional to $\cos\phi_\eta$ and to -0.84 for the integral over $\cos 2\phi_\eta$. We can therefore integrate Eq. (\ref{equ:dsigma_rho_cartesian}) over this range without any efficiency term. This leads to
\begin{eqnarray}
& {} &
\int_{(\pi-1)/2}^{(\pi+1)/2}\left( \frac{d\sigma}{d\Omega} (\theta_\eta, \phi_\eta) \right)_{pol}d\phi_\eta \nonumber \\
& = & \left( \frac{d\sigma}{d\Omega} (\theta_\eta) \right)_{unpol}  \left[1 +
 \frac{1}{4}p_{zz}\left( A_{yy}\left(\theta_\eta\right)0.16 + A_{xx}\left(\theta_\eta\right)1.84\right) \right]
\nonumber \\
& \approx &  \left( \frac{d\sigma}{d\Omega} (\theta_\eta) \right)_{unpol}  \left[1 +
\frac{1.84}{4}p_{zz} A_{xx}(\theta_\eta) \right] .
\label{equ:dsigma_average}
\end{eqnarray}
So the polarized cross section depends practically only on $A_{xx}$.
\begin{figure}[ht]
\begin{center}
\includegraphics[width=0.6\textwidth]{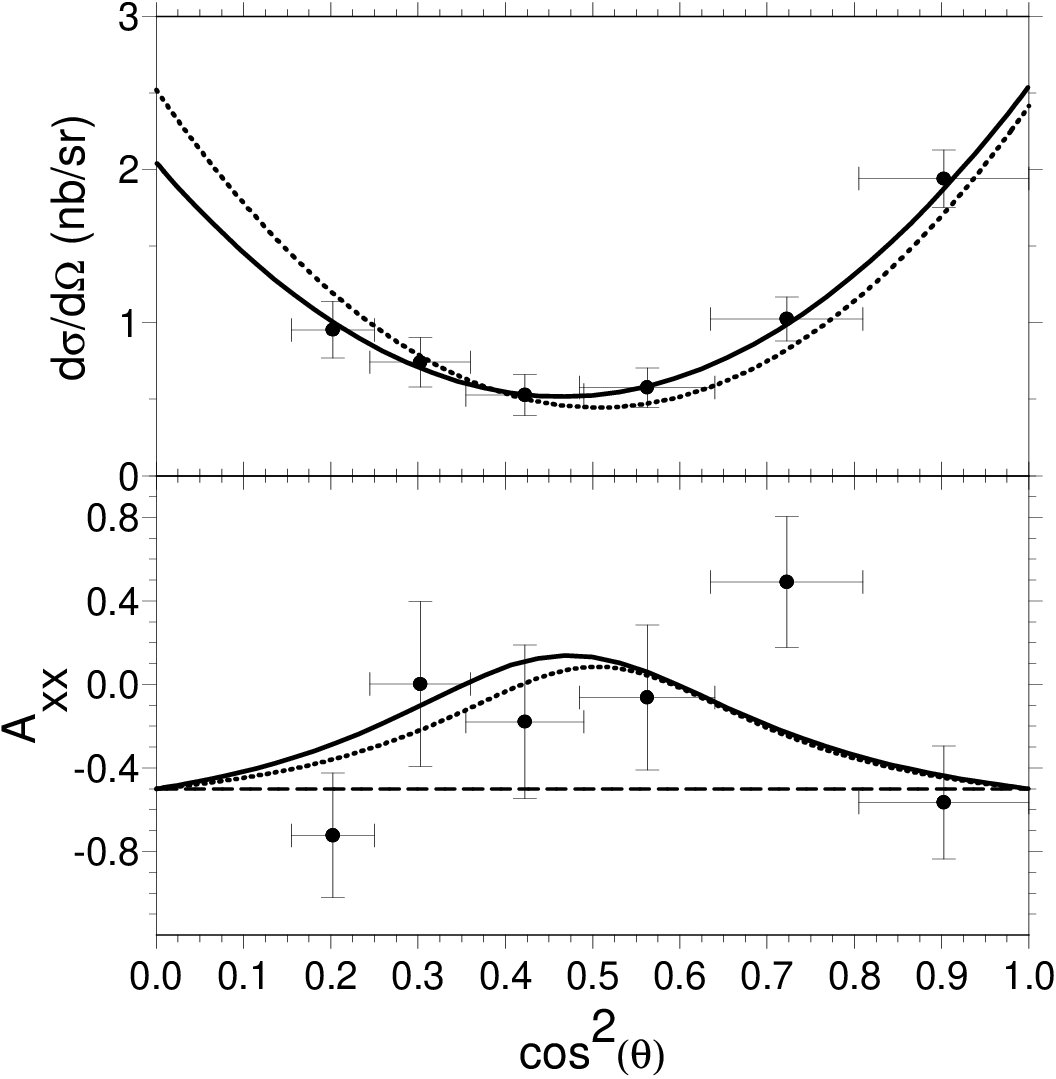}
\caption{Angular distributions of the unpolarised cross section and the analysing power $A_{xx}$ (from Ref. \cite{Budzanowski09b}). The solid
curves represent a fit with four partial waves with amplitudes $a_0-a_3$ (Table\ref{tab:dd-Amplitudes}); the dotted curves show fits with invariant amplitudes $A_0$, $A_2$, $B$ and $C$ Eqs. (\ref{INT}) and (\ref{equ:A0-A2}). The dashed line $A_{xx}=-1/2$ represents the case with $A=B=0$.}
\label{Fig:sig-A_xx}
\end{center}
\end{figure}
For the representation in tensor coordinates we have
\begin{eqnarray}
&{} &\int_{(\pi-1)/2}^{(\pi+1)/2}\left( \frac{d\sigma}{d\Omega} (\theta_\eta, \phi_\eta) \right)_{pol}d\phi_\eta \nonumber \\
& = & \left( \frac{d\sigma}{d\Omega} (\theta_\eta) \right)_{unpol}  \left[1 - \frac{1}{2}t_{20}T_{20}(\theta_\eta)+0.86\sqrt{\frac{3}{2}}t_{20}T_{22}(\theta_\eta)
 \right].
\label{equ:dsigma_average_tensor}
\end{eqnarray}

 The data could, therefore, \emph{in principle} fix the magnitudes of the amplitudes
$A_0$, $A_2$, $B$, and $C$, and the interference between $A_0$ and
$A_2$, while being completely insensitive to all of the other phases (see Eqs. \eqref{INT} to \eqref{equ:A0-A2}). Furthermore, the linear combination of Eqs.~(\ref{comb1}) and
(\ref{comb2}) shows that the fitting of $|B|$ and $|C|$ is decoupled
from that of $A_0$ and $A_2$. The parameters resulting from fitting the data in
this basis are given in Table~\ref{tab:MA_Fits}, with the fit curves
being shown in Fig.~\ref{Fig:sig-A_xx}. From this it is seen
that the amplitudes $A_0$ and $A_2$ are dominant, with  $C$ being consistent with
zero within error bars. If $|B|$ also vanishes, it would follow from
Eq.~(\ref{comb2}) that $A_{xx}=-\frac{1}{2}$ for all angles. On the
other hand, even the small contribution from the $B$ term changes the
angular dependence of $A_{xx}$, as is evident in
Fig.~\ref{Fig:sig-A_xx}.

\begin{table}[!h]
\caption{Fit results of invariant amplitudes squared of Eqs.~(\ref{comb1}) and
(\ref{comb2}) to the data of Ref. \cite{Budzanowski09b}. Since $|C|$ was found to be zero
within error bars, it was put exactly to
zero.\vspace{2mm}}\label{tab:MA_Fits}
\begin{center}
\begin{tabular}{|c|c|}
\hline
fit parameter &value (nb/sr) \\
\hline
$|A_0|^2$ & $\phantom{-1}6.6\pm1.7$ \\
$2\textit{Re}\,(A_0^*A_2)$ & $-25.0\pm9.5$ \\
$|A_2|^2$ & $\phantom{-1}48.4\pm14.5$ \\
$|B|^2$ & $\phantom{-1}9.3\pm5.1$ \\
$|C|^2$ & $\phantom{-1}0$ \\
\hline
\end{tabular}
\end{center}
\end{table}

The $s$ wave amplitude $f_s$ can now be extracted.
\begin{equation}\label{Eq:f_s}
\frac{d\sigma_s}{d\Omega}=\frac{p_\eta}{p_d}|f_s|^2 =
\frac{2p_\eta}{3p_d}|A_0|^2= \frac{1}{27} \frac{1}{4\pi}|a_0|^2.
\end{equation}
Using the values given in Table~\ref{tab:MA_Fits}, we find that
$|f_s|^2 = 4.4\pm 1.1$ nb/sr.

Very close to threshold only the $s$ wave contributes to the cross section. Then \begin{equation}\label{equ:threshold}
|f_s|^2= \frac{p_d}{p_\eta}\frac{\sigma}{4\pi}\,,
\end{equation}
is a good approximation. For the two highest energy measurements from Ref. \cite{Willis97} $d$ wave contributions have to be considered. It is a good approximation to assume the $d$ wave amplitudes $A_2$ and $B$ to depend on the $\eta$ momentum as $p_\eta^2$ and apply the results discussed here. This yields $|f_s|^2 = 13.8\pm 1.2$ nb/sr and $|f_s|^2 = 10.6\pm 1.3$ nb/sr for the momenta at 73 MeV/c and 91 MeV/c, respectively. For the Wronska result we find $|f_s|^2 = 14.3\pm 2.4$ nb/sr at 86 MeV/c.
We are now in a position to make a comparison of the world data for the $s$ wave amplitude.
\begin{figure}[h!]
\begin{center}
\includegraphics[width=0.6\textwidth]{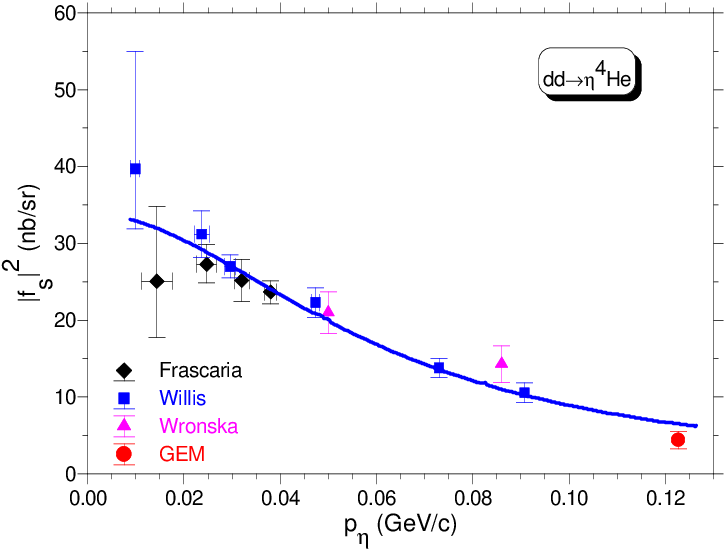}
\caption{The $s$ wave matrix element squared as function of $p_\eta$. The symbols have the same meaning as in Fig. \ref{Fig:exfu_dd_ae}. The data sources are the same as in Fig. \ref{Fig:exfu_dd_ae}. The solid curve is a fit with Eq. (\ref{equ:cross_section}) to the data.}
\label{Fig:exfu_f_s}
\end{center}
\end{figure}
This is done in Fig. \ref{Fig:exfu_f_s}. In a fit the production amplitude and the scattering length were fitted to the data yielding $a_{\eta\alpha}=[\pm (3.1\pm 0.5)+i(0.0\pm 0.5)]$ fm. This result corresponds to a bound state - if it exists - of $B_\eta=3.71\pm 0.09$ MeV and $\Gamma/2=0.0\pm 0.2$ MeV.

In Fig. \ref{Fig:Compare_he3-He4} we compare the excitation functions for the present reaction with the one for $d+p\to \eta+^3$He.
\begin{figure}[ht]
\begin{center}
\includegraphics[width=0.8\textwidth]{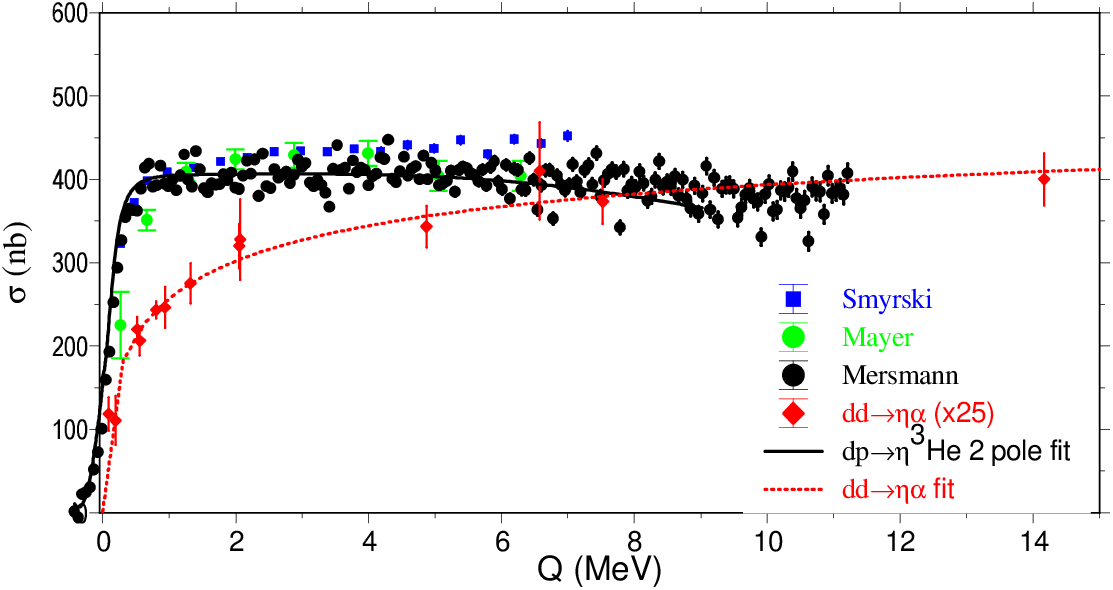}
\caption{Comparison of the excitation functions for the two reactions $d+p\to \eta+^3$He and $d+d\to \eta+^4$He. The total cross sections for the former reaction are from \cite{Mayer96} (diamonds), \cite{Smyrski07}(squares), and\cite{Mersmann07} (dots, five points together). The solid curve is the two pole fit. The data for the latter reaction (triangles up) are given in Fig. \ref{Fig:exfu_dd_ae}. The dashed curve is the scattering length fit to these data.}
\label{Fig:Compare_he3-He4}
\end{center}
\end{figure}
The latter reaction shows a much more rapid rise than the former.
This is an indication of the larger scattering length in case of the
lighter system.

\subsubsection{The Reaction $p+^6$Li$\to \eta+^7$Be}\label{sub:Reaction_6Li}

The next heavier system studied is the $^6$Li+$\eta$ system. In hadronic reactions it can only be reached via the $d+^4$He reaction. However, a $^4$He target is complicated to handle. Another possibility is the photoproduction on $^6$Li. Because of the small production cross section in photoproduction one needs a rather thick target and measures $\gamma+^6$Li$\to \eta+^6$Li$\to 2\gamma +^6$Li or to $6\gamma +^6$Li, depending in the decay $\eta\to 2\gamma$ or $\eta\to 3\pi^0$.

The reaction is dominated by the excitation of
the $S_{11}(1535)$ resonance via the $E{0+}$ multipole, which involves
a spin-flip of the participating nucleon. From this it follows that coherent $\eta$ production is practically forbidden for nuclei with spin $J = 0$ ground states. So candidates for coherent production where target nuclei with ground state spin $J$ and isospin $T$ different from zero. However, $^6\text{Li}$ has the same quantum numbers as the deuteron ($J=1$ and $T=0$) where the cross section was found to be very small \cite{Weiss01}. For the nucleus $^4\text{He}$ ($J=1$ and $T=0$) even only upper limits for the cross section have been extracted from the experiments \cite{Hejny99}. Therefore, experiments were performed on $^7$Li \cite{Maghrbi13}. The total cross section was measured to be below 20 mb. No unexpected threshold behaviour as in the case of the $^3$He was found. More details will be given in section \ref{sub:Photoproduction}.

Two experiments have been reported leading to the mirror nucleus $^7$Be. The experiments were performed at SATURNE
Saclay \cite{Scomparin93} and COSY J\"{u}lich \cite{Budzanowski10}. Both studies employed the reaction
\begin{equation}\label{equ:Be7}
p+{^6\text{Li}}\to \eta+{^7\text{Be}}\,.
\end{equation}
At Saclay the $\eta$ was measured through its two $\gamma$ decay at a beam energy of 683 MeV corresponding to a beam momentum of 1322 MeV/c or to an excess energy of $Q=19.13$ MeV.
\begin{figure}[h!]
\begin{center}
\includegraphics[width=0.6\textwidth]{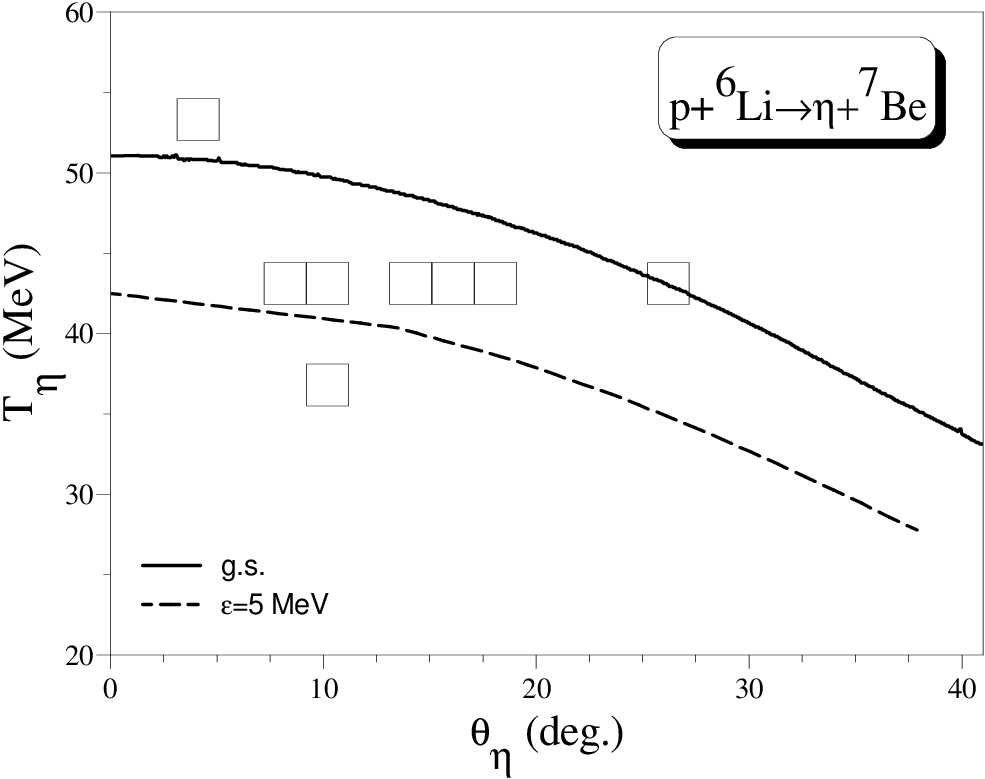}
\caption{Scatter plot of the measured events (open squares) for the indicated reaction as obtained at SATURNE \cite{Scomparin93}. The solid curve is the kinematical curve for the $^7$Be being in its ground state while the dashed curve is the one for a typical excitation of 5 MeV.}
\label{Fig:Scomparin}
\end{center}
\end{figure}
After applying all cuts eight events remain. They are shown in Fig. \ref{Fig:Scomparin} and are converted into a differential cross section of $d\sigma/d\Omega=4.6\pm 3.8$ nb/sr or into a total cross section of  $\sigma=57.8\pm 47.8$ nb. Also shown in the figure are kinematical curves for $^7$Be ground state and an excitation of 5~MeV. Clearly the events are scattered to the whole kinematical range. $^7$Be has in the range up to 6.73~MeV four states: two with $L=1$, a $J=3/2$ g.s. and a $J=1/2$ excited state at 0.429~MeV, and two with $L=3$, a $J=7/2$ at 4.57~MeV and a $J=5/2$ at 6.73~MeV. The exited states with $L=3$ are particle unstable. Al-Khalili et al. \cite{Al-Khalili93} have analysed these data. They assumed that the target nucleus consists of a deuteron and an $\alpha$ particle. The latter acts in the reaction as a mere spectator. The cross section is then
\begin{equation}\label{equ:Khalil}
\frac{d\sigma(p^6\text{Li}\to\eta ^7\text{Be})}{d\Omega}= \mathcal{C} \frac{p^*_\eta}{p^*_p}|f(pd\to\eta ^3\text{He})|^2
\sum_J  \frac{2J+1}{2}\mathcal{F}_j^2,
\end{equation}
with $J$ the total angular momentum of the final states in $^7$Be and $\mathcal{F}_j$ their form factors. $\mathcal{C}$ is the overlap of cluster wave functions, $p^*_\eta$ and $p^*_p$ the centre of mass momenta of the final and initial system, and $|f|$ the spin averaged matrix element of the underlying more elementary reaction $p+d\to ^3$He$+\eta$. Al-Khalili et al. derived the form factors from other reactions. Close to threshold they can be assumed to be constant.

The other experiment \cite{Budzanowski10} was performed at a beam energy of 673.1 MeV, corresponding to 1310 MeV/c momentum or an excess energy of $Q=11.28$ MeV. The recoiling $^7$Be nuclei were detected in the spectrograph Big Karl. Since the $L=3$ states are particle unstable only the two $L=1$ states contribute. The standard detectors in the focal plane were not adequate for this experiment since the recoiling particles have rather low energies of $\approx$100 MeV. The MWDC's were replaced by multi-wire avalanche-chambers to measure the track, followed by two layers of scintillation detectors one metre apart. They allow particle identification via TOF measurement. All these devices were housed in a large vacuum box made of stainless steel.
\begin{figure}[ht]
\begin{center}
\includegraphics[width=0.6\textwidth]{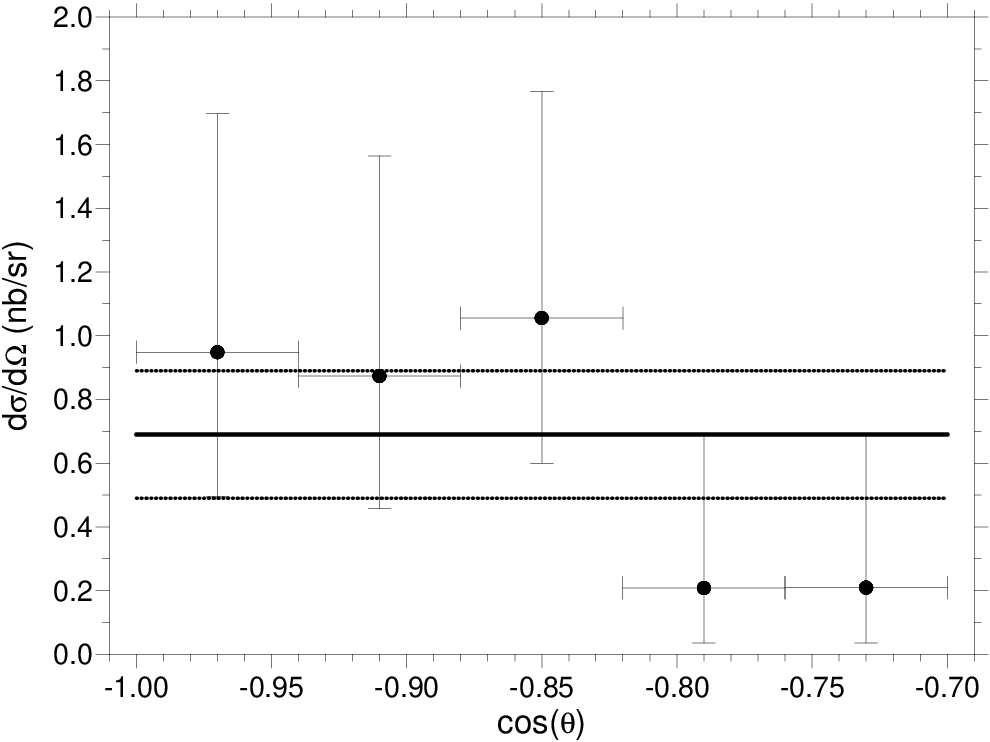}
\caption{Angular distribution of reconstructed $\eta$ mesons from the reactions $p+{^6\text{Li}}\to \eta + {^7\text{Be}}(E_x=0\text{MeV))}$ and $p+{^6\text{Li}}\to \eta + {^7\text{Be}}(E_x=0.429\text{MeV))}$\cite{Budzanowski10}. The horizontal line represent the isotropic cross section deduced with uncertainty.}
\label{Fig:Ang-dist}
\end{center}
\end{figure}
The $\eta$ meson events were identified via the missing mass technique. Finally the counts were converted to cross section. The angular distribution is shown in Fig. \ref{Fig:Ang-dist}. If the cross section is assumed to be isotropic
\begin{equation}
\frac{{d\sigma }}
{{d\Omega }} = (0.69 \pm 0.20 \text{( stat.)} \pm 0.20\text{ (syst.)}){\text{ nb/sr}}.
\end{equation}
was deduced and is also indicated in the figure with the statistical and systematical errors added in quadrature.

Together with the form factors from Ref. \cite{Al-Khalili93} and the
amplitude $f(pd\to\eta ^3\text{He})$ extracted from the two data sets discussed above, the cross section for the reaction leading to the $^7$Be ground state could be extracted.
\begin{figure}[h!]
\begin{center}
\includegraphics[width=0.8\textwidth]{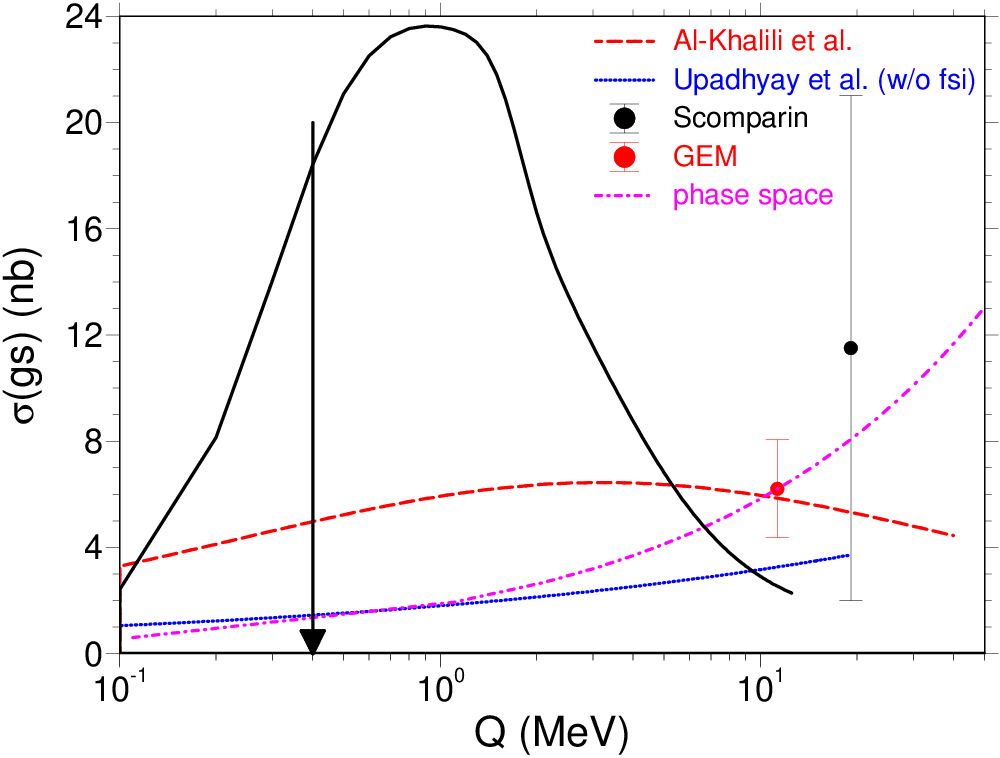}
\caption{Excitation function for the $p+^6\text{Li}\to\eta +
^7\text{Be(g.s.)}$ reaction. The two data points are from Refs.
\cite{Scomparin93} and \cite{Budzanowski10}. The dashed curve is the
Al-Khalili model and the dashed-dotted is the  phase space behaviour, both
normalised to the GEM data point. The solid and the dotted curves
are calculations \cite{Upadhyay09} with and without final state
interactions. The arrow indicates the region where only the ground state is involved in the reaction.}
\label{Fig:exfu_be7}
\end{center}
\end{figure}
The two data are shown in Fig. \ref{Fig:exfu_be7}. Also shown is the energy dependence of the Al-Khalili model normalised to the cross section of the GEM collaboration \cite{Budzanowski10}. Also the normalised phase space dependence is shown. In addition model predictions \cite{Upadhyay09} with and without \fsi are shown. A measurement even closer to threshold preferably below the first excited state could distinguish between the different models and could answer whether strong \fsi exists in this final channel. Upadhyay et al. \cite{Upadhyay09} got from $a_{\eta N} =(0.88+i0.41)$ fm  a value $a_{\eta ^7\text{Be}} = (-9.18+i8.53)$ fm.

\subsubsection{Photoproduction on Light Nuclei}
\label{sub:Photoproduction}
As is discussed above, $\eta$ mesons can also be produced by photons. In order to extract an $s$ wave contribution from an excitation function the final state has to be $\eta$ plus nucleus, i.e. $\gamma+A\to \eta+A$. Such a process is called coherent production in contrast to quasi-free production or inclusive production. For the measurement of coherent production the struck nucleon is measured in coincidence with the decay products of the $\eta$. In the latter case only the $\eta$ is measured irrespective of other reaction products.

Coherent $\eta$ production was studied so far only on {$^3$He} \cite{Pfeiffer04}, \cite{Pheron12} and on {$^7$Li} \cite{Maghrbi13}. In Ref. \cite{Pfeiffer04} production amplitudes for the two reactions $p+d\to\eta+{^3\text{He}}$ and $\gamma + {^3\text{He}}\to\eta+{^3\text{He}}$ were compared to each other with some modest agreement. Since then, newer and better data were published for both reactions. We therefore repeat this comparison making use of the new data. For the $p+d$ reaction we took the data from Ref. \cite{Mersmann07}  and for the photoproduction those of Ref. \cite{Pheron12} (see left panel in Fig. \ref{Fig:Gamma_p}). The cross sections of the photoproduction are already shown in Fig. \ref{Fig:Pheron}. In both cases we know that both are two body final states and we assume further $S$-wave production.
\begin{figure}[h!]
\begin{center}
\includegraphics[width=0.45\textwidth]{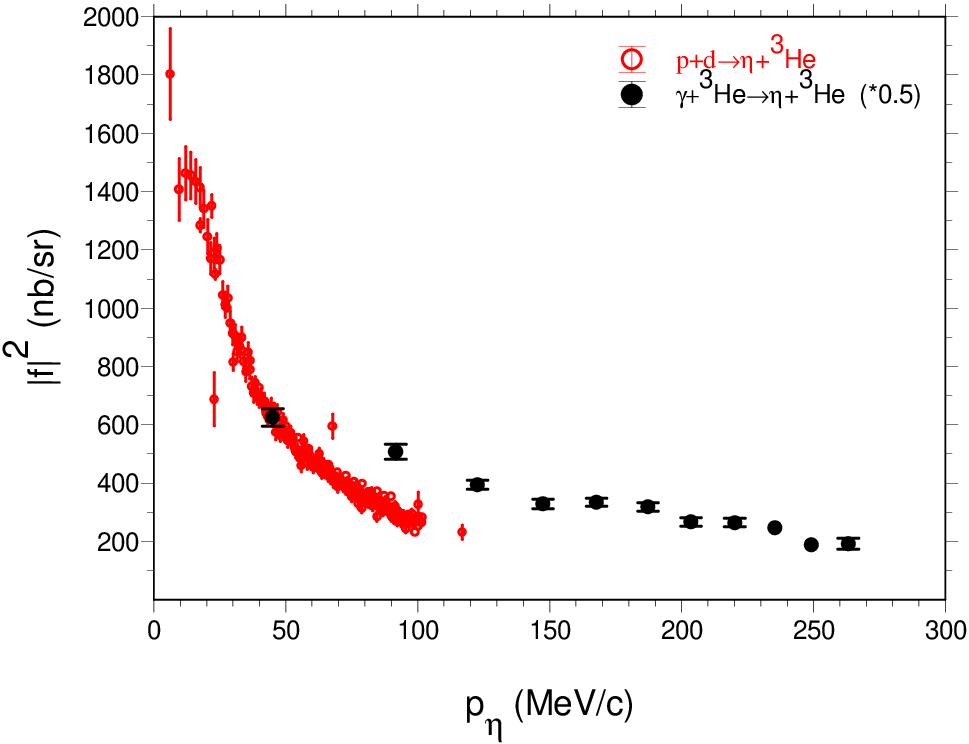}
\includegraphics[width=0.45\textwidth]{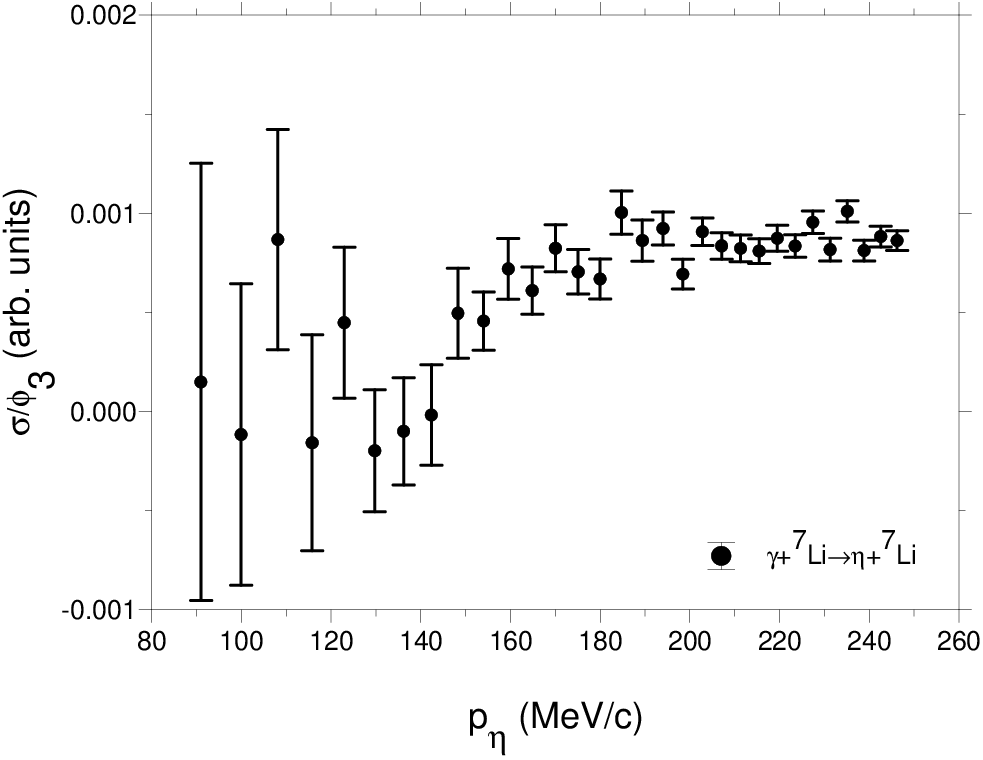}
\caption{Left panel: Comparison of the scattering amplitudes squared for the indicated reactions. The cross sections for the $p+d\to\eta+{^3\text{He}}$ are those given in Fig. \ref{Fig:Compare_he3-He4} and for $\gamma+{^3\text{He}}\to\eta+{^3\text{He}}$ those of \cite{Pheron12} have been used. The latter have been multiplied by 0.5 in order to reach the height of the previous data. Right panel: Same as left panel but for the $\gamma+{^7\text{Li}}\to\eta+{^7\text{Li}}$ reaction. Cross sections were taken from \cite{Maghrbi13}. }
\label{Fig:Gamma_p}
\end{center}
\end{figure}
The coherent  photoproduction data do not cover the range where the strong influence of the \fsi is visible, thus making this comparison not very useful.   It would now be natural to also compare the final {$\eta+^4$He} system from hadron and $\gamma$ induced reactions. As was stated earlier $\eta$ production via the $N^*$(1535) is a dominant channel. However, due to the quantum
numbers of {$^4$He} (S=0, I=0) photoproduction
involving the resonance will be strongly suppressed by angular momentum conservation. So coherent production is small compared to incoherent processes.

The next heavier system studied is coherent photoproduction on {$^7$Li}. We have extracted the scattering amplitude from data \cite{Maghrbi13}. It is also shown in the right frame of Fig. \ref{Fig:Gamma_p}. The momentum dependence of the amplitude squared is opposite to the one in the {$^3$He} case, i.e. it is rising while in the case of the lighter nucleus it is falling. Similar to the case of the {$^7$Be} final nucleus, there is also in {$^7$Li} a particle stable state at 0.478 MeV excitation energy with $J^\pi=1^-/2$ which can not be resolved from the ground state.

\section{Conclusions}\label{sec:Conclusions}

We have reviewed the theoretical as well as the experimental searches for (quasi) bound $\eta$ mesons in nuclei. Contrary to the pion, where the $\pi$ neutron strong interaction is repulsive and therefore also the $\pi$ nucleus interaction, the $\eta$ nucleon strong interaction is attractive. This led to the conjecture that $\eta$ mesons can be bound to nuclei. $\pi^-$ binding is only via the Coulomb force; they replace an electron. The orbit is outside the nuclear radius. On the contrary a bound $\eta$ which is uncharged will be embedded in the nuclear medium. It may change its properties like its mass in the medium. This is referred to as  chiral restoration.

The weak point in theoretical calculations is the input of the $\eta$ nucleon strong interaction in the form of the $\eta$ nucleon scattering length. Since it is impossible to measure elastic $\eta$ nucleon scattering, the derivation of the scattering length is based on production data $\pi+N\to\eta+N$ and $\gamma+N\to\eta+N$ and theory. This leads to a large range of the real and imaginary parts of the scattering length. However, the range below the production threshold is very sensitive to the scattering length and thus extremely model dependent. This threshold is very close to the pole position of the nucleon resonance $N^*_{11}(1535)$. Since this resonance is rather wide the $\eta$ production threshold overlaps with the resonance, the resonance can then play a dominant role in the existence of $\eta$ bound states.

On the experimental side different methods have been applied to search for such a state. Inclusive experiments searching for $\eta$-mesic
nuclei at BNL \cite{Chrien88} and LAMPF \cite{Lieb88}
by using a missing-mass technique in the ($\pi^+, p$)
reaction reached negative or inconclusive results.
Later it became clear that the peaks are not necessarily
narrow and that a better strategy of searching for
$\eta$-nuclei is required as for instance applied in
Ref. \cite{Budzanowski09}. Furthermore, the BNL experiment was in a region far from
the recoilless kinematics, so the cross section is
substantially reduced \cite{Hirenzaki07}.

Most spectacular example of the effective mass method was the
discovery of the $(J/\psi)$ meson: The effective mass of a pair of
particles emitted from the same point is obtained by measuring the
momentum of each of the particles $p_1$, and $p_2$, and the angles
$\theta_1$, and $\theta_2$, between their paths and the incident
beam direction, and by identifying the two particles simultaneously
so that their masses $m_1$ and $m_2$ can be determined. The
effective mass $m$ of the pair is defined by:
\begin{equation}
m=m_1^2+m_2^2+2[E_1E_2-p_1p_1\cos(\theta_1+\theta_2)]\,,
\end{equation}
where $E_i =$ total energy of the particle (from \cite{Ting76}). Such experiments searching for $\eta$ bound states were performed by the TAPS collaboration \cite{Pfeiffer04}, \cite{Pheron12} with different findings. Also a Lebedev-JINR group \cite{Sokol99} and \cite{Afanasiev13} performed this method. Only the work in the last citation shows some evidence for the existence of a quasi bound  state, although the weak significance raises some doubts. For the only experiment published, searching for the quasi bound eta mesic state the significance is even smaller.

One class of experiments produces the $\eta$-meson at rest in a quasi-free transfer reaction. In a second step the $\eta$ interacts with a nucleon thus forming a resonance $N^*(1535)$, which can decay back to its entrance channel or, with 50$\%$ probability, into a nucleon and a pion. Since the $\eta$ is at rest, these two final state particles are emitted almost back to back. The experiment by the GEM collaboration \cite{Budzanowski09} claimed a 5$\sigma$ effect in studying the $p+{^{27}\text{Al}}\to {^3\text{He}}+{^{25}\text{Mg}}\otimes\eta \to {^3\text{He}}+\pi^-+p+X$ reaction at a beam momentum for which the intermediate $\eta$ and then the quasi bound ${^{25}\text{Mg}}\otimes\eta$ is almost at rest.

Formation experiments give an upper limit for the existence of the bound state \cite{Adlarson13}. However, this limit does not exclude the existence of the state.

Another class of experiments searched for $\eta$-mesic nuclei
in final state interactions. Intensive studies were dedicated
esp. to the $p+d\rightarrow \eta+^3$He reaction
\cite{Smyrski07}, \cite{Mersmann07}, \cite{Mayer85}. It turns out that the new data do not fully agree with each other and that still not enough observables are known to extract the $s$ wave amplitudes. The situation is slightly better for the  $\eta^4$He final state, which has been studied in $d+d$ interactions making use of unpolarised
beams \cite{Wronska05}, \cite{Willis97}, \cite{Frascaria94} as well as
polarised beams \cite{Budzanowski09b}. The number of measured points is not sufficient yet to also extract the effective range with sufficient certainty from the data. The very large momentum
transfer tends to make direct production of $\eta$ mesons
more difficult with larger nuclei.
The heaviest system studied so far for final state interactions is  $\eta^7$Be produced in $p+^6$Li reactions
\cite{Scomparin93}, \cite{Budzanowski10}. In this case there are only two data points
at about 13 and 19 MeV above the threshold, so that no attempt for a final state interaction is successful yet.

With reasonable assumptions of the Watson-Migdal theory
\cite{Watson52}, \cite{Migdal55} final state studies can give
estimates for the imaginary value of the scattering length
and the absolute value of its real part \cite{Sibirtsev04a}.
However, the sign of the latter would be crucial as an
indication of a bound state.
Still, even $|a_r|$ could give indications of the
value of the binding energy, {\it provided it exists},
useful for experiments searching for such states.
Further useful information would be expectations of the
width of such states. This argumentation was brought further in \cite{Niskanen13}.

Search for decay products and reconstruction of the effective mass requires excellent position and energy resolution to identify the bound state. Transfer reactions in recoil free kinematics with additional measurement of the decay products seem to be the most favourable choice when the binding energy and width are deduced from the fast particle. Complex nuclear projectiles produce too much background, while nucleon transfer - although clean - suffers from small cross section. The reaction $\pi^++A\to n+(A-1)\otimes\eta$ seems favourable. Consider, for example  $A=\,^{12}$C. The recoil free beam momentum is 919 MeV/c which corresponds to an emerging neutron at zero degrees in the laboratory of 375 MeV. The total cross section for the elementary reaction $\pi^- p\to\eta n$ is $\approx$1~mb and probably with a flat angular distribution yielding $\approx 100\mu$b/sr \cite{Arndt05}. This has to be compared with backward $^3$He which has a differential cross section $d\sigma(pd\to\eta^3\text{He})/d\Omega<1$nb/sr (see \cite{Adlarson14}). Such an experiment, very much in the spirit of the one of Ref. \cite{Budzanowski09}, has been proposed \cite{Itahashi07}. The pion beam as well as other charged particles will be deflected by a dipole magnet behind the target, so the neutron can be detected at forward angles. The decay particles $\pi^-$ and $p$ will be detected by a magnetic cylindric detector set up surrounding the target.

In summary only one experiment has given so far a positive answer on the question whether a quasi bound $\eta$ nucleus system exists. The obtained binding energy and width is plotted in Figs. \ref{Fig:Predictions_BE} and \ref{Fig:Predictions_Gamma}.   The authors claim  a 5$\sigma$ evidence. However, the experiment should be repeated with higher statistics and on other target nuclei before a final decision on the existence of $\eta$ mesic nuclei can be made.

If the existence of $\eta$ mesic nuclei is confirmed one can think of even more exotica. The two nucleon transfer reaction $p+{^7\text{Li}}\to {^3\text{He}}+{^5\text{He}}\otimes\eta$  can lead to a Borromean system which is bound whereas the system ${^5\text{He}}$ is unbound. Scoccola and Riska \cite{Scocola98} have predicted a binding energy of 1 MeV for this system.

The activity in $\eta$ bound states has triggered searches for bound states of $\eta$'  \cite{Iwasaki09}, $\phi$ mesons \cite{Itahashi11}, and vector mesons \cite{Leupold10}, \cite{Metag12}. The situation with the $\bar{K}N$ interaction is similar to the $\eta N$ situation: it is strong and also an $s$ wave resonance $\Lambda(1405)$ plays a role. Recent reviews are \cite{Iwasaki09} and \cite{Gal13}. Also the binding of heavy quarkonium  $\eta_C$ to light nuclei was predicted \cite{Brodsky90}, \cite{Wasson91}, \cite{Luke92}, \cite{Teramond98} and \cite{Belyaev06}. Even the binding of the charmed $D^-$, $D^0$, and $\bar{D^0}$ mesons in {$^{208}$Pb} was proposed \cite{Tsushima99}.

\section{Acknowledgement}
Conversations with A. Gal, F. Hinterberger, K. Kilian, S. S. Kapoor,  V. Jha, J. Lieb, Lon-chan Liu, A. Magiera, P. Moskal, J. Niskanen, J. Ritman, and C. Wilkin are gratefully acknowledged. A. Magiera supplied a code calculating Clebsch Gordan coefficients for different partial waves. Finally, I would like to
acknowledge the hospitality of Institute of Nuclear Physics, Research Centre J\"{u}lich,  where part of this article was written.

\clearpage

\end{document}